\documentclass[twocolumn,resetfootnote]{aastex701}

\usepackage{etoolbox}
\patchcmd{\figure}{\setcounter{subfigure}{0}}{}{}{}

\usepackage{mathtools,amsmath} 
\usepackage{graphicx}
\usepackage{chngcntr} 
\usepackage{float}
\usepackage{xcolor}
\usepackage{txfonts}
\usepackage{caption}
\usepackage{subcaption}
\usepackage{tabularx} 
\usepackage{array}   
\usepackage{comment}
\usepackage{chngcntr} 
\counterwithin{figure}{section}
\counterwithin{table}{section}
\usepackage{float}
\usepackage{soul} 
\usepackage{booktabs}  
\usepackage{csvsimple}
\usepackage{xspace}
\usepackage{subcaption}

\defcitealias{ricci_tight_2023}{R23} 
\defcitealias{gupta_bass_2024}{G24} 

\usepackage{chngcntr}
\counterwithout{figure}{section}

\usepackage{chngcntr}
\counterwithout{figure}{section}
\counterwithout{table}{section}
\usepackage[labelfont=bf]{caption}

\begin{document}

\title{Nuclear Millimeter Continuum Emission in Radio-Quiet Active Galactic Nuclei: an ALMA Spectral and Variability Study}

\author[0009-0001-8342-7522]{Sophie M. Venselaar}
\affiliation{Department of Astronomy, University of Geneva, ch. d’Ecogia 16, 1290 Versoix, Switzerland}
\affiliation{Instituto de Estudios Astrofísicos, Facultad de Ingeniería y Ciencias, Universidad Diego Portales, Av. Ejército Libertador 441, Santiago, Chile}
\email[show]{\href{mailto:sophie.venselaar@unige.ch}{sophie.venselaar@unige.ch}}

\author[0000-0001-5231-2645]{Claudio Ricci}
\affiliation{Department of Astronomy, University of Geneva, ch. d’Ecogia 16, 1290 Versoix, Switzerland}
\affiliation{Kavli Institute for Astronomy and Astrophysics, Peking University, Beijing 100871, People’s Republic of China}
\affiliation{Instituto de Estudios Astrofísicos, Facultad de Ingeniería y Ciencias, Universidad Diego Portales, Av. Ejército Libertador 441, Santiago, Chile}
\email{..}

\author[0000-0002-7272-1136]{Yoshiyuki Inoue}
\affiliation{College of Systems Engineering and Science, Shibaura Institute of Technology, 307 Fukasaku, Minuma, Saitama 337-8570, Japan}
\affiliation{Interdisciplinary Theoretical \& Mathematical Science Program (iTHEMS), RIKEN, 2-1 Hirosawa, Wako, Saitama 351-0198, Japan}
\affiliation{Kavli Institute for the Physics and Mathematics of the Universe (WPI), UTIAS, The University of Tokyo, 5-1-5 Kashiwanoha, Kashiwa, Chiba 277-8583, Japan}
\email{..}

\author[0000-0001-9910-3234]{Chin-Shin Chang}
\affiliation{Department of Astronomy, University of Geneva, Chemin Pegasi 51, 1290 Versoix, Switzerland}
\email{..}

\author[0000-0003-4384-9568]{Akihiro Doi}
\affiliation{Institute of Space and Astronautical Science (ISAS), Japan Aerospace Exploration Agency (JAXA), 3-1-1 Yoshinodai, Chuo-ku, Sagamihara, Kanagawa 252-5210, Japan}
\affiliation{Department of Space and Astronautical Science, SOKENDAI, 3-1-1 Yoshinodai, Chuou-ku, Sagamihara, Kanagawa 252-5210, Japan}
\email{..}

\author[0000-0002-4180-174X]{Noemi Barnier}
\affiliation{Theoretical Astrophysics, Department of Earth and Space Science, Graduate School of Science, Osaka University, Toyonaka, Osaka 560-0043, Japan}
\affiliation{International Center for Hadron Astrophysics, Chiba University, Inage-ku, Chiba 263-8522, Japan}
\email{..}

\author[0000-0002-6808-2052]{Taiki Kawamuro}
\affiliation{Department of Earth and Space Science, Osaka University, 1-1 Machikaneyama, Toyonaka 560-0043, Osaka, Japan}
\email{..}

\author[0000-0002-5828-7660]{Susanne Aalto}
\affiliation{Department of Physics and Astronomy, Chalmers University of
Technology, 412 96 Gothenburg, Sweden}
\email{..}

\author[0009-0005-4575-7087]{Momoka Hamamoto}
\affiliation{Theoretical Astrophysics, Department of Earth and Space Science, Graduate School of Science, Osaka University, Toyonaka, Osaka 560-0043, Japan}
\email{..}

\author[0000-0003-2914-2507]{Elena Shablovinskaya}
\affiliation{Max-Planck-Institut für Radioastronomie, Auf dem Hügel 69, Bonn D-53121, Germany}
\email{..}

\author[0000-0001-7568-6412]{Ezequiel Treister}
\affiliation{Instituto de Alta Investigaci\'on, Universidad de Tarapac\'a, Casilla 7D, Arica, Chile}
\email{..}

\date{Received August 17, 2026; Accepted October 21, 2026}

\begin{abstract}

Unresolved millimeter emission is observed in radio-quiet (RQ) Active Galactic Nuclei (AGN), but its physical origin remains debated. 
Self-absorbed synchrotron emission from the compact X-ray corona is one of the leading candidates. 
Previous support for this coronal synchrotron origin has mainly been based on variability and spectral indices in a few RQ AGN, with a lack of (quasi-)simultaneous broadband data for larger samples. 
We present Atacama Large Millimeter/submillimeter Array (ALMA) observations of nine nearby ($z \sim 0.008$–0.064) RQ AGN, obtained quasi-simultaneously at 100, 145, and 233\,GHz. 
Based on power-law fits to the spectral energy distributions (SEDs), we measure millimeter spectral indices ranging from $-0.05^{+0.09}_{-0.09}$ to $0.88^{+0.12}_{-0.11}$.
Furthermore, across two 100\,GHz observing epochs per source, separated by 4--11\,days, we detect flux density variability of 30\%--56\% ($>3\sigma$ significance level) in three out of nine sources.
Spectral slopes alone are insufficient to constrain the origin of the millimeter emission; however, the observed short-timescale variability and inferred brightness temperature values are consistent with a nonthermal emission mechanism arising from a compact emitting region with a size of $R_{\rm mm}< 10^3$--$10^4$\,$R_{\rm g}$, such as the corona or a compact jet. 
Furthermore, we find a clear correlation between the 100/233\,GHz and X-ray flux in our sample, and establish a new linear correlation between 145\,GHz and X-ray emission.

\end{abstract}

\keywords{
\href{https://vocabs.ardc.edu.au/repository/api/lda/aas/the-unified-astronomy-thesaurus/current/resource.html?uri=http://astrothesaurus.org/uat/16}{Active galactic nuclei (16)};
\href{https://vocabs.ardc.edu.au/repository/api/lda/aas/the-unified-astronomy-thesaurus/current/resource.html?uri=http://astrothesaurus.org/uat/1663}{Supermassive black holes (1663)};
\href{https://vocabs.ardc.edu.au/repository/api/lda/aas/the-unified-astronomy-thesaurus/current/resource.html?uri=http://astrothesaurus.org/uat/1647}{Submillimeter astronomy (1647)}}

\section{Introduction}\label{Intro}

A key distinction within the Active Galactic Nuclei (AGN) population, objects powered by accretion onto supermassive black holes (SMBHs) and emitting across the electromagnetic spectrum \citep{Kellermann1989}, is between radio-loud (RL) and radio-quiet (RQ) systems. 
RQ AGN make up the vast majority of the AGN population in the local Universe \citep{Kellermann1989,Kellermann2016}. 
Both classes are generally understood to share a similar central structure, including an accretion disk that produces optical and ultraviolet (UV) emission, a compact hot corona responsible for X-ray radiation, and an obscuring dusty torus that reprocesses a significant fraction of the disk emission into the infrared (IR) band \citep[e.g.,][]{ramos_almeida_nuclear_2017,Ricci2026}. 
The division is typically defined by the presence or absence of powerful relativistic jets and extended radio structures \citep{Kellermann1989}.
When normalized by their optical luminosity, RQ AGN are generally three orders of magnitude fainter in radio power than their RL counterparts \citep{Kellermann1989}.
However, this does not imply that radio emission is entirely absent; radio-emitting parsec-scale jets can still be present in RQ AGN, though substantially weaker and less extended than those typically observed in RL systems \citep[e.g.,][]{panessa_origin_2019,Magno2025}. 
Despite this relative faintness in the radio band, RQ AGN systematically exhibit an excess of compact nuclear millimeter (mm) emission on spatial scales smaller than a few tens of parsecs \citep[e.g.,][]{Doi2011,behar_discovery_2015,behar_mm-wave_2018,panessa_origin_2019,kawamuro_bass_2022,ricci_tight_2023,Venselaar26}.

A promising origin for this unresolved millimeter emission is the X-ray corona, a region of hot, magnetized electron plasma located in the immediate vicinity of the central SMBH \citep[e.g.,][]{di_matteo_magnetic_1998, laor_origin_2008,inoue_unveiling_2014,raginski_agn_2016,inoue_detection_2018,Jana2025}.  
Microlensing and X-ray reverberation studies have indicated that the X-ray corona is highly compact, with a characteristic size of $R_{\rm c} <20$--30$\,R_{\rm g}$ \citep[e.g.,][]{Chartas2009,Reis2013,Wilkins2016}, where the gravitational radius is defined as $R_{\rm g} = GM_{\rm BH}/c^2$.
In the X-ray corona, magnetic reconnection has been proposed as a key heating mechanism \citep[e.g.,][]{di_matteo_magnetic_1998, Liu2002,Sridhar2021, Sridhar2023, Liu2024}. This would enable electrons to reach temperatures at which they can Comptonize optical/UV photons from the accretion disk via inverse Compton scattering, which produces the observed X-ray emission \citep[e.g.,][]{katz_nonrelativistic_1976, Haardt1991}.
Given the high electron densities and presumably strong magnetic fields in this region ($B=10$--60\,G; \citealp{del_palacio_millimeter_2025}), coronal electrons could emit synchrotron radiation observable in the millimeter regime \citep[e.g.,][]{laor_origin_2008, inoue_unveiling_2014}.

A key observational signature of coronal emission is synchrotron self-absorption (SSA).
In the optically thin regime, synchrotron emission follows a power-law spectrum ($S_{\nu} \propto \nu^{-\alpha}$) with a spectral index of $\alpha \sim 0.5$--1.0. 
This spectral slope is directly related to the energy distribution of the emitting electrons, $N(E)\propto E^{-p}$, through $p=2\alpha+1$ \citep[see e.g.,][]{RybickiLightman86}, where $p$ depends on the underlying electron acceleration mechanism.
At lower frequencies, however, the corona becomes optically thick due to SSA, producing a steeply rising spectrum with $\alpha \sim -2.5$, independent of $p$.
In the spectrum, the transition between the optically thick and thin regimes manifests itself as an SSA turnover or millimeter excess at $\nu_{\rm SSA}$.
The exact location of $\nu_{\rm SSA}$ has been modeled, and in a few cases observed, to be near $\nu_{\rm SSA} \sim 95$--300\,GHz \citep[e.g.,][]{raginski_agn_2016,inoue_detection_2018,Inoue2020, Michiyama2023, michiyama_alma_2024,shablovinskaya_joint_2024,del_palacio_millimeter_2025}.
The exact location of the spectral turnover depends on the size ($R$) of the millimeter-emitting region, the mass of the central SMBH ($M_{\rm BH}$) and the magnetic field strength ($B$), which can both be inferred through spectral modeling \citep[e.g.,][]{inoue_detection_2018,michiyama_alma_2024,del_palacio_millimeter_2025}. 
Therefore, observing this SSA turnover could provide a direct diagnostic of some of the fundamental properties of the coronal plasma.

The first direct evidence for this millimeter peak was reported by \cite{inoue_detection_2018}, who observed the two nearby RQ AGN IC\,4329A and NGC\,985 using observations from the Atacama Large Millimeter/submillimeter Array (ALMA) covering 90--250\,GHz, and at subarcsecond resolution. 
By combining these data with archival centimeter (cm) observations from the Karl G. Jansky Very Large Array (VLA) and the Australia Telescope Compact Array (ATCA), they constructed broadband radio-to-submillimeter spectral energy distributions (SEDs). 
A clear spectral peak was detected in both sources, while SED modeling with a hybrid thermal/nonthermal coronal model yielded coronal sizes of $R\sim80R_{\rm g}$ and magnetic field strengths ($B$) of $\sim 10$\,G.
These $B$ values were weaker than predicted for a purely magnetically reconnecting corona \citep{Liu2002,inoue_detection_2018}, but provided strong evidence for a nonthermal electron population contributing to the observed millimeter emission.

The corona may contain two electron populations, thermal and nonthermal \citep[e.g.,][]{Ghisellini1998, InoueTotaniUeda2008,Fabian2017}, with millimeter emission dominated by nonthermal electrons \citep[e.g.,][]{inoue_unveiling_2014} and X-rays produced by thermal electrons.
Recent evidence for neutrino emission from X-ray-bright Seyferts \citep{IceCube2022,Neronov2024,Abbasi2026,Kun2026} might also support the existence of such a nonthermal particle population within the compact corona. 
While neutrinos are the result of hadronic processes, specifically proton--proton ($pp$) or proton--photon ($p\gamma$) interactions, their detection could suggest the existence of an efficient acceleration mechanism that could also affect electrons \citep[e.g.,][]{inoue_high-energy_2019,Inoue2020}.

Additional constraints on the size of the millimeter-emitting region in RQ AGN come from millimeter variability. 
Rapid flux variability on timescales of $\leq$\,days has been reported in several objects \citep[e.g.,][]{baldi_milimetre-band_2015,behar_simultaneous_2020,petrucci_simultaneous_2023,michiyama_alma_2024,shablovinskaya_joint_2024}, indicating extremely compact millimeter-emitting regions with sizes $R\leq10$--100$R_{\rm g}$, i.e.\ on the order of light-days. 
Complementary, geometrical constraints on the millimeter-emitting region from gravitational microlensing provide independent evidence for this compact scale. 
By analyzing flux-ratio anomalies in multi-epoch ALMA observations of the lensed quasar RX\,J1131--1231, \cite{rybak_detection_2025} placed an upper limit on the millimeter-emitting half-light radius of $R\leq90R_{\rm g}\sim2.4\times10^{-4}$\,pc. 
These size constraints from both variability and microlensing effectively rule out an origin in larger-scale structures such as the dusty torus or diffuse star formation, confirming that the millimeter excess originates from a corona-scale region.

Additionally, an origin in a compact jet, rather than the corona, offers an equally plausible explanation for the millimeter emission in RQ AGN. 
Very Long Baseline Interferometry observations have established that subparsec radio cores are common in RQ AGN \citep[e.g.,][]{Giroletti2009,Panessa2013,Smith2016,Smith2020}, and more recent, higher-resolution Very Long Baseline Array (VLBA) campaigns have resolved several of these into core-jet structures \citep[e.g.,][]{Wang2023I,Wang2023II,Chen2023,Chen2025}, demonstrating that jet-like radio emission is not confined to RL AGN.
However, the radio cores in these samples typically exhibit flat spectra between 1--24 GHz \citep{Chen2023}, a signature expected for SSA emission from a stratified compact jet \citep{BlandfordKonigl1979}, but one that is equally consistent with self-absorbed coronal emission \citep[e.g.,][]{Chen2023}. 
Existing centimeter-wavelength VLBA data therefore cannot distinguish between a jet and a coronal origin, motivating the need for spectral coverage that extends closer to the predicted coronal turnover frequency.

However, it remains unclear how ubiquitous coronal millimeter emission is across the RQ AGN population, and whether the SSA spectral signature is a common feature or confined to a particular subset of sources. 
Systematic studies that combine broad millimeter frequency coverage, subarcsecond resolution, and multi-epoch monitoring are needed to disentangle coronal emission from star formation and potentially jet emission, and to constrain physical conditions such as size, magnetic field, and electron population.
Since millimeter emission is far less absorbed than X-rays, penetrating column densities up to $\log (N_{\rm H}/\rm cm^{-2})\sim27$ \citep[e.g.,][]{hildebrand_determination_1983} compared to $\log (N_{\rm H}/\rm cm^{-2})\sim24$ for X-rays \citep[e.g.,][]{ricci_compton-thick_2015}, it also offers a promising probe of the innermost regions in heavily obscured RQ AGN. This is enabled by the millimeter/X-ray relation observed across larger samples of RQ AGN \citep[e.g.,][]{behar_discovery_2015,kawamuro_bass_2022,ricci_tight_2023,Venselaar26}. 
Therefore, confirming the coronal origin of the millimeter emission would establish an important new window into highly obscured systems.

In this work, we present ALMA observations of nine nearby RQ AGN ($z\sim0.008$--0.064) to assess the presence of coronal magnetic activity across the sample. 
Using quasi-simultaneous 100\,GHz, 145\,GHz, and 233\,GHz observations (ALMA Bands\,3, 4, and 6), we construct broadband millimeter SEDs and constrain the nonthermal origin of the emission through spectral index analysis and brightness temperature estimates. 
Two epochs of 100\,GHz observations further allow us to probe variability and place upper limits on the physical size of the millimeter-emitting region.
This manuscript is structured as follows: in Section\,\ref{Sample} we present our AGN sample, and describe the ALMA observations.
Section\,\ref{Data} describes the imaging procedure, including self-calibration, as well as the detection of molecular emission in one of our sources.
Section\,\ref{Results} presents the computation of the SEDs, which allow for the determination of the spectral index. Furthermore, we discuss the variability observed in three sources at 100\,GHz.
In Section\,\ref{Discussion}, we discuss the origin of the millimeter emission based on the derived spectral indices, brightness temperatures and variability and constrain the size of the millimeter-emitting region. Furthermore, we discuss the differences between the intra- and broadband spectral indices and the location of our sources in the previously established millimeter/X-ray luminosity relations.
Finally, Section\,\ref{Conclusion} summarizes our conclusions.
Throughout the manuscript, we adopt standard cosmological parameters ($ H_0 = 70\ \rm km\,s^{-1}\,Mpc^{-1}$, $\Omega_{\rm m}=0.3$, $\Omega_{\Lambda}=0.7$), and all uncertainties presented are $1\sigma$ unless stated otherwise.

\begin{table*}[ht!]
\setlength{\tabcolsep}{18.5pt}
\centering
\caption{Source sample, parameters, and ALMA observation dates}
\begin{tabularx}{\textwidth}{ll|cccc}
\hline
\hline
(1) & (2) & (3) & (4) & (5) & (6) \\ 
Source            & Swift ID & $z$ & Observation   & Observation   & $\log(R_{\rm X})$ \\ 
                  &          &     & dates epoch 1 & dates epoch 2 &                   \\
\hline
\hline
Ark\,120          &SWIFT\,J0516.2--0009 & 0.033 & 12--10--2018 & 23--10--2018 & $-$5.36 \\
AX\,J1737.4--2907 &SWIFT\,J1737.5--2908 & 0.021 & 14--08--2019 & 18--08--2019 & $-$5.11 \\
IC\,4329A         &SWIFT\,J1349.3--3018 & 0.016 & 13--08--2019 & 18--08--2019 & $-$5.21 \\
IRAS\,05078+1626  &SWIFT\,J0510.7$+$1629& 0.018 & 12--08--2019 & 18--08--2019 & $-$6.01$\dagger$ \\
MR\,2251--178     &SWIFT\,J2254.1--1734 & 0.064 & 12--10--2018 & 22--10--2018 & $-$5.57 \\
Mrk\,509          &SWIFT\,J2044.2--1045 & 0.034 & 12--10--2018 & 22--10--2018 & $-$5.76 \\
Mrk\,926          &SWIFT\,J2304.8--0843 & 0.047 & 12--10--2018 & 22--10--2018 & $-$4.97 \\
NGC\,3783         &SWIFT\,J1139.0--3743 & 0.010 & 13--08--2019 & 18--08--2019 & $-$5.44 \\
NGC\,4593         &SWIFT\,J1239.6--0519 & 0.008 & 13--08--2019 & 18--08--2019 & $-$6.50 \\
\hline
\hline
\end{tabularx}
\par\vspace{1ex}
\parbox{1\textwidth}{\small
\textbf{Note:} 
(1) Source name, 
(2) Swift ID, 
(3) Redshift, 
(4) Observation date for the 100\,GHz epoch\,1, 145\,GHz, and 233\,GHz data, 
(5) Observation date for the 100\,GHz epoch\,2 data,  
(6) Value of the radio-loudness parameter based on the definition by \cite{Terashima2003}.
\\
$\dagger$ For IRAS\,05078+1626, the value of $\log(R_{\rm X})$ was determined based on the definition by \cite{teng_fermilat_2011}.
}
\label{tab:Sources1}
\end{table*}

\begin{table*}[ht!]
\setlength{\tabcolsep}{11.5pt}
\centering
\caption{Source sample and parameters}
\begin{tabularx}{\textwidth}{l|ccccccc}
\hline
\hline
(1) & (2) & (3) & (4) & (5) & (6) & (7) & (8)  \\ 
Source            &  $\log(M_{\rm BH})$ & $\log (L_{\rm bol})$ &  $\log(\lambda_{\rm Edd})$&$\log(N_{\rm H})$ &  $\rm SFR$ & $\log (L_{\rm 2-10keV})$ & $\log (L_{\rm 14-150keV})$ \\ 
                 &[$M_{\odot}$]&[$\rm erg\,s^{-1}$] & & [$\rm cm^{-2}$]&[$M_{\odot}\rm \,yr^{-1}$] & [erg\,s$^{-1}$] & [erg\,s$^{-1}$]\\
\hline
\hline
Ark\,120          & 8.07       &45.06 &$-$1.18 & 20.00   & 5.65 & 43.8 &  44.2\\
AX\,J1737.4--2907 & 7.84 &44.95 &$-$1.07 & $<$21.92& $<$1.24& 43.9& 44.0\\
IC\,4329A         & 7.65   &45.06 &$-$0.77 & 21.52   & 1.36 & 43.8& 44.2\\
IRAS\,05078+1626  & 6.88      &44.64 &$-$0.41 & 21.08   & 1.26 & 43.5 & 43.8\\
MR\,2251--178     & 8.19   &45.85 &$-$0.52 & 20.00   & 10.59& 44.6& 44.9\\
Mrk\,509          & 8.05       &45.27 &$-$0.96 & 20.00   & 10.76& 44.1& 44.4\\
Mrk\,926          & 7.98   &45.62 &$-$0.54 & 20.00   & 8.35 & 44.2& 44.7\\
NGC\,3783         & 7.37      &44.35 &$-$1.20 & 20.49   & 1.26 & 43.4 & 43.5 \\
NGC\,4593         & 6.88      &44.00 &$-$1.05 & 20.00   & 2.52 & 42.8& 43.1 \\
\hline
\hline
\end{tabularx}
\par\vspace{1ex}
\parbox{1\textwidth}{\small
\textbf{Note:} 
(1) Source name, 
(2) Black hole mass, with uncertainties of 0.4\,dex, from \cite{koss_bass_2022}, 
(3) Bolometric luminosity from \cite{koss_bass_2022}, 
(4) Eddington ratio from \cite{koss_bass_2022}, 
(5) Column density from \cite{ricci_bat_2017}, 
(6) Star formation rate from \cite{ichikawa_complete_2017,ichikawa_bat_2019}, 
(7) Averaged 2--10\,keV X-ray luminosity from \cite{ricci_bat_2017}, and 
(8) Averaged 14--150\,keV X-ray luminosity from \cite{ricci_bat_2017}.} 
\label{tab:Sources2}
\end{table*}

\section{Sample and Observations}\label{Sample}

\subsection{Sample selection}\label{Samplesel}

The RQ AGN studied in this work ($z=0.008$--0.064), were drawn from the 70-month, all-sky hard X-ray survey conducted with the Burst Alert Telescope\footnote{The full catalog and data products are publicly available: \url{https://swift.gsfc.nasa.gov/results/bs70mon/} \citep{baumgartner_70_2013}.} (BAT) on board the Neil Gehrels Swift Observatory \citep[Swift;][]{Gehrels2004}, covering $E\sim14$--195\,keV.
Our nine sources were selected according to four criteria.
First, sources were required to be unobscured, enabling precise determination of coronal properties from X-ray data.
Second, all sources lie at declination $<20$\,deg, ensuring observability by ALMA.
Third and fourth, the expected coronal millimeter fluxes must exceed potential contamination from both thermal dust emission and nonthermal jet or starburst activity.
Expected 100\,GHz, 145\,GHz and 233\,GHz coronal fluxes were originally estimated by scaling from the X-ray/radio flux ratio observed in IC\,4329A from \cite{inoue_detection_2018}. 
Dust contamination was assessed assuming an average dust temperature\footnote{Dust in the interstellar medium at temperatures of $\sim10$--30\,K is expected to peak at submillimeter frequencies \citep[e.g.,][]{Bracco2011}.} of 30\,K and an emissivity index of 1.5 at angular scales $<0.13$\,arcsec, yielding expected dust flux densities of $\leq1$\,mJy. 
Jet and starburst contamination was estimated by extrapolating archival centimeter radio flux densities from the NASA/IPAC Extragalactic Database\footnote{See \url{https://ned.ipac.caltech.edu}} (NED) to the ALMA bands assuming a spectral index of $0.70$, although this could also be flatter due to the presence of a flat-spectrum radio core. 
The resulting sample of nine RQ AGN, including their Swift IDs and column densities ($N_{\rm H}$; \citealt{ricci_bat_2017}), is presented in Tables\,\ref{tab:Sources1} and \ref{tab:Sources2}.
In addition, Table\,\ref{tab:Sources2} lists the values of $M_{\rm BH}$, bolometric luminosity ($L_{\rm bol}$), Eddington ratio ($\lambda_{\rm Edd}$), and star formation rate (SFR), as adopted from \cite{koss_bass_2022} and \cite{ ichikawa_complete_2017,ichikawa_bat_2019}.

We also computed the radio-loudness parameter ($R_{\rm X}$) for each source to confirm that all objects are RQ. 
Using the definition of \cite{Terashima2003}, we adopt $\log(R_{\rm X})=\log(L_{\rm 1.4GHz} / L_{\rm {2-10keV}})\leq-4.50$ as the threshold for an RQ source.
We obtain $\sim5$\,GHz radio luminosities for our sample from the NRAO VLA Archive\footnote{For the National Radio Astronomy Observatory (NRAO) VLA Archive, see \url{https://www.vla.nrao.edu/astro/nvas/\#search}.}, while the 2--10\,keV luminosities were taken from \cite{ricci_bat_2017}.
The resulting $\log(R_{\rm X})$ values range from $-6.50$ to $-4.97$, i.e. consistent with all nine sources being RQ \citep{Terashima2003}.
For IRAS\,05078+1626, $\sim5$\,GHz flux measurements were not available; therefore, we apply the definition of \cite{teng_fermilat_2011}, $\log(R_{\mathrm{X}})=\log(L_{1.4 \mathrm{GHz}} / L_{14-195 \mathrm{keV}}) \leq-4.70$ instead. 
For this object, we find $\log(R_{\mathrm{X}})=-6.01$, using $L_{14-195 \mathrm{keV}}$ from \cite{ricci_bat_2017} and $L_{1.4 \mathrm{GHz}}$ from the NRAO VLA Archive. 
The $\log(R_{\rm X})$ values for all sources in the sample are reported in Table\,\ref{tab:Sources1}.

\subsection{ALMA observations}

In this work, we focus on ALMA observations (Proposal ID 2018.1.00576.S; PI: Y.\,Inoue) which were carried out at 100\,GHz (Band\,3), 145\,GHz (Band\,4), and 233\,GHz (Band\,6) between October\,12, 2018, and August\,18, 2019. The specific observation dates for each source are listed in Table\,\ref{tab:Sources1}. 
These bands cover frequencies between 84--116\,GHz, 125--163\,GHz, and 211--275\,GHz, respectively.
The spectral configuration of the ALMA observations consisted of four spectral windows (SPWs) divided into 128 channels with a width of 15.63\,MHz each, resulting in a total covered bandwidth of $\sim8$\,GHz, and central frequencies of 97.5\,GHz\footnote{Throughout this work, we will denote this observation as the 100\,GHz observation.}, 145\,GHz and 233\,GHz. 
For these observations, a standard calibration strategy was adopted in which one bright quasar served as both the flux and bandpass calibrator, and a second quasar served as the phase calibrator. The calibration quality was verified using a check source.
The 100\,GHz, 145\,GHz, and 233\,GHz data were taken quasi-simultaneously, with each band observed only $\sim20$\,minutes apart.
In addition, a second epoch of 100\,GHz observations was carried out 4--11\,days after the initial epoch to investigate possible variability.
Finally, the first SPW of Ark\,120 covers the frequency range in which the CO(2--1) emission line might occur, and this spectral region must be excluded from the data used for continuum imaging. The detailed procedure for this CO(2--1) observation will be presented in Section\,\ref{Ark120}.

The 233\,GHz observations of all nine sources in our sample were previously analyzed by \cite{kawamuro_bass_2022} in their study of the millimeter/X-ray relation in RQ AGN; we revisit this relation in Section\,\ref{mmXray}. 
Furthermore, the ALMA observations of AX\,J1737.4--2907 were analyzed by \cite{michiyama_alma_2024}, who performed SED fitting; we compare our results for this source with theirs throughout this work.

\begin{table*}[ht!]
\setlength{\tabcolsep}{1.8pt}
\centering
\caption{Peak flux densities, resolutions, rms and luminosities at 100\,GHz in epochs\,1 and 2}
\begin{tabularx}{\textwidth}{l|cccc|cccc}
\hline
\hline
(1) & (2) & (3) & (4) & (5) & (6) & (7) & (8) & (9) \\ 
Source            &$ S^{\rm peak}_{\rm 100GHz, ep1}$ & $\theta^{\rm mean}_{\rm 100GHz, ep1}$ & $\sigma_{\rm 100GHz, ep1}$ & $ \log (\nu L_{\rm 100GHz, ep1})$ & $ S^{\rm peak}_{\rm 100GHz, ep2}$ & $\theta^{\rm mean}_{\rm 100GHz, ep2}$ &$\sigma_{\rm 100GHz, ep2}$&$ \log (\nu L_{\rm 100GHz, ep2})$\\
&[$\rm mJy\,beam^{-1}$] & & [$\rm mJy\,beam^{-1}$]& [$\rm erg\,s^{-1}$]&[$\rm mJy\,beam^{-1}$]& &[$\rm mJy\,beam^{-1}$]&[$\rm erg\,s^{-1}$]\\
\hline
\hline
Ark\,120          &1.54 & 0.434$\arcsec$/288\,pc & 0.1026 & 39.59 & 1.97 & 0.651$\arcsec$/432\,pc &0.0824 &39.69\\
AX\,J1737.4--2907 &1.65 & 0.258$\arcsec$/110\,pc & 0.0302 & 39.22 & 2.52 & 0.245$\arcsec$/105\,pc &0.0297 &39.40\\
IC\,4329A         &13.6 & 0.235$\arcsec$/77\,pc  & 0.0709 & 39.89& 12.4 & 0.260$\arcsec$/85\,pc  &0.0594 &39.85\\
IRAS\,05078+1626  &1.34 & 0.287$\arcsec$/106\,pc & 0.1040 & 38.99& 1.93 & 0.281$\arcsec$/103\,pc &0.0342 &39.15\\
MR\,2251--178     &3.74 & 0.379$\arcsec$/474\,pc & 0.0475 & 40.57& 3.17 & 0.678$\arcsec$/847\,pc &0.0422 &40.49\\
Mrk\,509          &2.64 & 0.371$\arcsec$/253\,pc & 0.0475 & 39.85& 2.50 & 0.644$\arcsec$/440\,pc &0.0352 &39.82\\
Mrk\,926          &5.60 & 0.387$\arcsec$/360\,pc & 0.0693 & 40.46& 4.87 & 0.719$\arcsec$/671\,pc &0.0555 &40.40\\
NGC\,3783         &2.18 & 0.277$\arcsec$/57\,pc  & 0.0336 & 38.69& 2.57 & 0.312$\arcsec$/64\,pc  &0.0363 &38.76\\
NGC\,4593         &1.23 & 0.248$\arcsec$/41\,pc  & 0.0307 & 38.24& 1.64 & 0.289$\arcsec$/48\,pc  &0.0315 &38.37 \\
\hline
\hline
\end{tabularx}
\par\vspace{1ex}
\parbox{1\textwidth}{\small
\textbf{Note:} (1) Source name, (2) Peak flux density in the 100\,GHz epoch\,1 observation (at $\nu_{\rm center}=100$\,GHz) where all SPWs are combined, (3) Geometric mean resolution in the 100\,GHz epoch\,1 observation, (4) Rms in the 100\,GHz epoch\,1 observation, (5) Luminosity in the 100\,GHz epoch\,1 observation, (6) Peak flux density in the 100\,GHz epoch\,2 observation (at $\nu_{\rm center}=100$\,GHz) where all SPWs are combined, (7) Geometric mean resolution in the 100\,GHz epoch\,2 observation, (8) Rms in the 100\,GHz epoch\,2 observation, and (9) Luminosity in the 100\,GHz epoch\,2 observation.}
\label{tab:originalbeam_3}
\end{table*}

\begin{table*}[ht!]
\setlength{\tabcolsep}{3pt}
\centering
\caption{Peak flux densities, resolutions, rms and luminosities at 145\,GHz and 233\,GHz}
\begin{tabularx}{\textwidth}{l|cccc|cccc}
\hline
\hline
(1) & (2) & (3) & (4) & (5) & (6) & (7) & (8) & (9) \\ 
Source            &$ S^{\rm peak}_{\rm 145GHz}$ & $\theta^{\rm mean}_{\rm 145GHz}$ & $\sigma_{\rm 145GHz}$ & $ \log (\nu L_{\rm 145GHz})$ & $ S^{\rm peak}_{\rm 233GHz}$ & $\theta^{\rm mean}_{\rm 233GHz}$ &$\sigma_{\rm 233GHz}$&$ \log (\nu L_{\rm 233GHz})$\\
&[$\rm mJy\,beam^{-1}$] & & [$\rm mJy\,beam^{-1}$]& [$\rm erg\,s^{-1}$]&[$\rm mJy\,beam^{-1}$]& &[$\rm mJy\,beam^{-1}$]&[$\rm erg\,s^{-1}$]\\
\hline
\hline
Ark\,120          &1.07 & 0.286$\arcsec$/190\,pc &0.0322 &39.59  & 0.996  & 0.175$\arcsec$/116\,pc& 0.0379 & 39.77\\
AX\,J1737.4--2907 &1.47 & 0.174$\arcsec$/74\,pc  &0.0274 &39.33  & 1.11 & 0.114$\arcsec$/49\,pc & 0.0325 & 39.41\\
IC\,4329A         &10.5 & 0.161$\arcsec$/53\,pc  &0.0577 & 39.94  & 9.51 & 0.101$\arcsec$/33\,pc& 0.0695& 40.11 \\
IRAS\,05078+1626  &1.19 & 0.189$\arcsec$/70\,pc  &0.0274 & 39.10  &  0.89 & 0.117$\arcsec$/43\,pc& 0.0278& 39.18\\
MR\,2251--178     &2.88 & 0.248$\arcsec$/309\,pc &0.0445 &40.61  & 2.34 & 0.152$\arcsec$/190\,pc  & 0.0510& 40.73  \\
Mrk\,509          & 2.05 & 0.272$\arcsec$/186\,pc &0.0441 & 39.90  & 1.99 & 0.190$\arcsec$/130\,pc& 0.0489& 40.09\\
Mrk\,926          &3.46 & 0.258$\arcsec$/241\,pc &0.0509 &40.42  & 2.17 & 0.174$\arcsec$/162\,pc& 0.0499& 40.42\\
NGC\,3783         &1.44 & 0.163$\arcsec$/34\,pc  &0.0262 &38.67  & 1.17 & 0.101$\arcsec$/21\,pc& 0.0299& 38.78\\
NGC\,4593         & 1.08 & 0.176$\arcsec$/30\,pc  &0.0288 &  38.35  &  0.72 & 0.105$\arcsec$/17\,pc & 0.0324& 38.38 \\
\hline
\hline
\end{tabularx}
\par\vspace{1ex}
\parbox{1\textwidth}{\small
\textbf{Note:} (1) Source name, (2) Peak flux density in the 145\,GHz observation (at $\nu_{\rm center}=145$\,GHz) where all SPWs are combined, (3) Geometric mean resolution in the 145\,GHz observation, (4) Rms in the 145\,GHz observation, (5) Luminosity in the 145\,GHz observation, (6) Peak flux density in the 233\,GHz observation (at $\nu_{\rm center}=233$\,GHz) where all SPWs are combined, (7) Geometric mean resolution in the 233\,GHz observation, (8) Rms in the 233\,GHz observation, and (9) Luminosity in the 233\,GHz observation.}
\label{tab:originalbeam_4_6}
\end{table*}

\section{Data reduction and imaging}\label{Data}

\subsection{Self-calibration}\label{selfcal}

Before imaging the final ALMA measurement sets using the method we describe in Section\,\ref{ALMAImaging}, phase-only self-calibration was performed on all visibilities, primarily for the 100\,GHz observations, to improve image fidelity and enable a robust assessment of any source variability.
Self-calibration was applied only to sources with signal-to-noise ratios (SNRs) exceeding 20; Ark\,120 was therefore excluded. Furthermore, for IRAS\,05078+1626, self-calibration significantly increased the rms due to low SNR in the separate scans; therefore, we did not perform self-calibration for this source either. 
For the remaining seven sources, an initial \texttt{tclean} model was generated down to a $5\sigma_{\rm rms}$ threshold, where $\sigma_{\rm rms}$ is the rms, to include only reliable source emission.
We then derived antenna-based phase solutions with \texttt{gaincal} (\texttt{solint} = inf). 
The solutions were inspected and applied to the visibilities using \texttt{applycal}, after which the corrected data were reimaged.
We obtained a slight improvement in SNR for each source, ensuring reliable flux measurements.

For IC\,4329A, the self-calibration procedure greatly improved the image quality.
The initial image exhibited substantial negative fluxes surrounding the central source, likely due to phase calibration errors. 
Phase-only self-calibration removed these negative features while leaving the peak flux density unchanged. The procedure thus enhanced the reliability of the measured fluxes and the overall image fidelity, which is presented in Figure\,\ref{fig:alma_ic4329a} (Appendix\,\ref{AppALMA}).

\subsection{ALMA imaging}\label{ALMAImaging}

Calibration and imaging of the ALMA observations were performed with Common Astronomy Software Applications (\textsc{CASA}), version 5.4.0--70 \citep{mcmullin_casa_2007}. 
Initially, dirty images of the observations were created to determine the $\sigma_{\rm rms}$ of each observation. Images were created for all four SPWs at each frequency combined and for each SPW separately, which will be used to derive intraband spectral indices (see Appendix\,\ref{intrabroad}).
We chose pixel sizes that satisfy the Nyquist sampling criterion to ensure the images captured sufficient details. 
Additionally, we determined the beam size ($\theta$) for each image. 
The $\sigma_{\rm rms}$ and $\theta$ values of all observations with combined SPWs are presented in Tables\,\ref{tab:originalbeam_3} and \ref{tab:originalbeam_4_6}. 

Regarding the separate SPW images, for MR\,2251--178, the final SPW of the 233\,GHz observation was fully flagged. 
Furthermore, the final SPWs in the 100\,GHz epoch\,1 and the 145\,GHz observations were taken with much shorter observation times, decreasing the obtained SNR by a factor $\sim4$ compared to the other SPWs in the respective bands.
For Mrk\,509 and Mrk\,926, the final SPWs in the 100\,GHz epoch\,1, 145\,GHz and 233\,GHz observations were fully flagged as well. Therefore, these SPWs were not taken into account by the spectral index study described in Appendix\,\ref{intrabroad}.
As discussed in more detail in Appendix\,\ref{intrabroad} as well, some SPWs exhibit significant scatter within individual ALMA bands \citep[see, e.g.,][for further details]{Francis2020}. 
We therefore use the combined SPW images, rather than treating each SPW separately, when computing our SEDs.

Since the goal of this work is to compare observations across different frequencies, the beam sizes of the images at the different frequencies were matched to ensure an accurate comparison. 
For this reason, the largest beam size among all SPWs of a given source was adopted as the common beam and used when generating the final images. 
In addition to beam matching, uv-clipping was applied to match the different baselines sampled at each frequency, restricting the images to comparable spatial scales and removing scales that are unevenly sampled. The adopted uv-ranges and matched beam sizes for each source are listed in Table\,\ref{tab:Sources_matched}.

After obtaining the $\sigma_{\rm rms}$ values, we used the task \texttt{tclean} in multifrequency synthesis (mfs) mode to reconstruct the final source images. 
We adopted Briggs weighting with a robust parameter of $+0.5$, which ensures the best trade-off between sensitivity and resolution.
As previously mentioned, we adopted the largest beam size for each source as the restoring beam in each observation (Table\,\ref{tab:Sources_matched}).
In the task \texttt{tclean}, we used auto-masking by multiple thresholds (\texttt{auto-multithresh}) to speed up the cleaning process and to ensure that these results can easily be reproduced. 
In auto-multithresh mode, we adopted a side-lobe threshold of $2.0$, a noise threshold of $5.0$, a low noise threshold of $1.5$, a minimum beam fraction of $0.2$, and a negative threshold of $0.0$\footnote{These values were adopted from the table in the CASA Automasking Guide (\url{https://casaguides.nrao.edu/index.php/Automasking_Guide_CASA_6.6.1}).}. 
We investigated the residuals obtained when cleaning with thresholds of 1$\sigma_{\rm rms}$, 2$\sigma_{\rm rms}$, 3$\sigma_{\rm rms}$, and 5$\sigma_{\rm rms}$, and generally selected a 2$\sigma_{\rm rms}$ threshold to ensure that no source emission remained in the residuals while avoiding over-cleaning.
However, in practice, the variations in peak flux across the different detection thresholds are negligible. 
For instance, in the 100\,GHz epoch\,1 observation of AX\,J1737.4--2907, the measured peak flux densities are 1.728, 1.726, 1.721, and 1.713\,mJy, for the different thresholds, respectively.

Then, the primary beam corrections were applied to the images, after which the peak flux densities of the sources could be obtained.
Across most sources, both the original beam images and the self-calibrated, uv-clipped, and $\theta$-matched images yield peak-to-integrated flux density ratios of 0.90--1.0, consistent with predominantly unresolved emission. 
The smallest ratios are measured for AX\,J1737.4--2907 across all frequencies, which has already been confirmed to host extended emission \citep{michiyama_alma_2024} and is further discussed in Sections\,\ref{results_var} and \ref{synchr_origin}.
In addition, Ark\,120, IRAS\,05078$+$1626, Mrk\,509, NGC\,3783, and NGC\,4593 show peak-to-integrated flux density ratios of roughly 0.75–0.85 in the 233\,GHz original-beam data, suggesting the presence of extended emission; this is examined further in Section\,\ref{spectralanalysis}.
The final flux densities in each band, measured with matched $\theta$ and uv-wave ranges, are presented in Table\,\ref{tab:Sources_matched} and will be used in our SEDs.
The flux densities obtained by preserving the original $\theta$ are presented in Tables\,\ref{tab:originalbeam_3} and \ref{tab:originalbeam_4_6}.
Furthermore, we adopt a flux uncertainty of
\begin{equation}
    \sigma_{S_{\nu}}=\sqrt{(\sigma_{\rm sys})^2+(\sigma_{\rm stat})^2},
    \label{eq:fluxunc}
\end{equation}
with $\sigma_{\rm sys}$ the systematic uncertainty and $\sigma_{\rm stat}$ the statistical uncertainty of the observation.
Here, the $\sigma_{\rm sys}$ equals to 5\% of the total peak flux density at 100\,GHz and 145\,GHz, and 10\% at 233\,GHz, according to the ALMA guidelines\footnote{ALMA Cycle 6 Technical Handbook \url{https://almascience.nrao.edu/documents-and-tools/cycle6/alma-technical-handbook}} for Cycle\,6.
The value of $\sigma_{\rm stat}$ equals the rms of the observation, as listed in Tables\,\ref{tab:originalbeam_3} and \ref{tab:originalbeam_4_6}.
For the majority of the sources, the $\sigma_{\rm stat}$ is negligible compared to $\sigma_{\rm sys}$; however, for Ark\,120 and IRAS\,05078+1626, the $\sigma_{\rm stat}$, i.e. the rms contribution, must be taken into account.
Based on the derived fluxes, we obtained luminosities through the following relation:

\begin{equation}
\begin{split}
  \left(\frac{\nu_{\rm } L_{\nu_{\rm }}}{\mathrm{erg\,s^{-1}}}\right)
  = {} & 4\pi \left(\frac{D_{\rm L}^{2}}{\mathrm{cm}^{2}}\right)
    \left(\frac{\nu_{\rm }}{\mathrm{Hz}}\right)
    \left(\frac{\nu_{\rm }}{\nu'}\right)^{-\alpha_{\rm mm}} \\
    & \times \left(\frac{S_{\nu '}}{\mathrm{erg\,s^{-1}\,cm^{-2}\,Hz^{-1}}}\right)
    (1+z)^{(\alpha_{\rm mm}-1)},
\end{split}
\label{eq:Lmm}
\end{equation}
where $D_{\rm L}$ is the luminosity distance, $S_{\rm \nu '}$ is the flux density at the observed frequency $\nu'$, $\alpha_{\rm mm}$ is the spectral index, defined as $S_\nu \propto \nu^{-\alpha_{\rm mm}}$ (see Section\,\ref{spectralanalysis}). 
The luminosity $\nu L_{\nu}$ is evaluated at $\nu = 100$\,GHz, 145\,GHz or 233\,GHz, while the factor $(1+z)^{\alpha_{\rm mm}-1}$ is the $k$-correction \citep[e.g.,][]{novak2017, kawamuro_bass_2022}.

\begin{table*}[ht!]
\centering
\setlength{\tabcolsep}{8.5pt}
\caption{Parameters for matched resolution ALMA images}
\begin{tabularx}{\textwidth}{l|ccccccc}
\hline
\hline
(1) & (2) & (3) & (4) & (5) & (6) & (7) & (8)  \\ 
Source            & $\theta_{\rm matched}$ & $\theta_{\rm matched}$ & uv-clipping range & $ S_{\rm peak, \theta_{matched}}^{\rm 100GHz, ep1}$ & $ S_{\rm peak, \theta_{matched}}^{\rm 100GHz, ep2}$ & $ S_{\rm peak, \theta_{matched}}^{\rm 145GHz}$ & $ S_{\rm peak, \theta_{matched}}^{\rm 233GHz}$   \\ 
                  &   &  & [$\rm k\lambda$]  & [$\rm mJy\,beam^{-1}$] & [$\rm mJy\,beam^{-1}$]& [$\rm mJy\,beam^{-1}$]& [$\rm mJy\,beam^{-1}$]   \\
\hline
\hline
Ark\,120          & 0.700$\arcsec$ & 465\,pc & 9.19 -- 437.72   &1.58&1.97&1.18&1.15  \\
AX\,J1737.4--2907 & 0.335$\arcsec$ & 143\,pc & 27.69 -- 1246.67 &1.73&2.69&1.67&1.39   \\
IC\,4329A         & 0.319$\arcsec$ & 104\,pc & 30.66 -- 1255.93 &13.9&12.4&11.0&10.4    \\
IRAS\,05078+1626  & 0.341$\arcsec$ & 125\,pc & 25.38 -- 1065.96 &1.35&1.94&1.31&1.18 \\
MR\,2251--178     & 0.799$\arcsec$ & 998\,pc & 11.36 -- 476.32  &3.79&3.28&3.14&2.69   \\
Mrk\,509          & 0.683$\arcsec$ & 467\,pc & 10.21 -- 468.76  &2.84&2.57&2.49&2.49  \\
Mrk\,926          & 0.874$\arcsec$ & 815\,pc & 11.05 -- 454.73  &5.91&5.36&3.80&2.89   \\
NGC\,3783         & 0.451$\arcsec$ & 93\,pc & 29.97 -- 1039.69  &2.54&2.87&1.96&1.74  \\
NGC\,4593         & 0.394$\arcsec$ & 65\,pc & 29.65 -- 1217.67  &1.33&1.73&1.29&1.39   \\
\hline
\hline
\end{tabularx}
\par\vspace{1ex}
\parbox{1\textwidth}{\small\textbf{Note:} (1) Source name, (2) Largest major axis beam ($\theta_{\rm maj}$) across the different ALMA bands to which all resolutions were matched in arcseconds, (3) in pc, (4) The uv-clipping range to match the different baselines between observations in the different bands, (5) Peak flux density at 100\,GHz epoch\,1 (at $\nu_{\rm center}=97.5$\,GHz) obtained after matching beams and self-calibration, (6) Peak flux density at 100\,GHz epoch\,2 (at $\nu_{\rm center}=97.5$\,GHz) obtained after matching beams and self-calibration, (7) Peak flux density at 145\,GHz (at $\nu_{\rm center}=145.0$\,GHz) obtained after matching beams and self-calibration, and (8) Peak flux density at 233\,GHz (at $\nu_{\rm center}=233.0$\,GHz) obtained after matching beams and self-calibration.} 
\label{tab:Sources_matched}
\end{table*}

\subsection{CO(2--1) emission in Ark\,120}\label{Ark120}

In the 233\,GHz observation of Ark\,120 ($z\sim0.0327$), we detect molecular gas emission in the first SPW. The emission is attributed to CO(2--1), which has a rest-frame frequency of $\nu^{\rm rest}_{\rm CO(2-1)} = 230.538$\,GHz, and appears as two peaks at approximately 223.2\,GHz and 223.3\,GHz, consistent with rotating or outflowing molecular gas. 
The full spectrum and the inferred gas mass ($M_{\rm gas}$) and gas surface density ($\Sigma_{\rm gas}$) are presented in Appendix\,\ref{AppMol}. 
Since our primary goal is to characterize the peak continuum millimeter emission, these spectral lines were flagged and removed from the SPW prior to reimaging the continuum and obtaining the continuum flux.

\begin{figure*}[ht!]
    \centering
    \begin{subfigure}[b]{0.49\textwidth}
        \includegraphics[width=\textwidth]{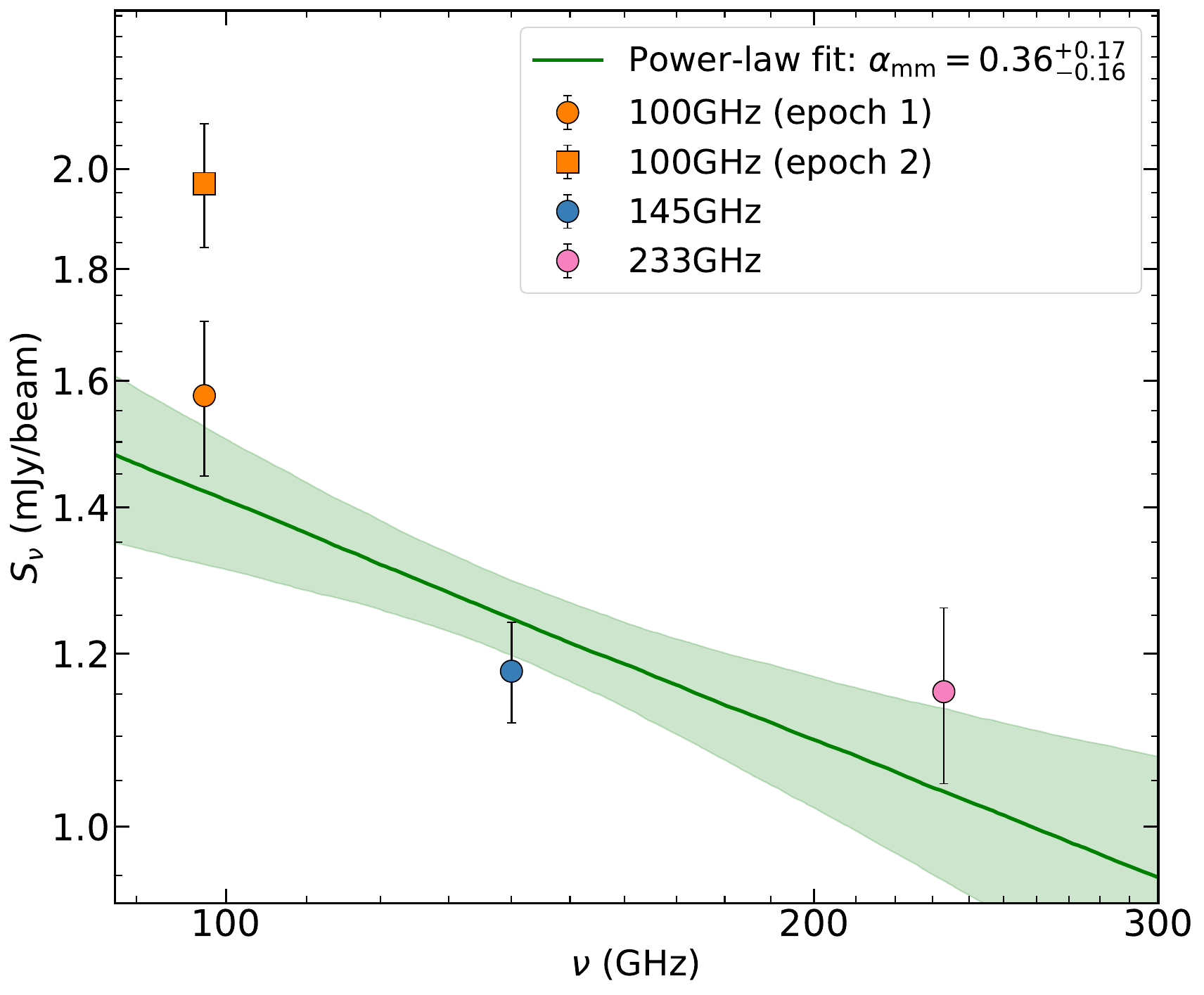}
        \caption{Ark\,120}
        \label{fig:SED_Ark120}
    \end{subfigure}
    \hfill
    \begin{subfigure}[b]{0.49\textwidth}
        \includegraphics[width=\textwidth]{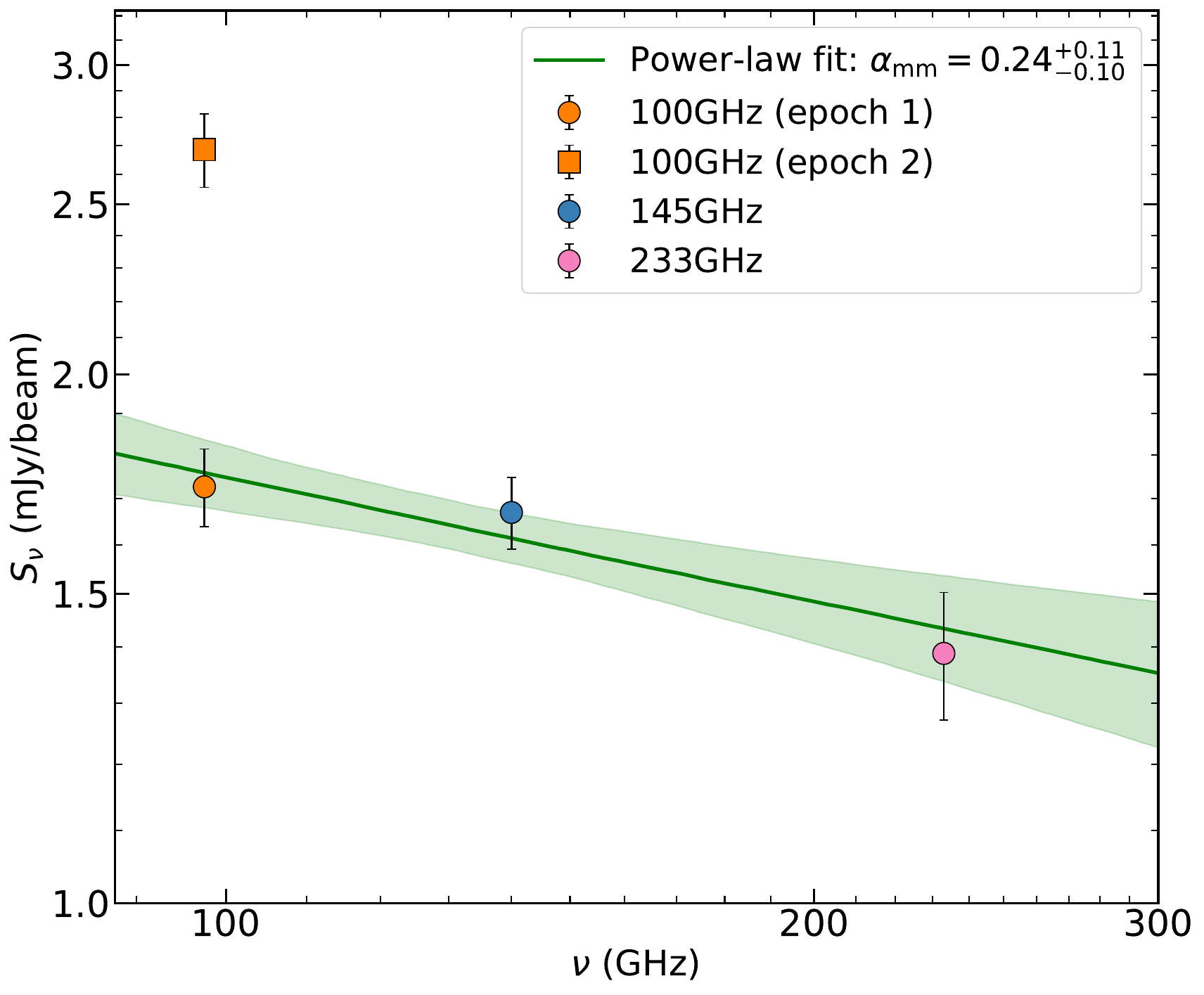}
        \caption{AX\,J1737.4--2907}
        \label{fig:SED_AX}
    \end{subfigure}

    \vspace{0.3cm}

    \begin{subfigure}[b]{0.49\textwidth}
        \includegraphics[width=\textwidth]{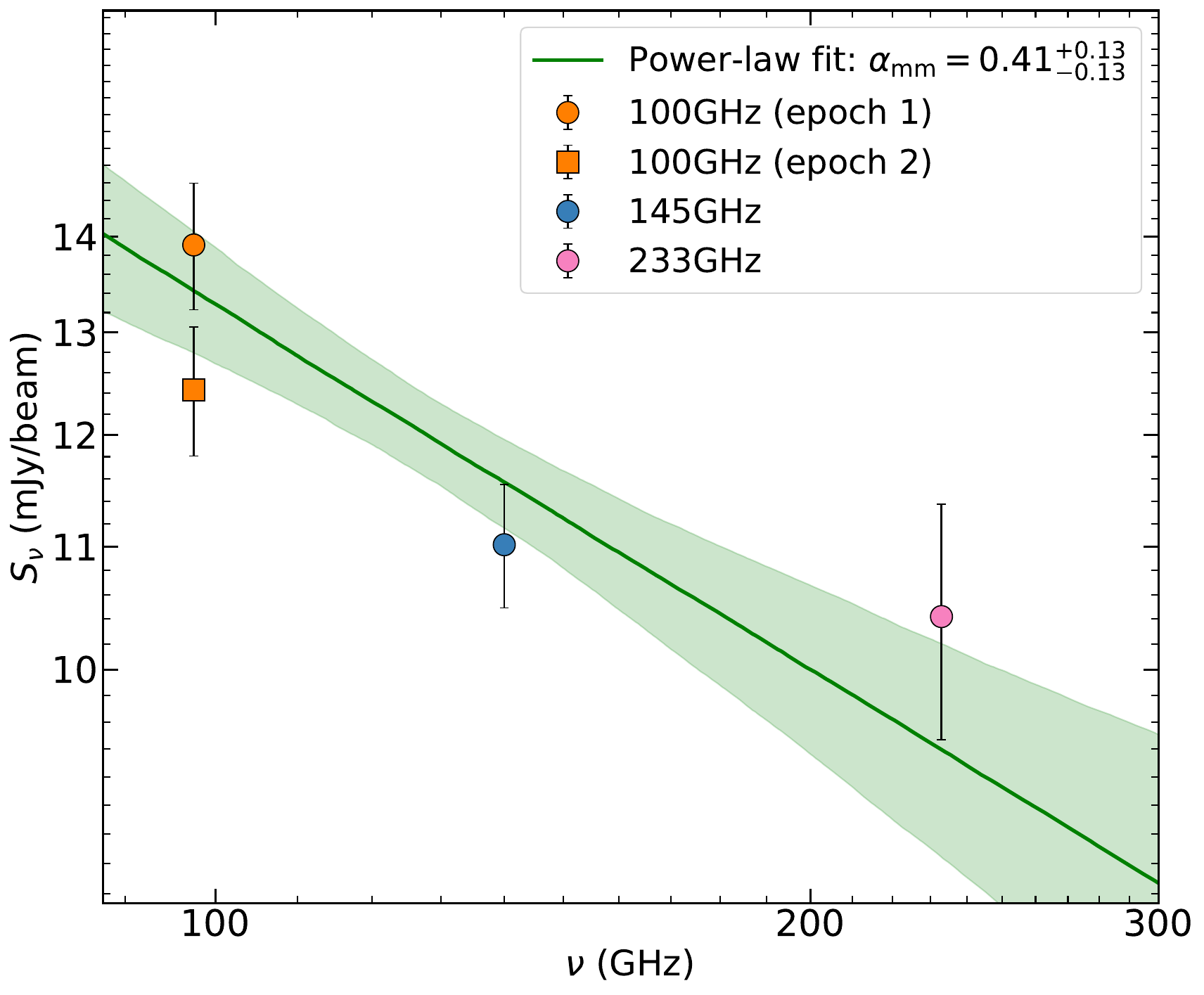}
        \caption{IC\,4329A}
        \label{fig:SED_IC}
    \end{subfigure}
    \hfill
    \begin{subfigure}[b]{0.49\textwidth}
        \includegraphics[width=\textwidth]{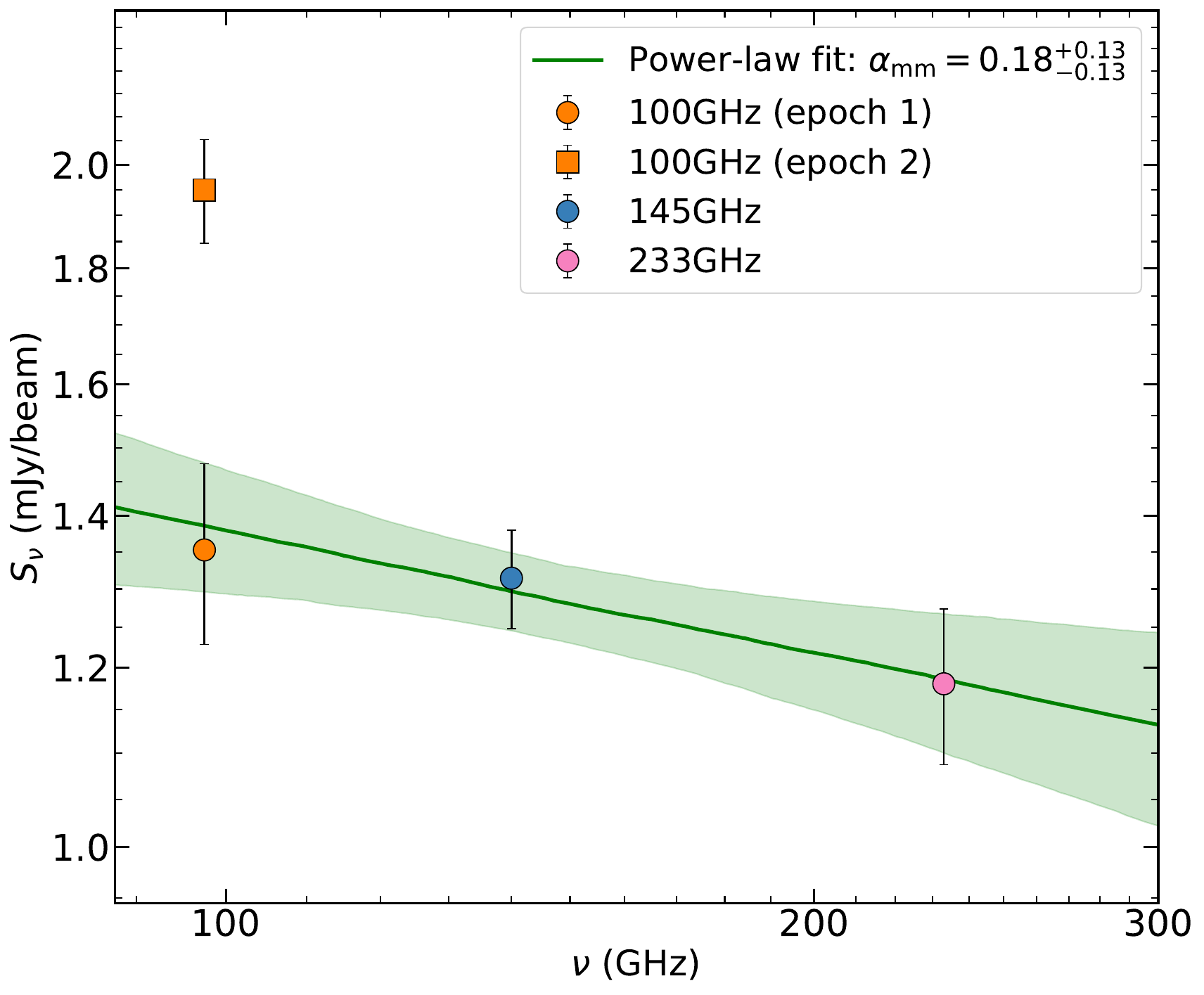}
        \caption{IRAS\,05078+1626}
        \label{fig:SED_IRAS}
    \end{subfigure}
    \vspace{0.3cm}

    \begin{subfigure}[b]{0.49\textwidth}
        \includegraphics[width=\textwidth]{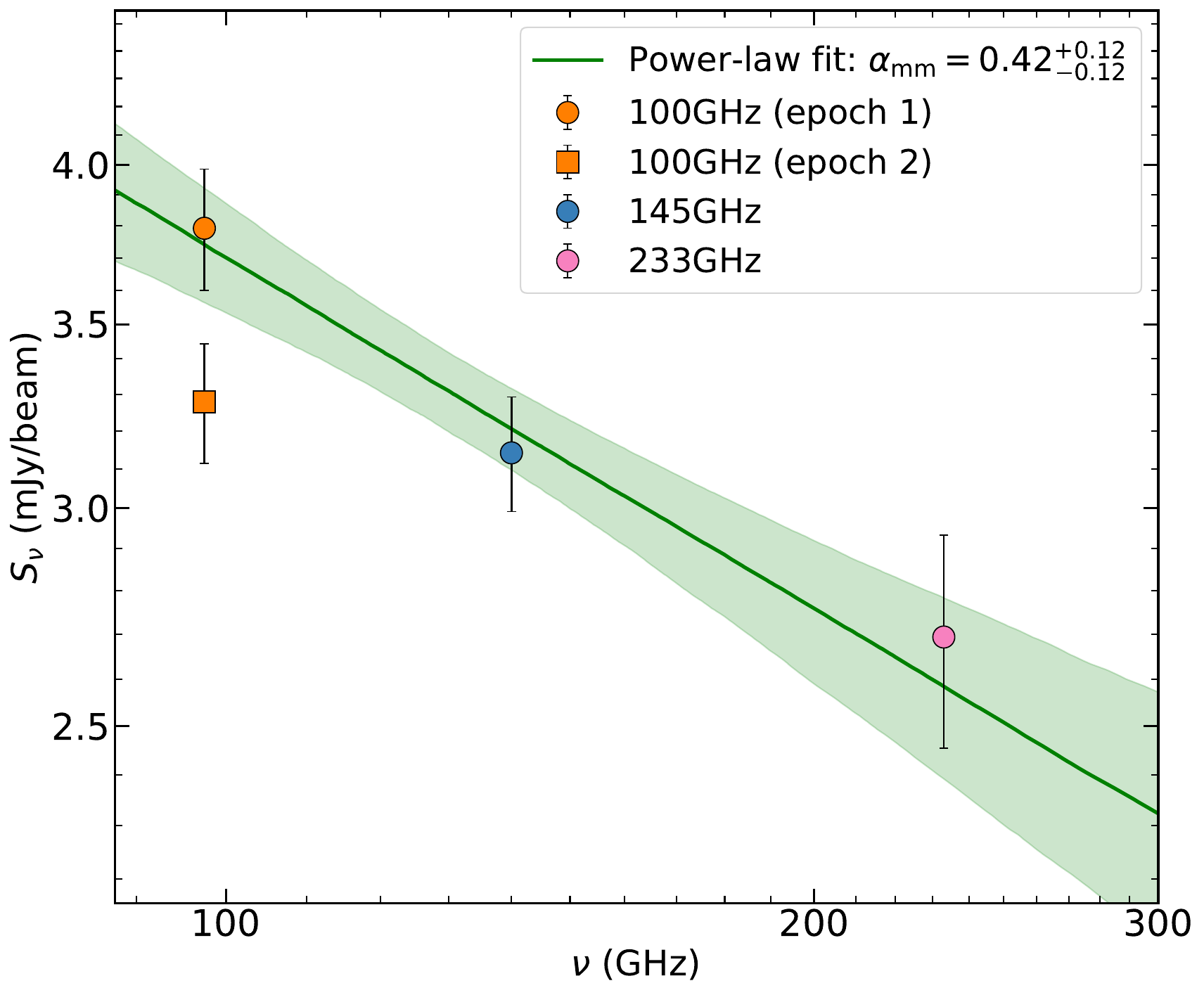}
        \caption{MR\,2251--178}
        \label{fig:SED_MR}
    \end{subfigure}
    \hfill
    \begin{subfigure}[b]{0.49\textwidth}
        \includegraphics[width=\textwidth]{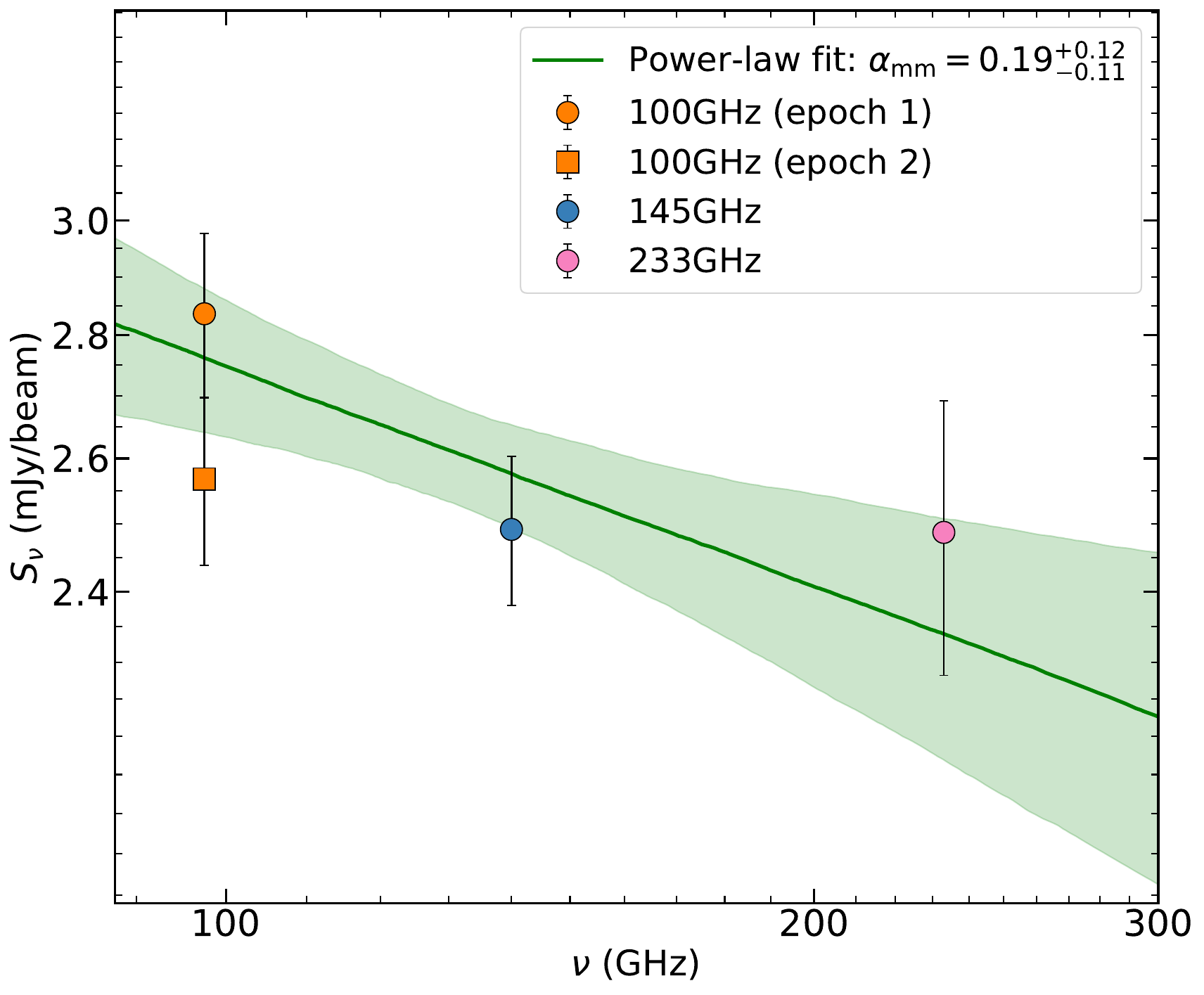}
        \caption{Mrk\,509}
        \label{fig:SED_Mrk509}
    \end{subfigure}
\caption*{(Part 1 of 2)}
\end{figure*}
\begin{figure*}[ht!]
    \ContinuedFloat
    \centering
    \begin{subfigure}[b]{0.49\textwidth}
        \includegraphics[width=\textwidth]{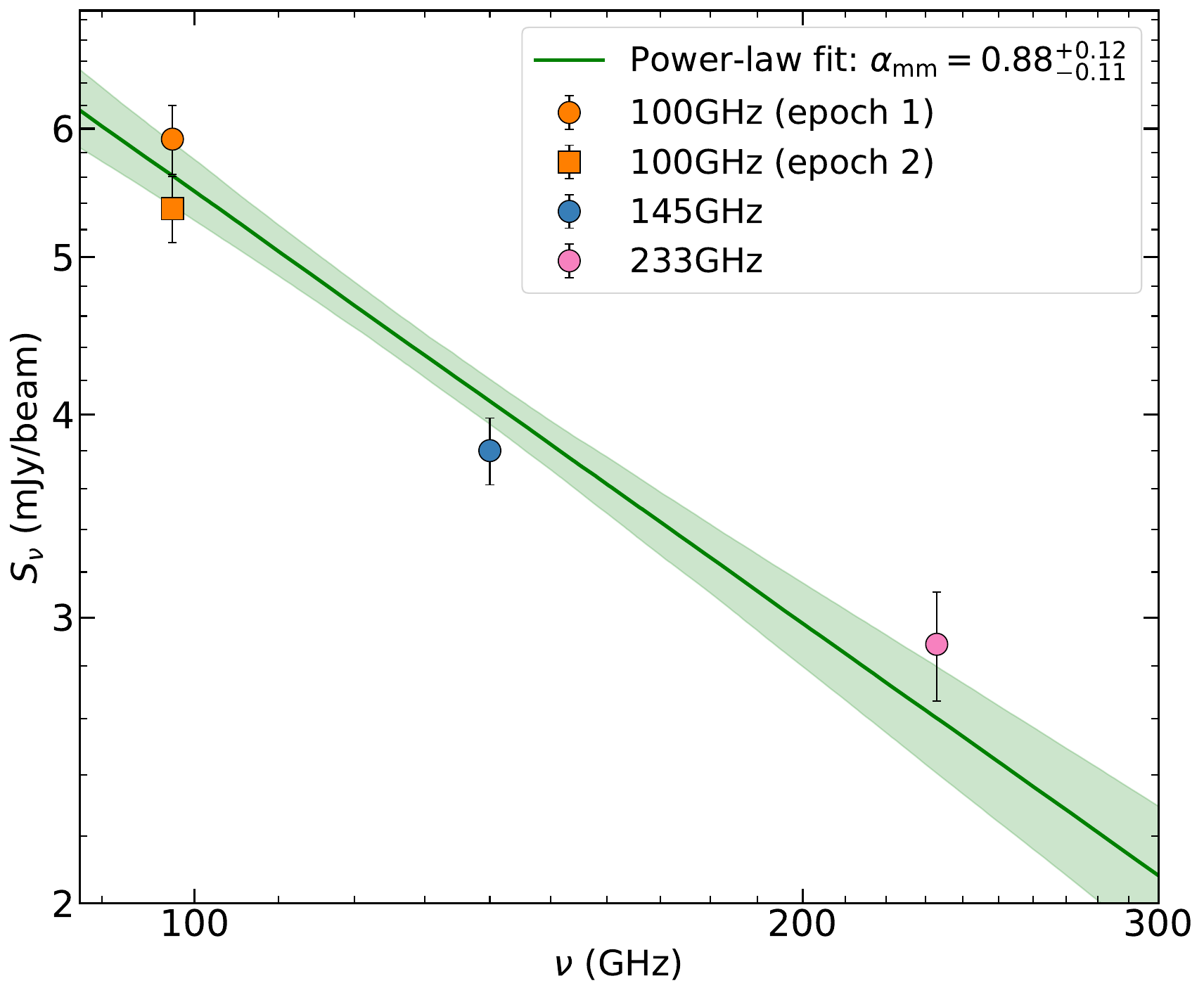}
        \caption{Mrk\,926}
        \label{fig:SED_Mrk926}
    \end{subfigure}
    \hfill
    \begin{subfigure}[b]{0.49\textwidth}
        \includegraphics[width=\textwidth]{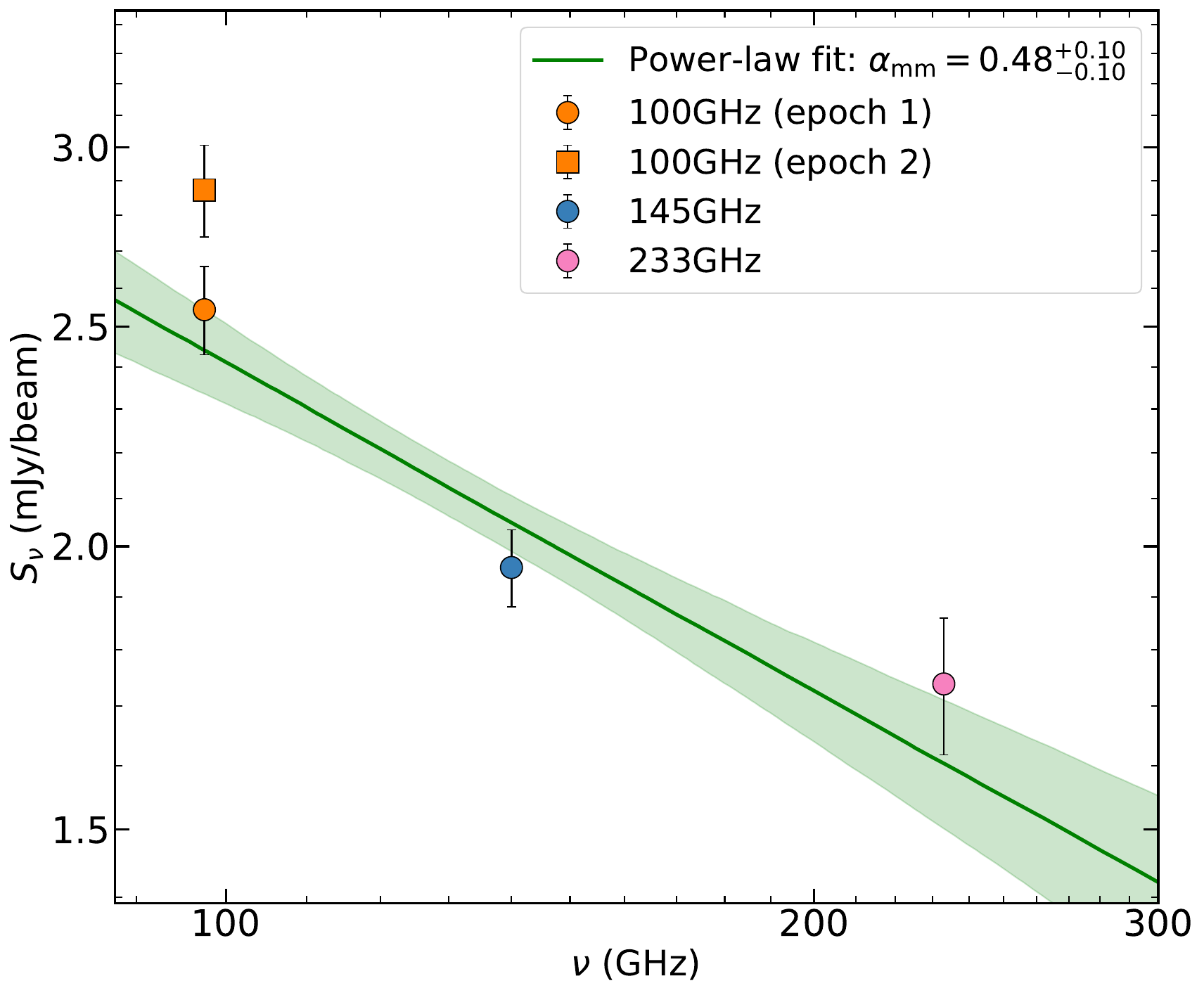}
        \caption{NGC\,3783}
        \label{fig:SED_NGC3783}
    \end{subfigure}

    \vspace{0.3cm}

    \begin{subfigure}[b]{0.49\textwidth}
        \includegraphics[width=\textwidth]{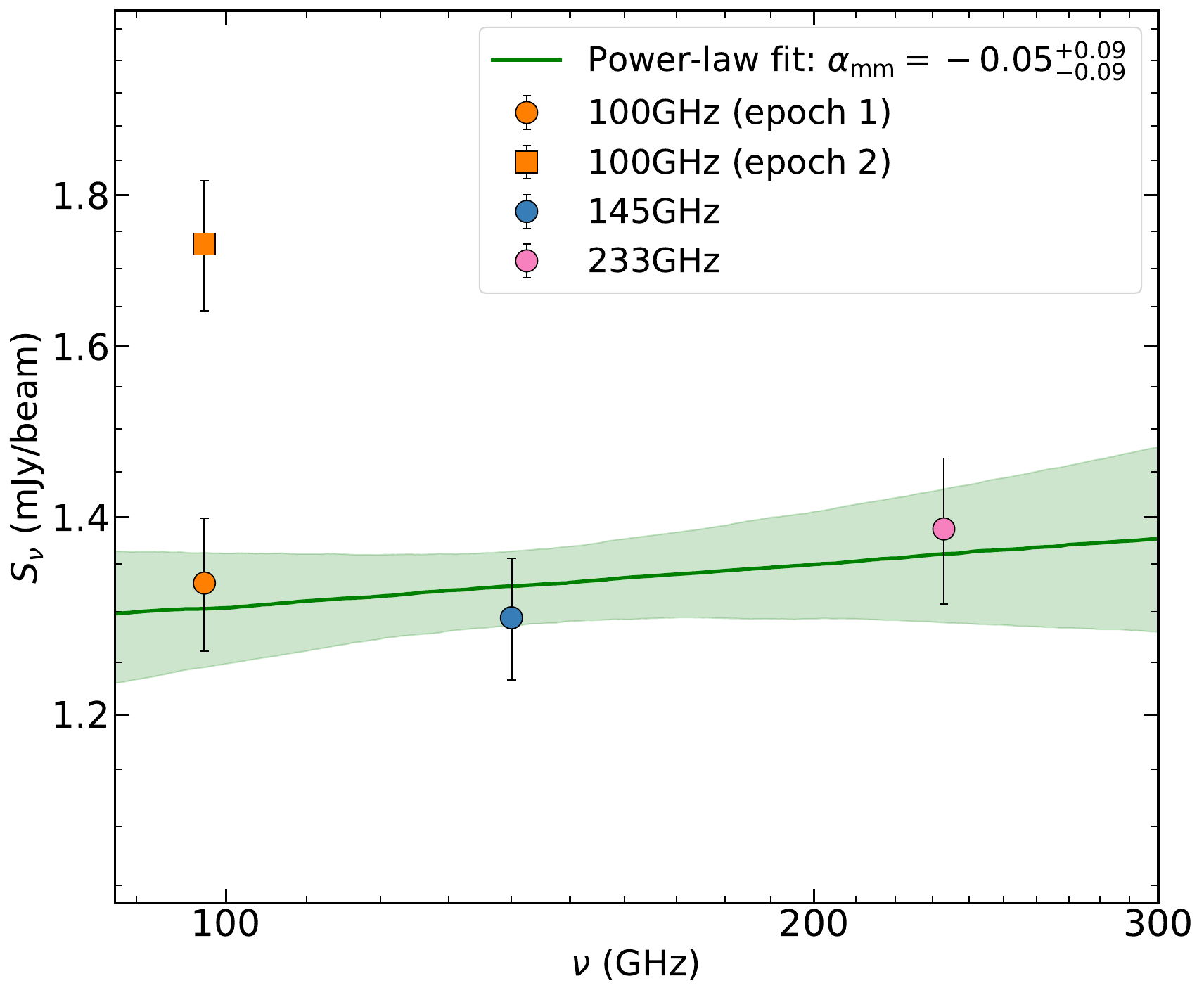}
        \caption{NGC\,4593} 
        \label{fig:SED_NGC4593}
    \end{subfigure}
\caption{(Part 2 of 2) The SEDs of the nine RQ AGN in our sample are shown, with flux density in mJy plotted as a function of frequency in GHz. Orange, blue, and pink circles represent the 100\,GHz, 145\,GHz, and 233\,GHz measurements from epoch\,1, respectively. The orange squares indicate the 100\,GHz data from epoch\,2. The error bars include a 5\% flux uncertainty at 100\,GHz and 145\,GHz, and a 10\% flux uncertainty at 233\,GHz, added in quadrature with the $\sigma_{\rm rms}$ contribution. 
For each source, the observations at all three frequencies and a second 100\,GHz epoch were convolved to a common angular resolution, presented in Table\,\ref{tab:Sources_matched}. 
We fitted a power-law of the form $S_{\nu}\propto \nu^{-\alpha}$ to the epoch\,1 data.
In this figure, the 100\,GHz observation from the second epoch is included to illustrate variability, but it is not incorporated into the SED fit.}
\label{fig:SEDs}
\end{figure*}

\section{Results}\label{Results}

We obtained peak flux densities from quasi-simultaneous ALMA observations at 100\,GHz, 145\,GHz, and 233\,GHz of nine RQ AGN. 
Additionally, second epoch 100\,GHz observations were obtained, enabling an assessment of variability and, if present, the derivation of upper limits on the size of the millimeter-emitting region over which such a change could occur.

\subsection{Spectral analysis}\label{spectralanalysis}

To establish whether nonthermal emission is a common feature among our sample of nine RQ AGN, we characterize the spectral index ($\alpha_{\rm mm}$) of the observed millimeter emission to assess its synchrotron origin.

As described in Section\,\ref{Intro}, around $\nu\sim95$--300\,GHz, an SSA peak would be expected from coronal synchrotron emission \citep[e.g.,][]{del_palacio_millimeter_2025}.
The SSA model \citep[e.g.,][]{michiyama_alma_2024} is defined as:
\begin{equation}
S_{\nu} = S_{\rm \nu,SSA} \left( \frac{\nu}{\nu_{\rm SSA}} \right)^{5/2} \left[1 - \exp\left( \left( -\frac{\nu}{\nu_{\rm SSA}}\right)^{-5/2-\alpha_{\rm NT}}\right) \right],
\label{eq:SSA}
\end{equation}
where $S_{\rm \nu,SSA}$ is the normalization, $\nu_{\rm SSA}$ is the frequency of the SSA peak, and $\alpha_{\rm NT}$ is nonthermal spectral index. 
Unfortunately, this SSA model has three free parameters, while observations are available at only three frequencies, yielding zero degrees of freedom and preventing a meaningful assessment of the goodness of fit.
Fitting the SSA model to our data would therefore require fixing one of these three free parameters, as was, for example, done for AX\,J1737.4--2907 by \citet{michiyama_alma_2024}, who adopted $\alpha_{\rm NT}=0.85$. 
Since the electron energy distribution $N(E)\propto E^{-p}$ is generally not well constrained, and because the spectral index $\alpha_{\rm NT}$ is tied to its slope through $p=2\alpha_{\rm NT}+1$, any adopted value of $\alpha_{\rm NT}$ lacks strong justification.
If we adopt $\alpha_{\rm NT}=0.85$ as in \cite{michiyama_alma_2024}, only AX\,J1737.4--2907 yields a reasonable SSA fit, while IRAS\,05078+1626 would require a steeper index of $\alpha_{\rm NT}\sim0.65$. 
However, without an independent constraint on $\alpha_{\rm NT}$, the SSA turnover frequency cannot be reliably determined based on our data.

Therefore, instead of fitting the SSA model, which would require assumptions on $\alpha_{\rm NT}$, we fit a simple power-law model of the form: 
\begin{equation}
S_{\nu} = A \left(\frac{\nu}{\nu_0}\right)^{-\alpha_{\rm mm}},
\label{eq:powerlaw}
\end{equation}
where $A$ is the flux density normalization.
This yields the broadband millimeter spectral index ($\alpha_{\rm mm}$), which we will use to characterize the origin of the observed millimeter emission. 
The power-law fits are performed using Markov Chain Monte Carlo sampling (emcee; \citealp{Foreman-Mackey13}). 
The values we report for $\alpha_{\rm mm}$ and the $1\sigma$ uncertainties correspond to the 16th, 50th, and 84th percentiles of the posterior distributions.

\begin{figure}[ht!]
\centering
\includegraphics[width=1\columnwidth]{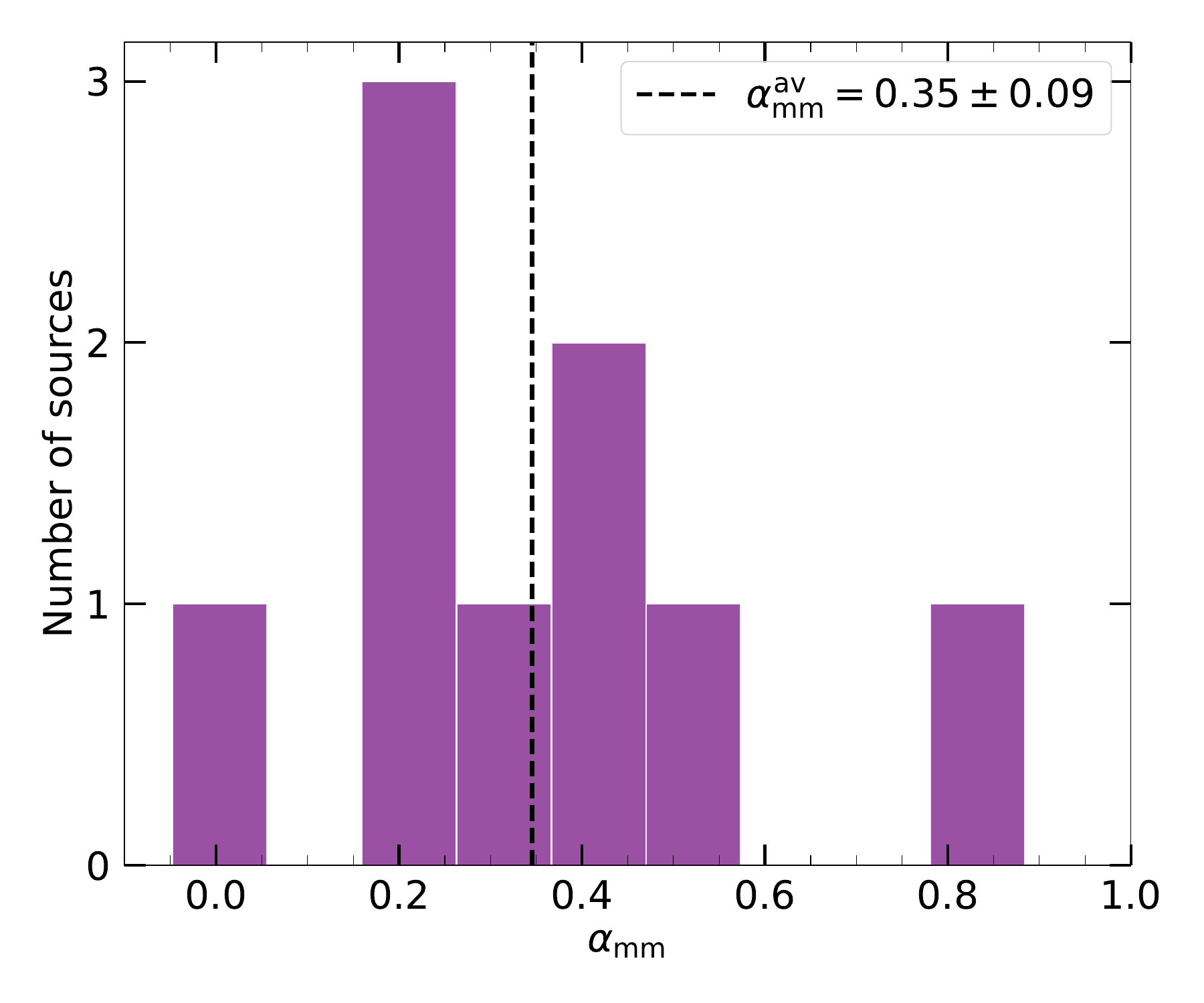}
\caption{Histogram of the $\alpha_{\rm mm}$ values obtained by fitting a power-law (see Equation\,\ref{eq:powerlaw}) to the observations in epoch\,1. We obtain values between $\alpha_{\rm mm}=-0.05^{+0.09}_{-0.09}$ and $0.88^{+0.12}_{-0.11}$ and an average of $\alpha_{\rm mm}^{\rm av}=0.35\pm0.09$. 
Only one of the nine sources shows a spectral slope consistent with optically thin synchrotron emission, whereas the others display flatter slopes, consistent with a spectral flattening around $\nu_{\rm SSA}$.}
\label{fig:alpha_histogram}
\end{figure}

Figure\,\ref{fig:SEDs} presents the SEDs computed for our nine RQ AGN with a power-law model fitted to the three epoch\,1 data points (circles).
We observe spectral indices between $\alpha_{\rm mm}=-0.05^{+0.09}_{-0.09}$ and $0.88^{+0.12}_{-0.11}$ for the nine sources, with an average of $\alpha_{\rm mm}^{\rm av}=0.35\pm0.09$. 
An overview of the $\alpha_{\rm mm}$ values obtained in this work is presented in Figure\,\ref{fig:alpha_histogram} and Table\,\ref{tab:alpha}.
Furthermore, in addition to $\alpha_{\rm mm}$, we fitted power-laws separately to the 100--145\,GHz and 145--233\,GHz data to test for curvature across the three frequencies. This also helps quantify a possible extended-emission contribution at 233\,GHz in Ark\,120, Mrk\,509, and NGC\,4593, which is suggested by their apparently higher flux densities in Figure\,\ref{fig:SEDs} and by the extended features described in Section\,\ref{ALMAImaging}.
These spectral slopes derived from the 100--145\,GHz and 145--233\,GHz observations are reported in Table\,\ref{tab:alpha} as well.

\begin{table}[ht!]
\centering
\caption{Spectral index results}
\setlength{\tabcolsep}{9pt}
\begin{tabularx}{\columnwidth}{l|ccc}
\hline
\hline
(1) & (2) & (3) & (4) \\
Source & $\alpha_{\rm mm}$ & $\alpha_{100-145}$ & $\alpha_{145-233}$ \\ 
\hline
\hline 
Ark\,120          & $0.36^{+0.17}_{-0.16}$ & $0.78^{+0.27}_{-0.27}$ & $0.06^{+0.22}_{-0.20}$   \\
AX\,J1737.4--2907 & $0.24^{+0.11}_{-0.10}$ & $0.08^{+0.19}_{-0.19}$ & $0.38^{+0.19}_{-0.18}$  \\
IC\,4329A         & $0.41^{+0.13}_{-0.12}$ & $0.64^{+0.19}_{-0.19}$ & $0.12^{+0.21}_{-0.19}$   \\
IRAS\,05078+1626  & $0.18^{+0.13}_{-0.13}$ & $0.06^{+0.28}_{-0.30}$ & $0.22^{+0.19}_{-0.18}$    \\
MR\,2251--178     & $0.42^{+0.12}_{-0.12}$ & $0.51^{+0.19}_{-0.19}$ & $0.32^{+0.21}_{-0.19}$   \\
Mrk\,509          & $0.19^{+0.11}_{-0.11}$ & $0.35^{+0.18}_{-0.18}$ & $0.02^{+0.19}_{-0.18}$   \\
Mrk\,926          & $0.88^{+0.12}_{-0.11}$ & $1.21^{+0.19}_{-0.19}$ & $0.55^{+0.18}_{-0.18}$    \\
NGC\,3783         & $0.48^{+0.10}_{-0.10}$ & $0.72^{+0.16}_{-0.17}$ & $0.24^{+0.16}_{-0.15}$   \\
NGC\,4593         & $-0.05^{+0.09}_{-0.09}$ & $0.07^{+0.20}_{-0.19}$ & $-0.13^{+0.15}_{-0.14}$   \\
\hline
\hline 
\end{tabularx}
\par\vspace{1ex}
\parbox{1\columnwidth}{\small
\textbf{Note:} (1) Source name, (2) Broadband millimeter spectral index ($\alpha_{\rm mm}$) inferred from SED fitting, (3) Spectral index based on a power-law fit between the 100\,GHz epoch\,1 and 145\,GHz observations, and (4) Spectral index based on a power-law fit between the 145\,GHz and 233\,GHz observations.}
\label{tab:alpha}
\end{table}

\begin{table}[ht!]
\setlength{\tabcolsep}{1.4pt}
\centering
\caption{Variability results}
\begin{tabularx}{\columnwidth}{l|cccccc}
\hline
\hline
(1) & (2) & (3) & (4) & (5) & (6) \\
Source & $\Delta S_{\rm100GHz}$ & $\Delta S_{\rm100GHz}$ & $\Delta t$    & $R_{\rm mm}^{\rm ul}$ & $R_{\rm mm}^{\rm ul}$  \\                    
       & [\%]                   &  [$\sigma$]            & [days]        & [pc]                  & [$R_{\rm g}$]\\
\hline
\hline 
Ark\,120          
& $+$25\%         & 2.1$\sigma$          & 11 & \nodata & \nodata \\ 
AX\,J1737.4--2907 
& $+$56\%         & 5.8$\sigma$          & 4  &  $3.4\times10^{-3}$ & $1.0\times10^3$ \\ 
IC\,4329A         
& $-$11\%         & 1.6$\sigma$          & 5  & \nodata & \nodata \\ 
IRAS\,05078+1626  
& $+$44\%         & 3.7$\sigma$          & 6  &  $5.0\times10^{-3}$ & $1.4\times10^4$ \\ 
MR\,2251--178     
& $-$14\%         & 2.0$\sigma$          & 10 & \nodata & \nodata \\ 
Mrk\,509          
& $-$10\%         & 1.4$\sigma$          & 10 & \nodata & \nodata \\ 
Mrk\,926          
& $-$9\%          & 1.4$\sigma$          & 10 & \nodata & \nodata \\ 
NGC\,3783         
& $+$13\%         & 1.7$\sigma$          & 5  & \nodata & \nodata \\ 
NGC\,4593         
& $+$30\%         & 3.4$\sigma$          & 5  &  $4.2\times10^{-3}$ & $1.2\times10^4$ \\ 
\hline
\hline 
\end{tabularx}
\par\vspace{1ex}
\parbox{1\columnwidth}{\small
\textbf{Note:} (1) Source name, (2) The flux difference observed between the 100\,GHz epoch\,1 and epoch\,2 observations in \%, (3) The significance of the variability observed between the 100\,GHz epoch\,1 and epoch\,2 observations in $\sigma$, (4) The time period over which the flux variations are observed, (5) The upper limit on the radius of the variable millimeter component in parsec for sources in which we detected variability at $>3\sigma$, and (6) The upper limit on the radius of the variable millimeter component in $R_{\rm g}$ for sources in which we detected variability at $>3\sigma$.}
\label{tab:SEDs}
\end{table}

\subsection{Variability}\label{results_var}

In addition, Figure\,\ref{fig:SEDs} shows the 100\,GHz epoch\,2 measurements (squares) for each source, which we use to assess any 100\,GHz variability in the sources.  
Equation\,\ref{eq:fluxunc} previously stated the computation of the flux density uncertainties, consisting of a systematic and statistical (i.e. $\sigma_{\rm rms}$) component.
We regard variations between the two 100\,GHz fluxes ($|\Delta S_{\rm 100GHz}|$) that exceed the flux uncertainty by more than 
$3\sigma$ as significant variability.
Here, $|\Delta S_{\rm 100GHz}|\gtrsim\sqrt{(\sigma_{S_{\rm 100GHz,ep1}})^2 + (\sigma_{S_{\rm 100GHz,ep2}})^2}=7\%$ for seven out of nine sources, where the $\sigma_{\rm stat}$ term is negligible compared to the $\sigma_{\rm sys}$ of 5\%.
For Ark\,120 and IRAS\,05078+1626, $\sigma_{\rm stat}$ contributes significantly to the total uncertainty, and including it in the expression for $|\Delta S_{\rm 100GHz}|$ raises the minimum fractional variability threshold to $\gtrsim10$\%.  
The variability detected in the majority of sources (six out of nine) is statistically insignificant, i.e. on the order of 1--2$\sigma$. 
In the remaining three sources, AX\,J1737.4--2907, IRAS\,05078$+$1626 and NGC\,4593, we detect variability between 30--55\% with significance levels on the order of 3--6$\sigma$. 
The variability observed in these three sources, along with their associated timescales and significances, is summarized in Table\,\ref{tab:SEDs}.
We note that the variability of AX\,J1737.4--2907 measured in this work ($\sim55$\%) differs from that reported by \cite{michiyama_alma_2024} ($\sim70$\%). We attribute this discrepancy to differences in the analysis procedure: (i) a slightly different imaging strategy, (ii) a smaller matched synthesized beam, and (iii) the flux estimator adopted. Although we report peak fluxes taken as the maximum pixel value, \cite{michiyama_alma_2024} derive peak fluxes from a two-dimensional Gaussian fit to the source using the CASA task \texttt{imfit}.
Furthermore, as described in Section\,\ref{ALMAImaging}, AX\,J1737.4--2907 exhibits extended emission. To verify that the measured variability is associated with the central component rather than the extended structure, we applied a multicomponent Gaussian model to the two 100\,GHz epochs using the \textsc{CASA} task \texttt{imfit}. The resulting fits indicate that the flux density of the extended emission is consistent between the two observations, whereas the central peak varies. This conclusion is also in agreement with the results of \cite{michiyama_alma_2024}.

Furthermore, to ensure that the observed $>3\sigma$ variability in three out of nine sources is intrinsic to our sources, we examine the calibrators and check sources in these ALMA observations for signs of flux changes. 
The flux/bandpass calibrators were monitored multiple times per week during our observation periods and exhibit flux density variations of $<5$\,\% across these epochs, consistent with the assumed calibration uncertainties at 100\,GHz. 
Both the phase calibrators and check sources remained stable, although these primarily serve to verify calibration and do not directly affect the target flux densities.
Further details are presented in Figure\,\ref{fig:calibrators_3sources} (Appendix\,\ref{AppALMA}).
We therefore conclude that the variability observed in our targets is not driven by any significant changes in the calibrators.

\begin{figure}[h!]
    \centering
    \includegraphics[width=1\columnwidth]{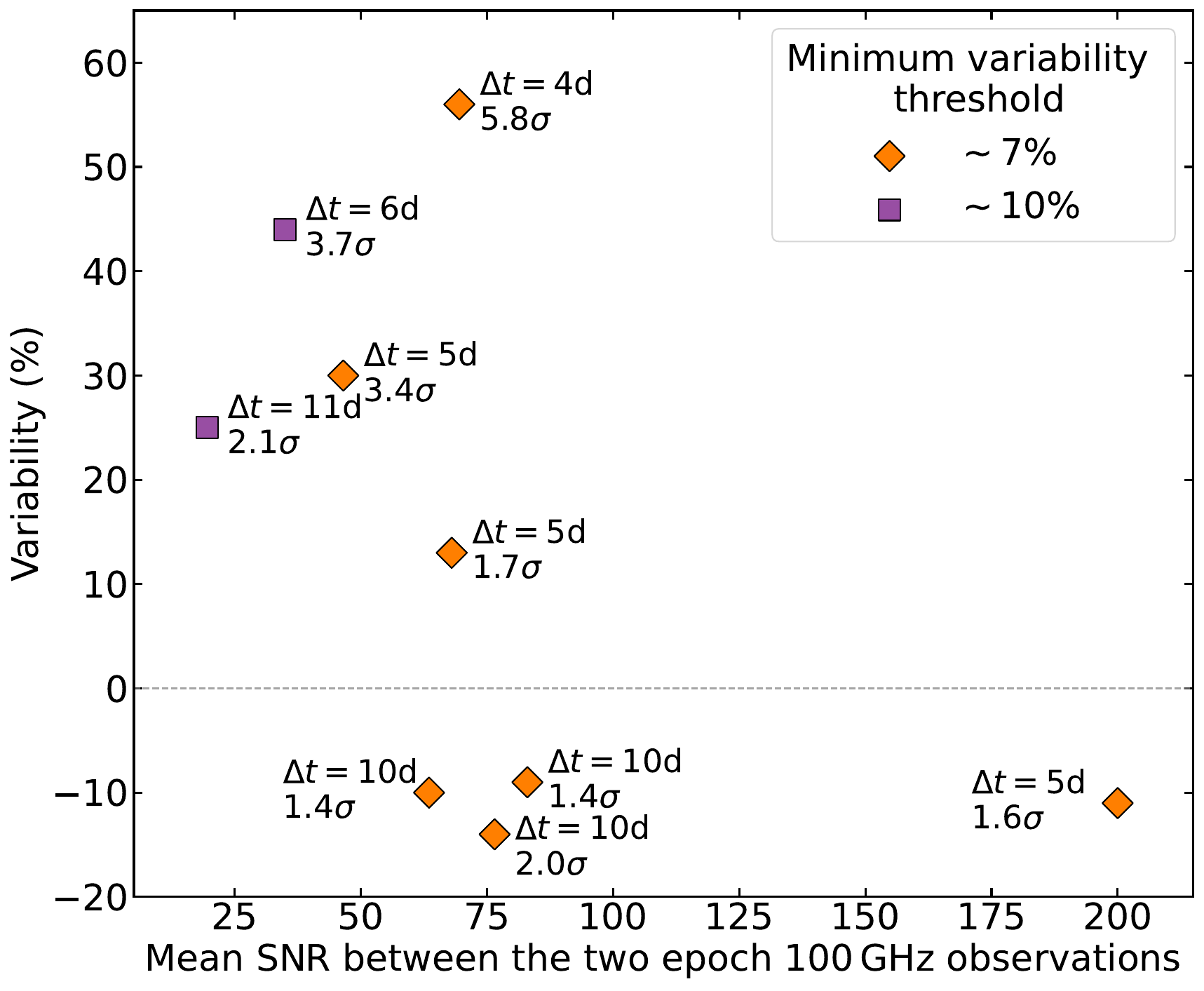}
    \caption{Fractional variability measured between the two 100\,GHz epochs as a function of the mean SNR.
    The minimum variability threshold for each source is determined based on the total flux uncertainty, as described in Section\,\ref{results_var}.
    Variability exceeding 7\% (or 10\% for Ark\,120 and IRAS\,05078+1626) is classified as tentative or significant depending on its significance level, which is indicated in the figure for each source together with the time period in days over which the variability was observed. Variability detections with a significance of $>3\sigma$ are considered significant.}
    \label{fig:var_vs_SNR}
\end{figure}

Furthermore, J.S.\,Elford et al. (2026, in preparation) studied 40 RQ AGN, detecting millimeter variability on timescales of hours to years, and found that the detection of variability depends on SNR. However, the ability to detect variability is primarily compromised at $\rm SNR<20$, where our self-calibrated observations have $\rm SNR \sim 24$--208.
Nevertheless, we test whether our sources show any trend with SNR. 
Figure\,\ref{fig:var_vs_SNR} shows the observed fractional variability as a function of the average SNR between the two epochs of 100\,GHz observations.
We do not observe a decline of variability with SNR.

\section{Discussion}\label{Discussion}

We constructed SEDs for all nine RQ AGN using quasi-simultaneous ALMA observations at 100\,GHz, 145\,GHz, and 233\,GHz, from which we derived the spectral indices ($\alpha_{\rm mm}$) of the millimeter emission. 
Additionally, we detected variability over timescales of 4--6\,days and in the range of 30--56\%, with a statistical significance exceeding $3\sigma$, in three out of nine sources.

\subsection{Spectral shape and Brightness temperature}\label{synchr_origin}

Because fitting an SSA model to the data would require assumptions on the slope of the electron energy distribution ($p$), we instead fit the data using a simple power-law model.
We measure millimeter spectral indices ranging from $\alpha_{\rm mm}=-0.05^{+0.09}_{-0.09}$ to $0.88^{+0.12}_{-0.11}$, see Figure\,\ref{fig:alpha_histogram} and Table\,\ref{tab:alpha}. 
As described in Section\,\ref{Intro}, optically thin synchrotron emission typically exhibits spectral indices of $\alpha_{\rm synchr}^{\rm opt.thin}=0.5$--1.0 \citep[e.g.,][]{behar_discovery_2015}, whereas optically thick synchrotron emission is characterized by $\alpha_{\rm synchr}^{\rm opt.thick}=-2.5$, due to SSA. In contrast, free--free emission produces much flatter spectra with $\alpha_{\rm ff}\approx0.1$.
Of our nine sources, only Mrk\,926 shows a spectral index clearly consistent with optically thin synchrotron emission ($\alpha_{\rm mm}=0.88^{+0.12}_{-0.11}$). 
The remaining sources have flatter spectra that could be consistent with a contribution from, e.g., free--free emission. 
Therefore, the spectral slopes alone cannot conclusively identify the origin of the millimeter emission. 
However, for the three significantly variable sources, we can place additional constraints indicating a nonthermal origin for the millimeter emission and thereby disfavor alternative mechanisms such as free--free radiation or thermal dust emission.

The brightness temperature ($T_{\rm b}$) can provide further insights into this origin.
Following \cite{RybickiLightman86}, we define $T_{\rm b}$ as
\begin{equation}
    T_{\mathrm{b}} \equiv \frac{I_\nu c^2}{2 k \nu^2},
    \label{eq:brightnessT}
\end{equation}
where $I_\nu$ is the specific intensity, $c$ is the speed of light, $k$ is the Boltzmann constant of $1.38\times10^{-16}\,\mathrm{erg\,K^{-1}}$, and $\nu$ is the rest-frame frequency. 
For an unresolved source, the brightness temperature associated with the observed flux-density variability can be expressed as
\begin{equation}
    T_{\rm b} =
    \frac{2 \ln 2}{\pi k}
    \frac{D_{\mathrm{L}}^2 \Delta S_\nu}
    {\nu_{\mathrm{obs}}^2 \Delta t^2 (1+z)^4},
\end{equation}
where $\Delta S_\nu$ is the difference between the flux densities measured at the two 100\,GHz epochs, $D_{\mathrm{L}}$ is the luminosity distance, $\nu_{\mathrm{obs}}$ is the observing frequency, and $\Delta t$ is the observed time interval between the two epochs \citep{Kang2021}. 
This expression follows from $I_\nu = \Delta S_\nu/\Omega$, where the solid angle of the unresolved source is approximated by that of a circular Gaussian,
\begin{equation}
    \Omega = \frac{\pi \theta^2}{4\ln 2},
\end{equation}
with the angular size estimated from the light-travel distance associated with the variability timescale:
\begin{equation}
    \theta = \frac{c\Delta t}{D_{\rm A}(1+z)},
\end{equation}
where $D_{\rm A}=D_{\rm L}/(1+z)^2$ is the angular-diameter distance.  
A detailed derivation of the variability $T_{\rm b}$ can be found in, e.g., \cite{Kang2021}. 
For the three variable sources in this work: AX\,J1737.4--2907, IRAS\,05078$+$1626, and NGC\,4593, we derive $T_{\rm b}=1.86\times10^9$\,K, $3.81\times10^8$\,K, and $7.52\times10^7$\,K, respectively. Because $c\Delta t$ is constrained only by upper limits, as will be discussed in Section\,\ref{disc_var} as well, these estimates should be regarded as lower limits on $T_{\rm b}$.
Free--free emission from photoionized gas ($T_{\rm e} \sim 10^4$\,K) and thermal dust emission cannot produce values of $T_{\rm b} \sim 7.52\times10^7–1.86\times10^9$\,K \citep[e.g.,][]{Condon1992}, whereas synchrotron emission can reach up to $T_{\rm b} \sim 10^{11}–10^{12}$\,K \citep{Kellermann1969}. The variable component therefore requires at least mildly relativistic electrons, as expected for synchrotron emission from either a nonthermal population or the hot ($\sim10^9$\,K) thermal electrons of the corona itself \citep[e.g.,][]{inoue_unveiling_2014}.

Furthermore, free--free emission could originate from ionized gas associated with the dusty torus ($R_{\rm torus}\sim \rm few$ parsecs; e.g., \citealt{Carilli2019}), potentially producing the relatively flat spectral indices observed in some of our sources. 
However, variability from such an extended region is expected only on timescales of $\Delta t\sim R_{\rm torus}/\rm c$, corresponding to years \citep{Baskin2021}, whereas the millimeter variability observed in our sample occurs on much shorter timescales of $\sim$days (see Section\,\ref{results_var}). 
Similarly, thermal dust emission is expected to vary on long timescales and is therefore unlikely to dominate the observed emission.
In addition, the measured spectral indices are inconsistent with thermal optically thin dust emission, which is typically expected to be $\alpha_{\rm dust}\approx-3.5$ \citep[e.g.,][]{condon_nrao_1998,mullaney_defining_2011}. 
We also find no correlation between $\alpha_{\rm mm}$ and the SFR (Figure\,\ref{fig:alpha_vs_params} in Appendix\,\ref{alpha_vs}), indicating that star-formation-related emission processes do not dominate the millimeter continuum.
The short time-scale variability also rules out Anomalous Microwave Emission (AME), which is thought to arise from rapidly rotating dust grains and can produce a spectral peak resembling an SSA turnover \citep[e.g.,][]{Battistelli2019}.

Therefore, based on the inferred $T_{\rm b}$ estimates and the variability detected in three RQ AGN, we favor a nonthermal explanation for the millimeter emission in these three sources.
The spectral flattening seen in AX\,J1737.4--2907 and IRAS\,05078$+$1626, showing a concave-downward behavior when comparing $\alpha_{100-145}$ with $\alpha_{145-233}$ (see Table\,\ref{tab:alpha}), could arise if the SSA turnover frequency lies close to 100\,GHz.

\subsection{Dependence of $\alpha_{\rm mm}$ on AGN and host-galaxy properties}

Section\,\ref{synchr_origin} already stated the lack of correlation observed between $\alpha_{\rm mm}$ and SFR, suggesting that star formation does not dominate the observed millimeter emission.
Although our source sample is small and any observed relations might hold low statistical power, Figure\,\ref{fig:alpha_vs_params} (Appendix\,\ref{alpha_vs}) further presents potential correlations between $\alpha_{\rm mm}$ and the AGN parameters $L_{\rm bol}$, $\lambda_{\rm Edd}$ and $M_{\rm BH}$.
Since the sources in this work are part of the BAT survey, as described in Section\,\ref{Samplesel}, $L_{\rm bol}$, $\lambda_{\rm Edd}$ and $M_{\rm BH}$ are available through the BAT AGN Spectroscopic Survey (BASS) as published by \cite{koss_bass_2022}. 
The values for each source and the uncertainties $\log(M_{\rm BH})$ are reported in Table\,\ref{tab:Sources2}.
Based on the Spearman--$r$ correlation test, where $p<0.01$ indicates a significant correlation, we do not find any correlations between these parameters and $\alpha_{\rm mm}$, obtaining $p$--values of 0.16, 0.89, and 0.25, respectively. 
The scatter in $\alpha_{\rm mm}$ appears uniform across the full sampled ranges of $\log (L_{\rm bol} /\rm erg\,s^{-1}) \sim 44.00$ to 45.85, $\log (\lambda_{\rm Edd }) \sim-0.41$ to $-1.20$, and $\log (M_{\rm BH} / M_{\odot}) \sim 6.88$ to 8.19.
Nevertheless, our sample could be too small to confirm the presence or absence of these correlations and their physical implications. 
Future studies with larger samples will be able to shed light on the presence of possible trends.

\subsection{The size of the variable millimeter-emitting region}\label{disc_var}

Section\,\ref{synchr_origin} argued that the detected millimeter emission, at least in the three sources exhibiting statistically significant variability, is most consistent with a nonthermal origin. 
To further understand the physical process producing this emission, we aim to place constraints on the typical size of the millimeter-emitting region, since such bounds can help distinguish emission scenarios.
As described in Section\,\ref{results_var}, at 100\,GHz, we observe significant $>$3\,$\sigma$ variability ranging between 30\% and 56\% in three RQ AGN: AX\,J1737.4--2907, IRAS\,05078$+$1626 and NGC\,4593.
The observed flux variability can be used to constrain upper limits on the physical size of the variable millimeter-emitting region ($R_{\rm mm}$) through: 
\begin{equation}
    R_{\rm mm} \leq c \times \Delta t,
\end{equation}
where $\Delta t$ is the period over which variability is observed, and $c$ is the speed of light. 
The resulting upper limits on the size of the varying millimeter-emitting region ($R_{\rm mm}^{\rm ul}$) are presented in Table\,\ref{tab:SEDs}, in units of parsec and the gravitational radius. 
These sizes range from $3.4\times10^{-3}$\,pc to $5.0\times10^{-3}$\,pc, corresponding to $1.0\times10^3$--$1.4\times10^4$\,$R_{\rm g}$, based on the $M_{\rm BH}$ estimates from BASS (see Table\,\ref{tab:Sources2}) and should be regarded as conservative upper limits.

These obtained radii are significantly larger than the typical order-of-magnitude size of the X-ray corona, which is estimated to be $R_{\rm c}\sim10-100\,R_{\rm g}$ (see Section\,\ref{Intro}).
However, it remains plausible that the sources in this sample can exhibit variability on timescales shorter than those considered here. 
For instance, we find only $1.6\sigma$ evidence for variability in IC\,4329A, while \citet{shablovinskaya_joint_2024} report significant millimeter variability on one-day timescales in this object.
Furthermore, our upper limits are consistent with the constraint on the millimeter-emitting region size of $\leq10^4\,R_{\rm g}$ as reported by \cite{Hankla2025}, who argues that the millimeter emission might arise from an extended region outside the X-ray corona, though magnetically linked to it.
Additionally, the size of the millimeter-emitting region in AX\,J1737.4--2907 was previously studied by \cite{michiyama_alma_2024}, who derived an upper limit of $R_{\rm mm}<200\,R_{\rm g}$, which is smaller than our estimate of $R_{\rm mm}\lesssim1.0\times10^3\,R_{\rm g}$. 
This discrepancy is due to to the different $M_{\rm BH}$ assumptions. In particular, \citet{michiyama_alma_2024} adopted $\log(M_{\rm BH}/M_{\odot})=8.5$, inferred from the broad H$\alpha$ emission line \citep{Tortosa2017}, whereas we adopt $\log(M_{\rm BH}/M_{\odot})=7.84$, corresponding to a $M_{\rm BH}$ that is smaller by a factor of $\approx 4.5$.
By modeling the SED using Equation\,\ref{eq:SSA} with an assumed electron energy distribution index of $p=2.7$, and including a jet component at centimeter wavelengths, \citet{michiyama_alma_2024} placed an even tighter constraint on the coronal size, finding $R_{\rm c}=20\,R_{\rm g}$. 
Section\,\ref{synchr_origin} also demonstrated that our observed variability timescales restrict the millimeter-emitting region's size to well within the torus.
Nevertheless, our results provide constraints on the variability and, consequently, set upper limits on the size of the millimeter-emitting region in three RQ AGN of $R_{\rm mm}\leq 10^4\,R_{\rm g}$.

\subsection{Possible jet and dust contributions}\label{DiscJet}

\begin{figure*}[ht!]
    \centering
    \includegraphics[width=0.90\linewidth]{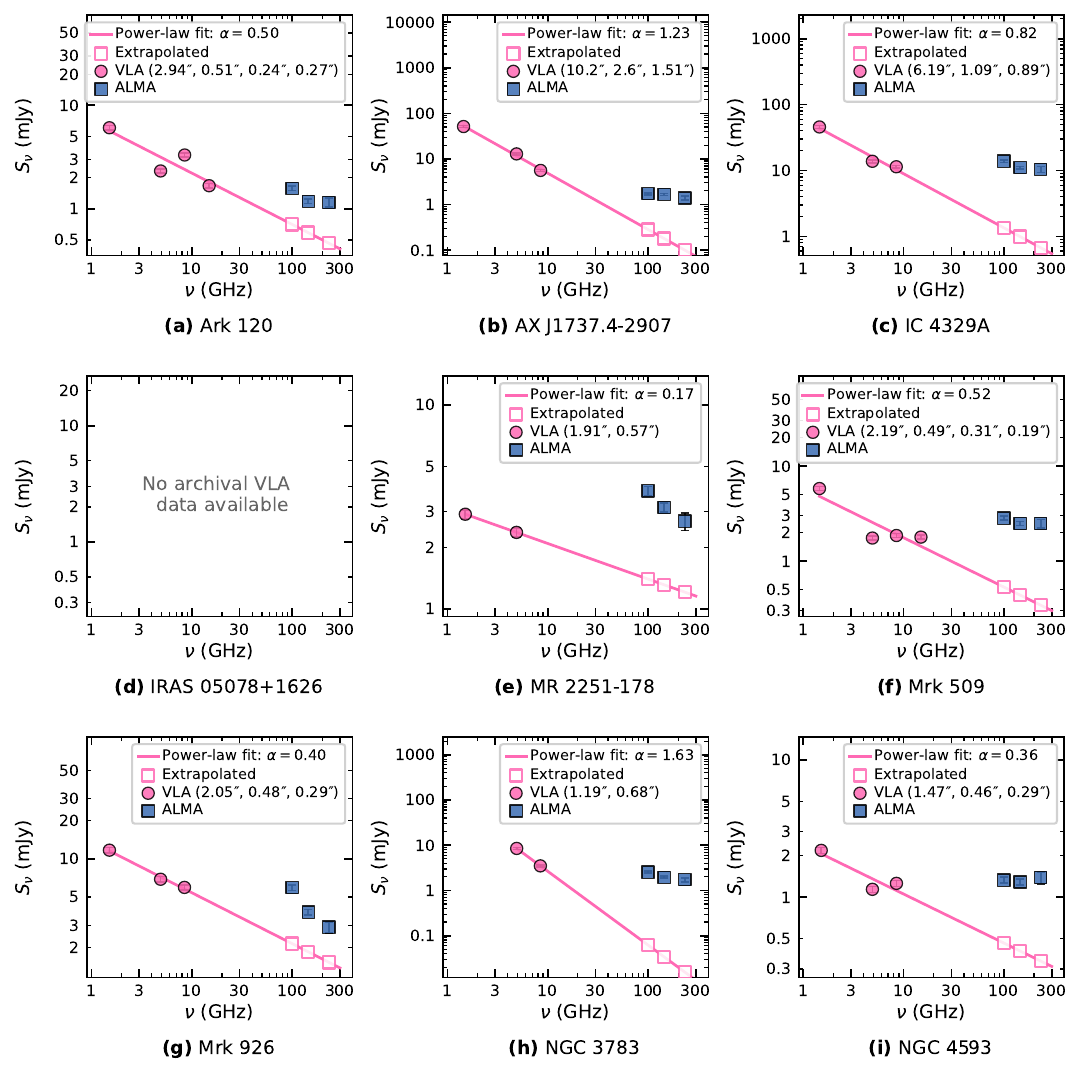}
    \caption{Archival VLA centimeter-band flux densities retrieved from the NRAO VLA Archive for each source in our sample. The pink lines show power-law fits to these data, extrapolated to the frequency range covered by our ALMA observations (100--233\,GHz). Furthermore, we present our ALMA observations as previously shown in Figure\,\ref{fig:SEDs}. The legend provides the values of the full width at half maximum (FWHM) of the major axis of the beams for the archival VLA data. Higher radio frequencies correspond to smaller beams, i.e., higher angular resolution.
    }
    \label{fig:radio}
\end{figure*}

Jet or dust emission, either nuclear or extended, could, however, still contribute to the observed millimeter emission.
As can be seen from Figure\,\ref{fig:alma_images} (Appendix\,\ref{AppALMA}), most of our sources appear unresolved; with the exception of AX\,J1737.4--2907, which, as previously noted by \citet{michiyama_alma_2024}, shows an extended structure in the 100\,GHz observations, potentially pointing to e.g., jet contamination. 
To assess whether optically thin synchrotron emission from a compact jet, which may be present in RQ AGN (see Section\,\ref{Intro}), contributes to the observed millimeter emission, we present archival VLA centimeter-band observations of our sources in Figure\,\ref{fig:radio}.
We fit the radio spectrum with a power-law, $S_\nu \propto \nu^{-\alpha}$, and extrapolate it to the ALMA frequencies to estimate the contribution of optically thin synchrotron emission from e.g., a jet to the observed millimeter flux density.
We obtain radio spectral indices of $\alpha_{\rm radio}=0.17$--1.63 over the 1.5--14.9\,GHz range.
An extrapolation to 100--233\,GHz yields a contribution that is small relative to the measured millimeter emission, with the caveats that it does not account for variability or for the different resolutions between the VLA and ALMA observations. 
On average, the mean observed/expected flux ratio is $\sim5$--7 across the three ALMA frequencies, excluding the outlier NGC\,3783, for which the ratios are much higher, $\sim40$--110.
Furthermore, we find no evidence for a flat-spectrum radio core (see Section\,\ref{Intro}), which could otherwise contribute to the flatter spectral slopes observed in our ALMA data.
However, the VLA's arcsecond resolution can potentially blend the nuclear emission with steeper-spectrum extended emission. We therefore cannot rule out a compact, flat- or inverted-spectrum core on milliarcsecond scales, which could contribute to the flatter slopes in our ALMA data.
For AX\,J1737.4--2907 and IC\,4329A more detailed broadband SED analysis has been performed \citep{inoue_detection_2018,michiyama_alma_2024}.

Furthermore, seven of the nine sources show a steeper spectral slope between 100--145\,GHz than between 145--233\,GHz (see Table,\ref{tab:alpha}). 
Individually, these differences are not significant ($<2\sigma$), but the sign of the change is the same in all seven sources. 
This concave-upward spectrum is not what a SSA turnover near 100\,GHz would produce, as mentioned in Section\,\ref{synchr_origin}, and instead suggests an additional component, possibly thermal dust emission, contributing at 233\,GHz.

\begin{figure*}[ht!]
    \centering
    \includegraphics[width=\textwidth]{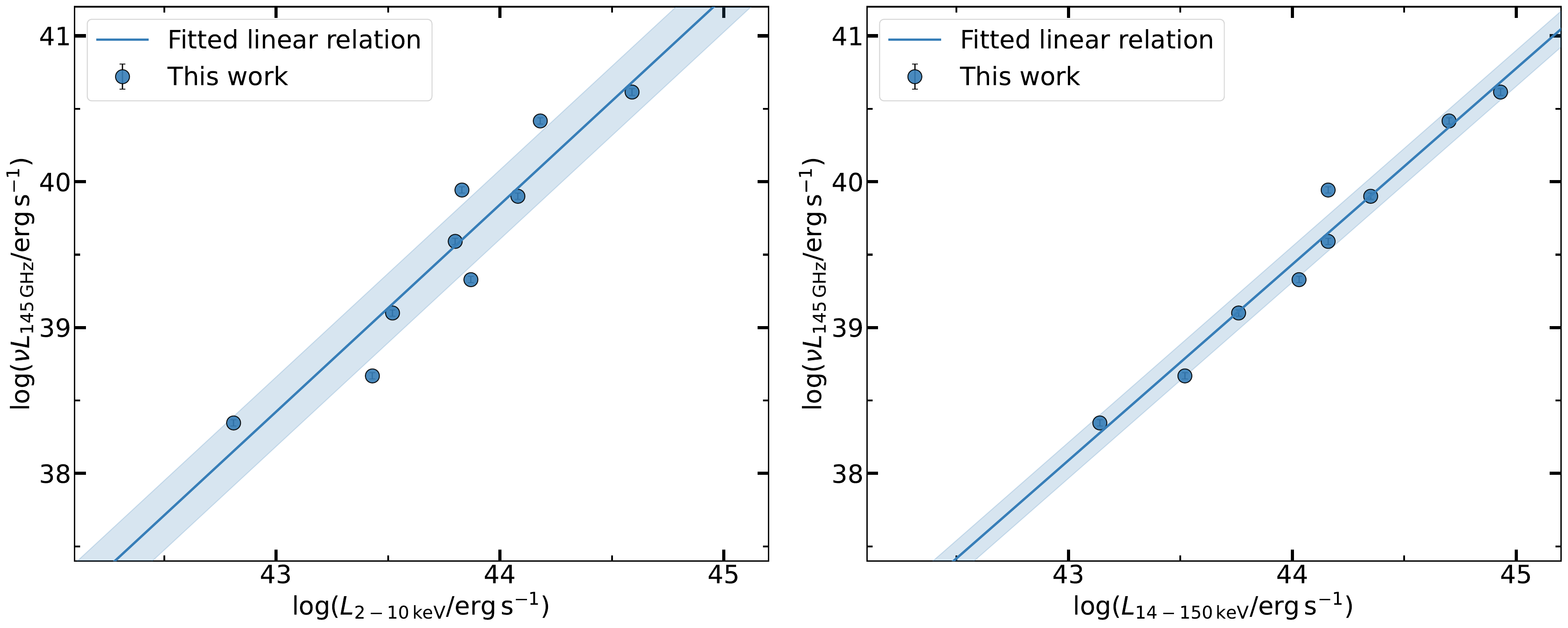}
    \caption{The 145\,GHz millimeter emission versus 2--10\,keV and 14--150\,keV X-ray emission. We obtained the first relations between the 145\,GHz and X-ray emission, presented in Equations\,\ref{eq:L145_210} and \ref{eq:L145_14150}. Our 100\,GHz and 233\,GHz data overplotted with the existing relations from \cite{ricci_tight_2023}, \cite{Venselaar26} and \cite{kawamuro_bass_2022} are presented in Figure\,\ref{fig:mmXray} (Appendix\,\ref{AppmmXray}).}
    \label{fig:band4}
\end{figure*}

\subsection{The millimeter/X-ray relation}\label{mmXray}

A compact origin of the millimeter emission in either the corona or a compact jet, as discussed in Sections\,\ref{synchr_origin} and \,\ref{disc_var}, is supported by the tight linear relation observed between millimeter and X-ray emission. 
Recent studies found such a tight correlation between the coronal 2--10\,keV and 14--150\,keV X-ray emission (observed by Swift, averaged over several years; \citealt{ricci_bat_2017}) and the nuclear millimeter emission measured with ALMA at 100\,GHz \citep{ricci_tight_2023} and 233\,GHz \citep{kawamuro_bass_2022} in nearby RQ AGN.
Though the millimeter and X-ray data were not obtained (quasi-)simultaneously, which may introduce additional scatter due to source variability, these correlations supported the proposed shared origin between these two types of emission. 
Furthermore, using new quasi-simultaneous millimeter and X-ray observations, \cite{Venselaar26} recently reported that the 100\,GHz/X-ray luminosity ratio increases for more luminous AGN ($\log (L_{\rm bol}/\rm erg\,s^{-1})= 45.3$--46.3). This trend is potentially driven by enhanced coronal cooling at higher $L_{\rm bol}$ or additional millimeter emission originating from outflows that are expected at higher $\lambda_{\rm Edd}$.

With the ALMA data in this work, we can further explore the existence of this millimeter/X-ray correlation.
Here, we adopt the time-averaged X-ray luminosities at 2--10\,keV ($L_{2-10}$) and 14--150\,keV ($L_{14-150}$) from \cite{ricci_bat_2017} obtained from the BAT survey, see Table\,\ref{tab:Sources2}.
Figures\,\ref{fig:band4} and \ref{fig:mmXray} (Appendix\,\ref{AppmmXray}) present the millimeter/X-ray relation for all three ALMA bands in this work (100\,GHz, 145\,GHz, and 233\,GHz) and the two previously described X-ray energy ranges (2--10\,keV and 14--150\,keV).

For the 100\,GHz data (Figure\,\ref{fig:band3}), we have added the relations obtained by \cite{ricci_tight_2023} and \cite{Venselaar26} to our data points. 
Furthermore, Figure\,\ref{fig:band3} also presents the values for $z$, $\lambda_{\rm Edd}$, $L_{\rm bol}$, the bolometric correction ($\kappa_{2-10}$), $L_{2-10}$, $L_{14-150}$ and the resolution ($\theta$) for the three different samples.
The sources in this work seem to closely follow the shift between the relations of \cite{ricci_tight_2023} and \cite{Venselaar26} as one moves to higher luminosities.

At 145\,GHz, the millimeter/X-ray relation has not previously been studied. Therefore, we present the first comparison between 145\,GHz millimeter and X-ray emission in Figure\,\ref{fig:band4}, fitting tight linear relations of the form:
\begin{equation}
\begin{split}
    \log(L_{\rm 145GHz}/\rm erg\,s^{-1}) &= (1.42\pm0.19) \log(L_{\rm 2-10}/\rm erg\,s^{-1})  \\
    &\quad - (22.67\pm8.26),
\end{split}
\label{eq:L145_210}
\end{equation}
with an intrinsic scatter of 0.24 and a $p$--value of $5.0\times10^{-4}$, and
\begin{equation}
\begin{split}
    \log(L_{\rm 145GHz}/\rm erg\,s^{-1}) &= (1.34\pm0.09)\,\log(L_{\rm 14-150}/\rm erg\,s^{-1}) \\
    &\quad - (19.69\pm3.90),
\end{split}
\label{eq:L145_14150}
\end{equation}
with an intrinsic scatter of 0.12 and a $p$--value of $1.3\times10^{-5}$.
While we find a strong linear correlation, larger and less strictly selected samples of RQ AGN will be needed to test how generally this relation holds and to derive robust limits on its intrinsic scatter. 
Simultaneous millimeter/X-ray observations may also help to decrease the measured scatter.

For completeness, in Figure\,\ref{fig:band6} (Appendix\,\ref{AppmmXray}), we also present our 233\,GHz data overplotted with the relation reported by \cite{kawamuro_bass_2022}, given that this relation was originally derived from the 233\,GHz observations of 98 AGN, including the nine sources in our sample.

Based on the tight relation with coronal X-ray emission, a nonthermal origin in the corona or a compact jet, as discussed in Sections\,\ref{synchr_origin} and \ref{disc_var}, both remain viable.
With the currently available data, we are unable to discriminate between these two interpretations; they may even represent the same underlying physical structure, as suggested for X-ray binaries \citep[e.g.,][]{Markoff2005}. 
Resolving this would require simultaneous millimeter/X-ray monitoring.

\section{Conclusions}\label{Conclusion}

This work presents the first quasi-simultaneous, three-band (100--233\,GHz) millimeter SED study of RQ AGN, combined with a few-day variability analysis at 100\,GHz.

\begin{enumerate}
    \item From the power-law fits to the SEDs, we derive spectral indices ($\alpha_{\rm mm}$) between $-0.05^{+0.09}_{-0.09}$ and $0.88^{+0.12}_{-0.11}$, with a mean value of $\alpha_{\rm mm}^{\rm av}=0.35\pm0.09$. 
    This range could be compatible either with optically thin synchrotron radiation or (a contribution of) free--free emission (Section\,\ref{synchr_origin}).
    The spectral index alone is insufficient to confirm the emission mechanism of the observed millimeter emission.
    However, for the three sources exhibiting substantial variability, the inferred brightness temperatures ($T_{\rm b} \sim 7.52\times10^7$--$1.86\times10^9$\,K), together with the $\sim$days variability timescales, are indicative of a nonthermal origin for the millimeter emission.

     \item We observe no correlation between $\alpha_{\rm mm}$ and the parameters $L_{\rm bol}$, $\lambda_{\rm Edd}$, $M_{\rm BH}$ and SFR (Section\,\ref{synchr_origin} and Appendix\,\ref{alpha_vs}). Although future studies with larger samples are needed to statistically confirm this. 

     \item We examined whether the intraband spectral indices at each frequency are representative of the broadband spectral slope (Appendix\,\ref{intrabroad}). In four sources, significant scatter between the agreement of $\alpha_{\rm broadband}$ and $\alpha_{\rm intraband}$ is observed, potentially due to the lower SNR of the observations. 
     This suggests that the intraband spectral index should be used with caution, particularly for $\rm SNR<20$, as a proxy for the broadband spectral slope.

     \item Extrapolating archival VLA data to 100--233\,GHz, we find that the optically thin synchrotron emission that dominates at low frequencies (e.g., from a jet) contributes only a small fraction to the observed ALMA flux densities (Figure\,\ref{fig:radio} and Section\,\ref{DiscJet}). The average observed/expected flux ratios are $\sim$5--7 across the three ALMA frequencies, excluding the outlier NGC\,3783, for which the ratios are up to $\sim$110. 
    
    \item We detect variability at the 30--55\% level, with a significance of $>3\sigma$, in three of the nine sources.
    This leads to upper-limit constraints on the size of the variable millimeter-emitting region of $R_{\rm mm}\leq1.0\times10^3$--$1.4\times10^4$\,$R_{\rm g}$ (Section\,\ref{disc_var}).

    \item We present the millimeter/X-ray relation in all three ALMA bands (Section\,\ref{mmXray} and Appendix\,\ref{AppmmXray}).  
    Based on our nine sources, we obtained the first relation between 145\,GHz and X-ray emission, fitting a linear relation (Equations\,\ref{eq:L145_210} and \ref{eq:L145_14150}) with an intrinsic scatter of 0.24 and 0.12\,dex, respectively.

\end{enumerate}

In this work, we were unable to pinpoint the location of the SSA peak in the spectrum. 
However, future 40\,GHz (ALMA Band\,1) observations could help determine this feature, thereby providing tighter constraints on both the size of the millimeter-emitting region and the magnetic field strength through spectral modeling. 
A stratified coronal model, applied to the observations analyzed in this work together with available radio observations of IC\,4329A, will be presented in M.\,Hamamoto et al. (2026, in preparation) and will also quantify any contributions from jets or star formation.
In addition, future monitoring of these sources on hour-long timescales, similar to that carried out for IC\,4329A by \cite{shablovinskaya_joint_2024}, could uncover even faster variability, providing tighter observational limits on the millimeter-emitting region.

\section*{Acknowledgments}

We thank the anonymous referee for the helpful comments, which improved the quality of this work.

S.V. acknowledges support from SNSF Consolidator grant F01--13252 and the China-Chile joint research fund. 
C.R. acknowledges support from SNSF Consolidator grant F01--13252, Fondecyt Regular grant 1230345, ANID BASAL project FB210003 and the China-Chile joint research fund.
Y.I. is supported by JST FOREST Program (Grant Number JPMJFR256L, Japan), MEXT Supporting Pioneering Research through AI for 1,000 Discovery challenges Program (SPReAD) Japan Grant Number JPMXP1726291302, and JSPS KAKENHI grant Nos. JP22K18277, JP26H00604, and JP26K22331.
N.B. is supported by JSPS KAKENHI grant No. JP23F23773, JP24K00673 and JP23H04892.
T.K. is supported by JSPS KAKENHI grant No. 24K00673. 
S.A. thankfully acknowledges support from Knut and Alice Wallenberg (KAW) foundation (KAW 2023.0298 Scholar). 
E.S. acknowledges support by the Alexander von Humboldt Foundation through a Humboldt Research Fellowship.
E.T. acknowledges support from the ANID CATA-BASAL program FB210003, and FONDECYT Regular 1241005 and  1250821.

We also thank the internal Max-Planck-Institut f\"ur Radioastronomie reviewer, Aditya Tamar, for his helpful comments.

This paper makes use of the following ALMA data: ADS/JAO.ALMA\#2018.1.00576.S. 
ALMA is a partnership of ESO (representing its member states), NSF (USA), and NINS (Japan), together with NRC (Canada), MOST and ASIAA (Taiwan), and KASI (Republic of Korea), in cooperation with the Republic of Chile. 
The Joint ALMA Observatory is operated by ESO, AUI/NRAO, and NAOJ.
The National Radio Astronomy Observatory and Green Bank Observatory are facilities of the U.S. National Science Foundation operated under cooperative agreement by Associated Universities, Inc.

Facilities: ALMA.

\bibliographystyle{aasjournal}
\bibliography{ref.bib}

\appendix

\counterwithin{figure}{section}
\renewcommand{\thefigure}{\thesection\arabic{figure}}

\section{ALMA observations}\label{AppALMA}

We have imaged two epochs of ALMA 100\,GHz (Band\,3) observations and one epoch of 145\,GHz (Band\,4) and 233\,GHz (Band\,6) observations.

\begin{figure*}[ht!]
    \centering
    \begin{subfigure}[b]{0.49\textwidth}
        \includegraphics[width=\textwidth]{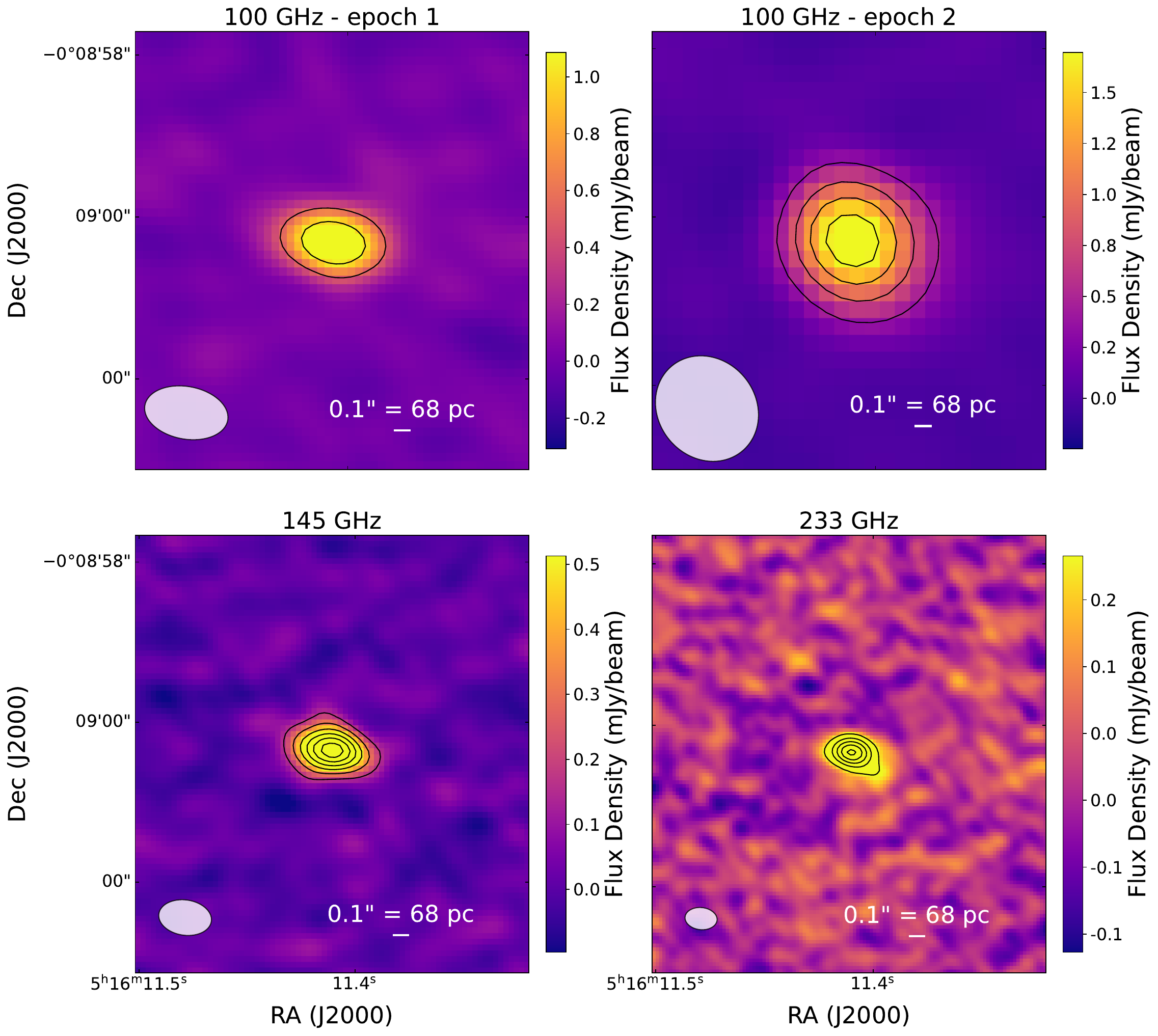}
        \caption{\textbf{Ark\,120}: $\mathrm{N}\times 5\sigma_{\rm rms}$ level contours.}
        \label{fig:alma_ark120}
    \end{subfigure}
    \hfill
    \begin{subfigure}[b]{0.49\textwidth}
        \includegraphics[width=\textwidth]{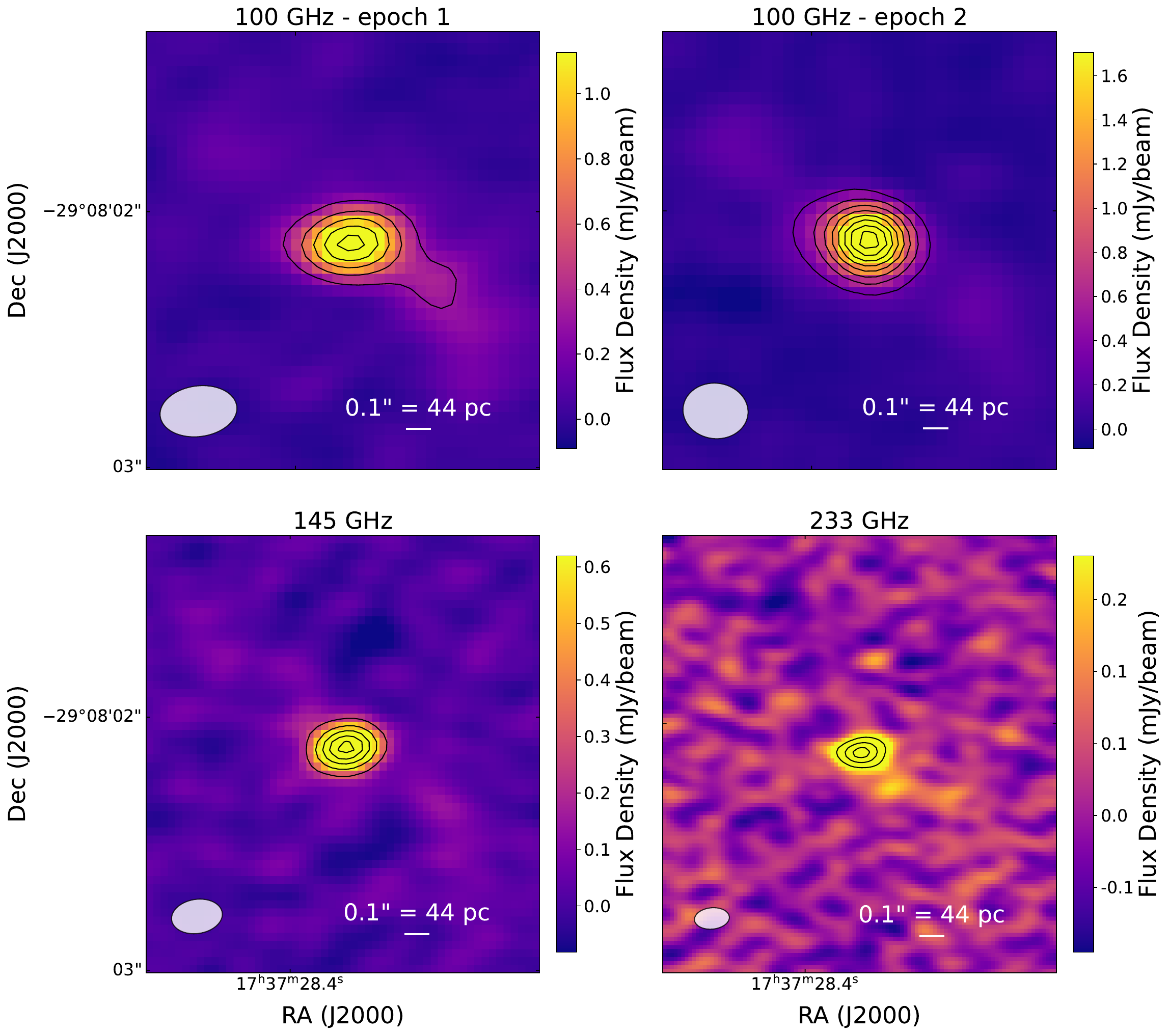}
        \caption{\textbf{AX\,J1737.4--2907}: $\mathrm{N}\times 10\sigma_{\rm rms}$ level contours.}
        \label{fig:alma_ax}
    \end{subfigure}

    \vspace{0.1cm}

    \begin{subfigure}[b]{0.49\textwidth}
        \includegraphics[width=\textwidth]{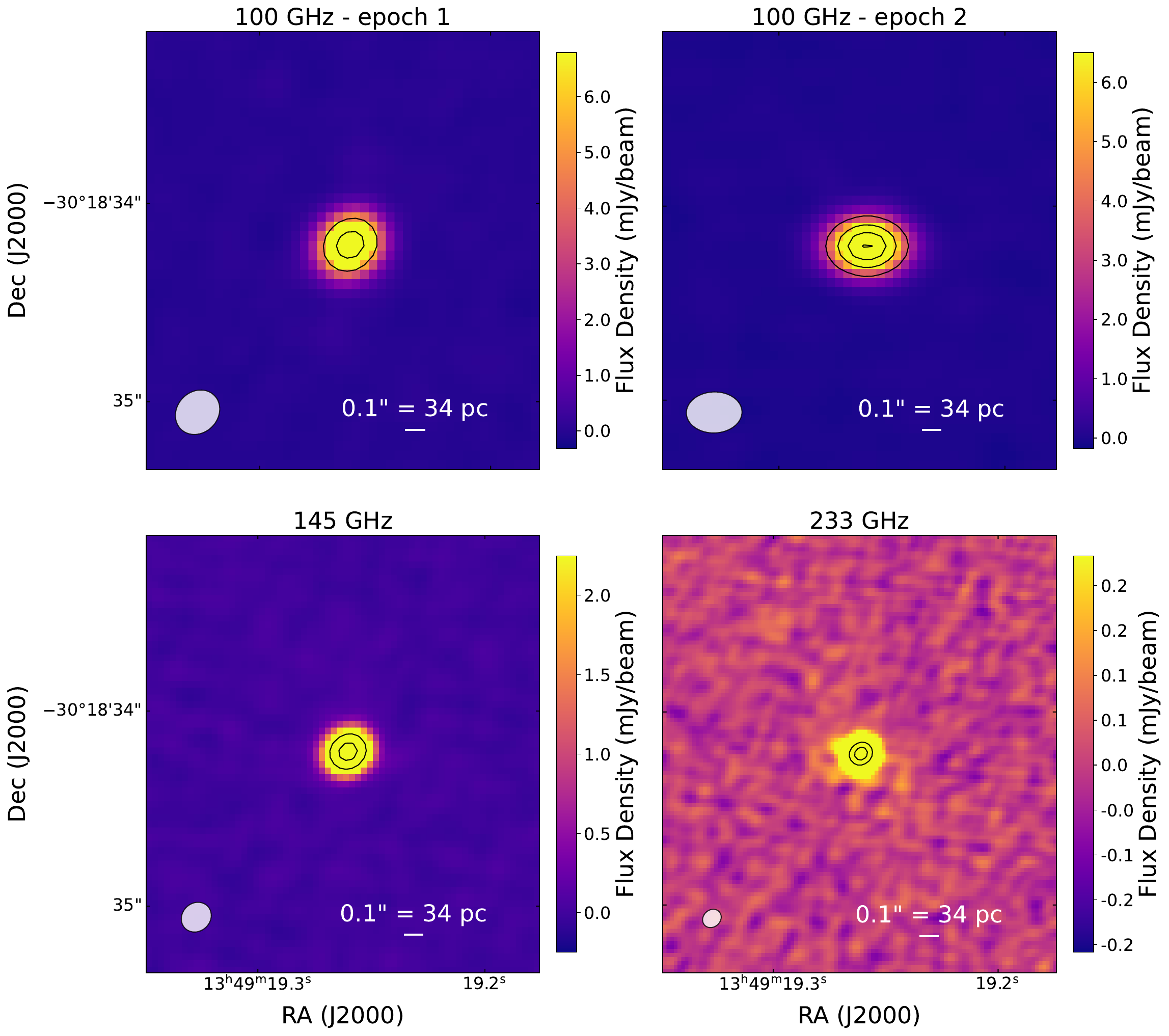}
        \caption{\textbf{IC\,4329A}: $\mathrm{N}\times 50\sigma_{\rm rms}$ level contours.}
        \label{fig:alma_ic4329a}
    \end{subfigure}
    \hfill
    \begin{subfigure}[b]{0.49\textwidth}
        \includegraphics[width=\textwidth]{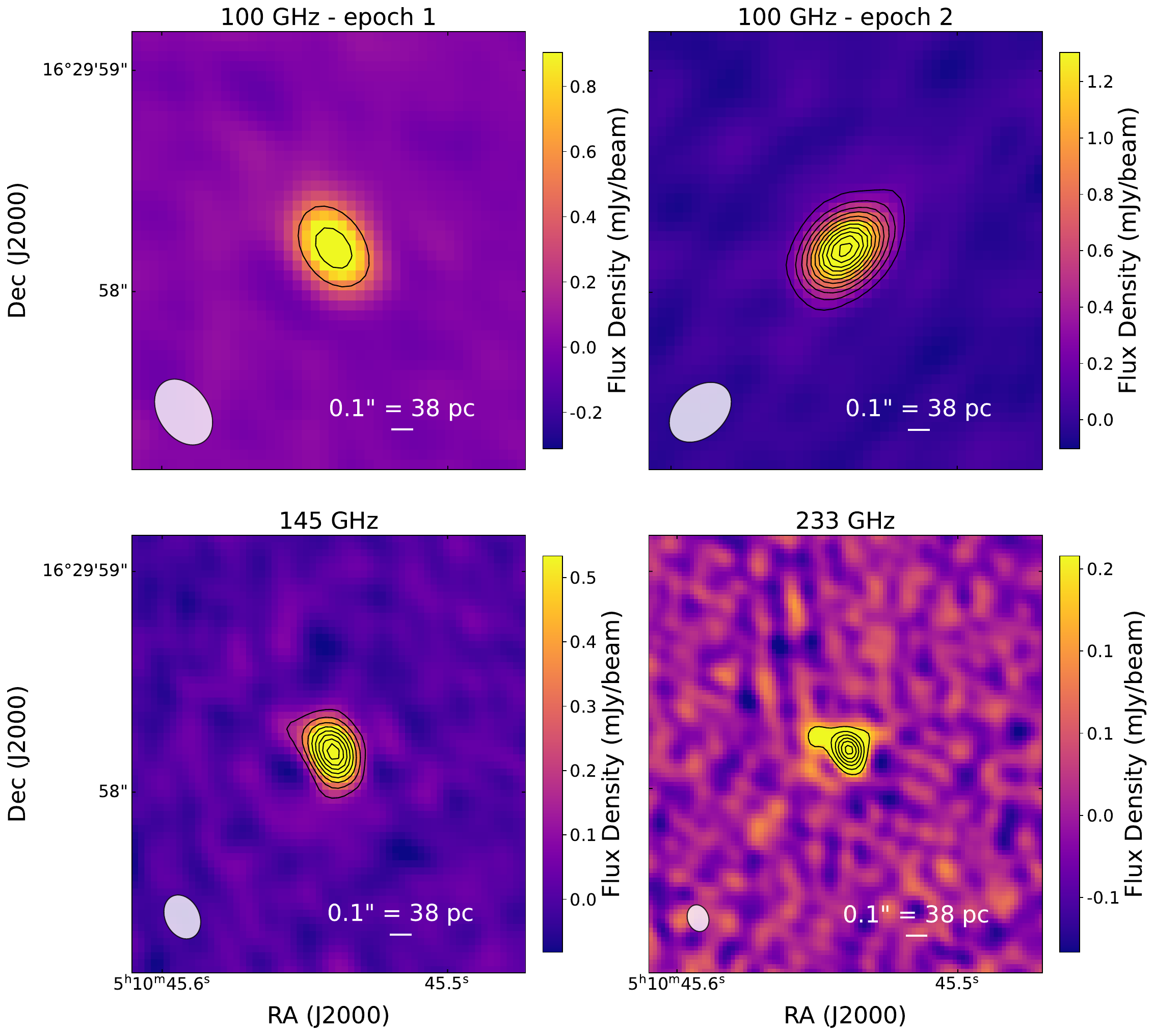}
        \caption{\textbf{IRAS\,05078+1626}: $\mathrm{N}\times 5\sigma_{\rm rms}$ level contours.}
        \label{fig:alma_iras}
    \end{subfigure}
    \caption{(Part 1 of 3) ALMA 100\,GHz (two epochs), 145\,GHz, and 233\,GHz images. Contours are overlaid. The beam is shown in the lower-left corner of each panel.}
    \label{fig:alma_images}
\end{figure*}
\begin{figure*}[ht!]
    \ContinuedFloat
    \centering
    \begin{subfigure}[b]{0.49\textwidth}
        \includegraphics[width=\textwidth]{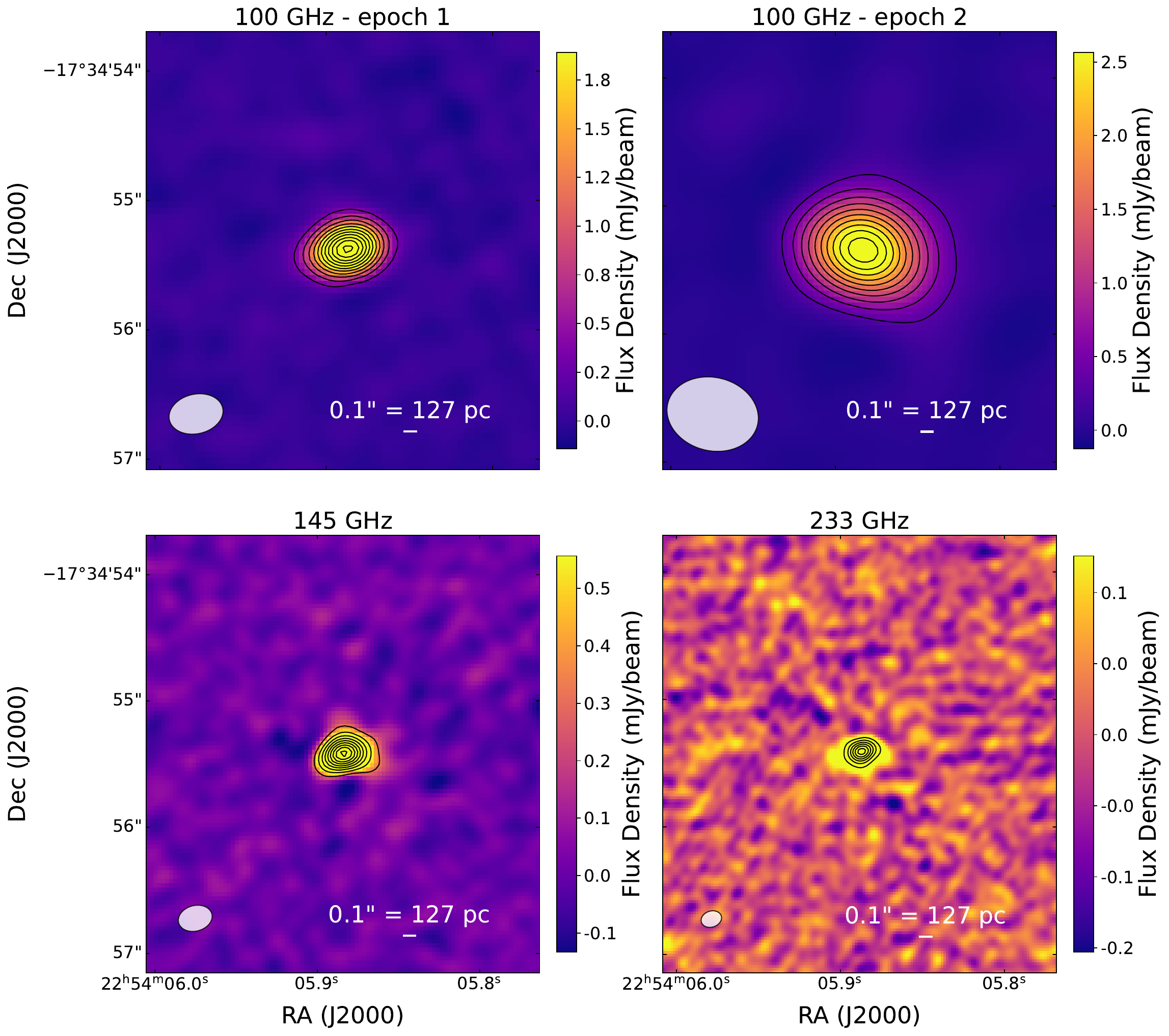}
        \caption{\textbf{MR\,2251--178}: $\mathrm{N}\times 7\sigma_{\rm rms}$ level contours.}
        \label{fig:alma_mr2251}
    \end{subfigure}
    \hfill
    \begin{subfigure}[b]{0.49\textwidth}
        \includegraphics[width=\textwidth]{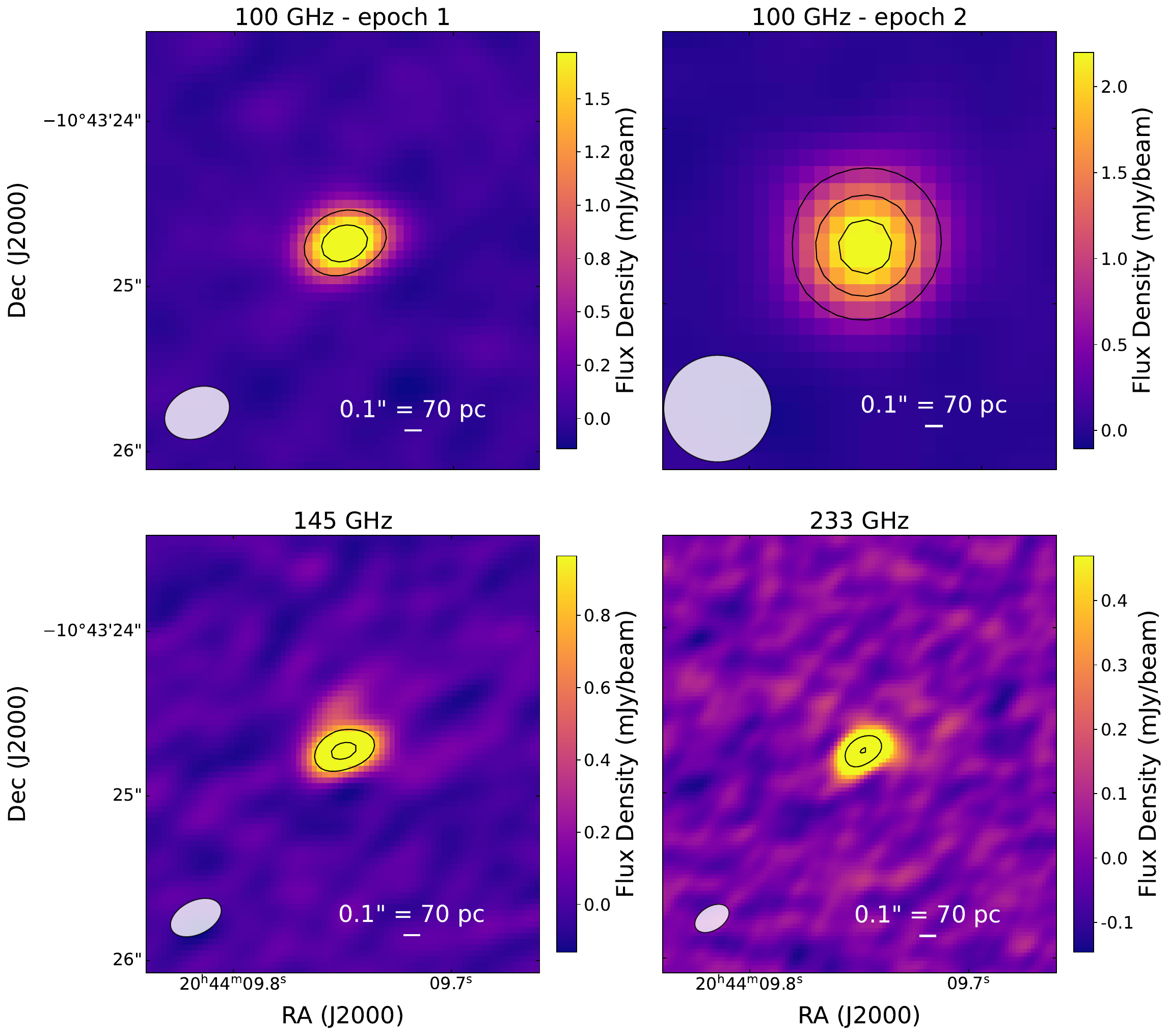}
        \caption{\textbf{Mrk\,509}: $\mathrm{N}\times 20\sigma_{\rm rms}$ level contours.}
        \label{fig:alma_mrk509}
    \end{subfigure}

    \vspace{0.1cm}

    \begin{subfigure}[b]{0.49\textwidth}
        \includegraphics[width=\textwidth]{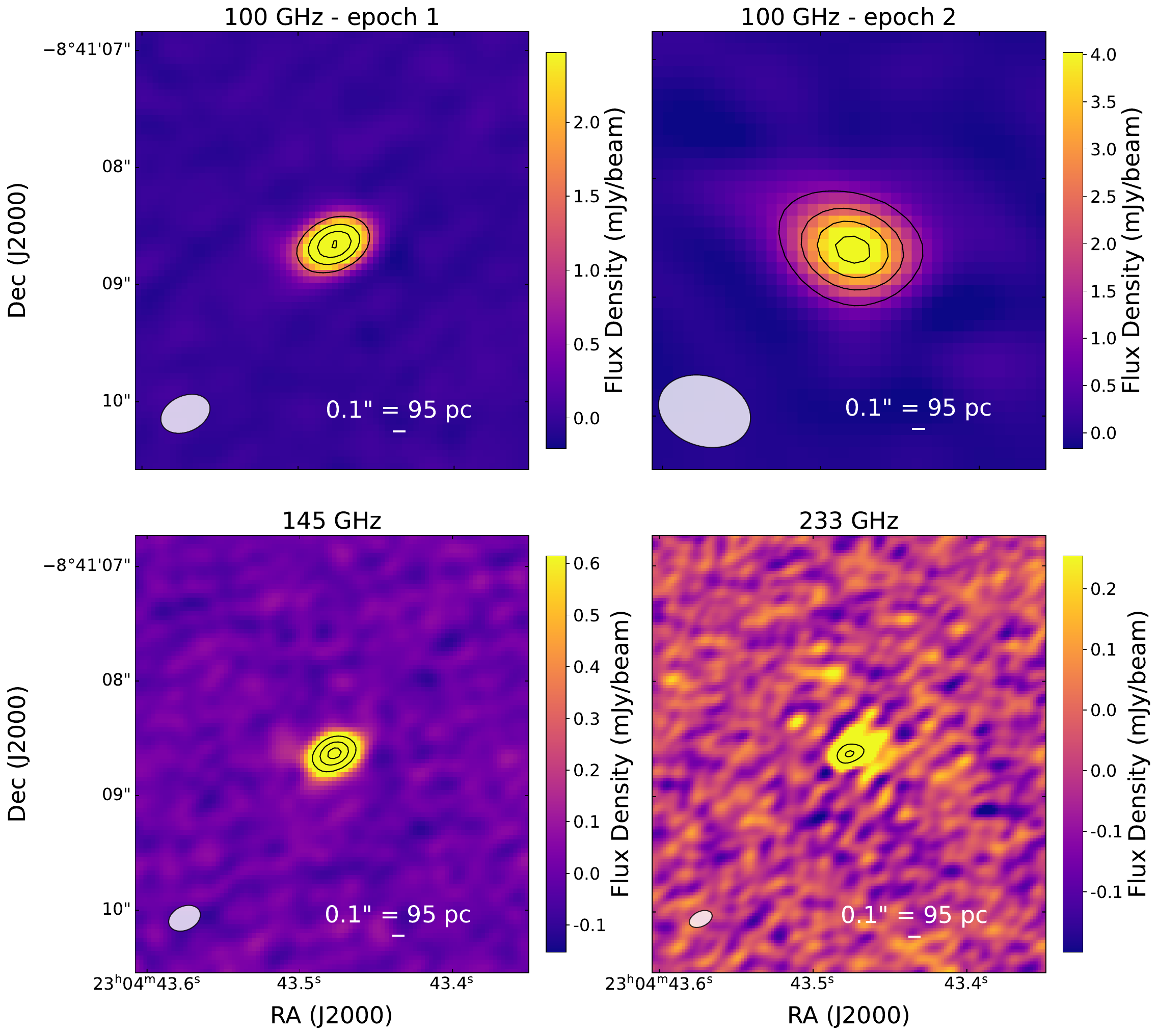}
        \caption{\textbf{Mrk\,926}: $\mathrm{N}\times 20\sigma_{\rm rms}$ level contours.}
        \label{fig:alma_mrk926}
    \end{subfigure}
    \hfill
    \begin{subfigure}[b]{0.49\textwidth}
        \includegraphics[width=\textwidth]{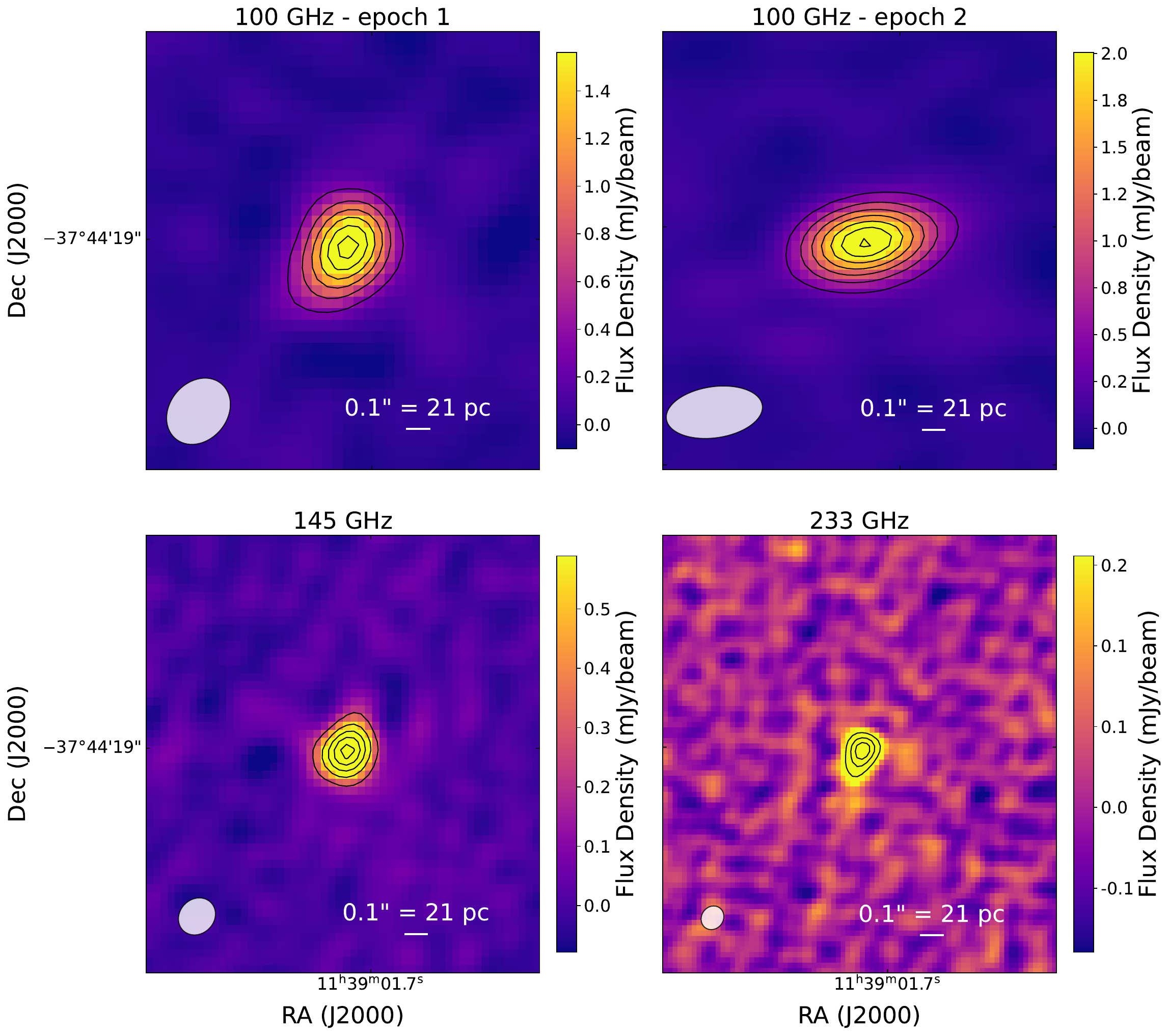}
        \caption{\textbf{NGC\,3783}: $\mathrm{N}\times 10\sigma_{\rm rms}$ level contours.}
        \label{fig:alma_ngc3783}
    \end{subfigure}
    \caption{(Part 2 of 3) ALMA 100\,GHz (two epochs), 145\,GHz, and 233\,GHz images. Contours are overlaid. The beam is shown in the lower-left corner of each panel.}
    \label{fig:alma_images}
\end{figure*}
\begin{figure*}[ht!]
    \ContinuedFloat
    \centering   
    \begin{subfigure}[b]{0.49\textwidth}
        \includegraphics[width=\textwidth]{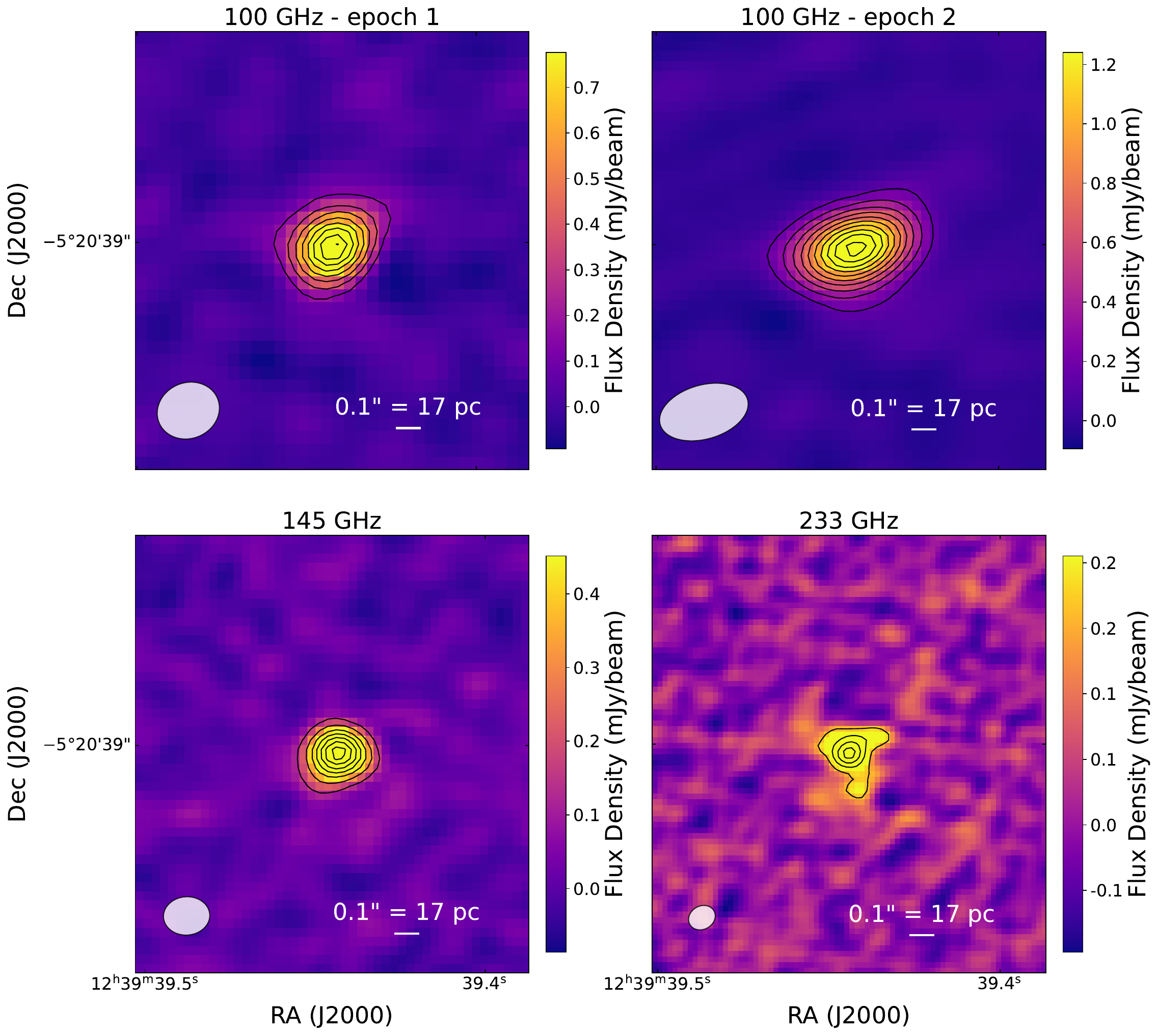}
        \caption{\textbf{NGC\,4593}: $\mathrm{N}\times 10\sigma_{\rm rms}$ level contours.}
        \label{fig:alma_ngc4593}
    \end{subfigure}
    \caption{(Part 3 of 3) ALMA 100\,GHz (two epochs), 145\,GHz, and 233\,GHz images. Contours are overlaid. The beam is shown in the lower-left corner of each panel.}
    \label{fig:alma_images_cont}
\end{figure*}

\newpage

Figure\,\ref{fig:alma_images_cont} presents the continuum images obtained after the \textsc{CASA} cleaning procedure described in Section\,\ref{ALMAImaging}, including the respective beams (as presented in Tables\,\ref{tab:originalbeam_3} and \ref{tab:originalbeam_4_6}) and contours based on the rms (as presented in Tables\,\ref{tab:originalbeam_3} and \ref{tab:originalbeam_4_6} as well).

\newpage

Furthermore, as discussed in Section\,\ref{results_var}, we analyzed the calibrators associated with sources showing significant variability to verify that the variability is intrinsic to our target sources.
In Figure\,\ref{fig:calibrators_3sources}, we show the bandpass calibrators, phase calibrators, and check sources for the three variable sources. In each panel, the flux densities are normalized to the epoch\,1 value of the corresponding source, so that any change in flux density between the two epochs of 100 GHz observations is directly visible. 
All four spectral windows are displayed.
For AX\,J1737.4--2907, an equivalent presentation has been reported previously by \cite{michiyama_alma_2024} using the same methodology.

\begin{figure*}[hb!]
    \centering
    \begin{subfigure}[b]{0.49\textwidth}
        \includegraphics[width=\textwidth]{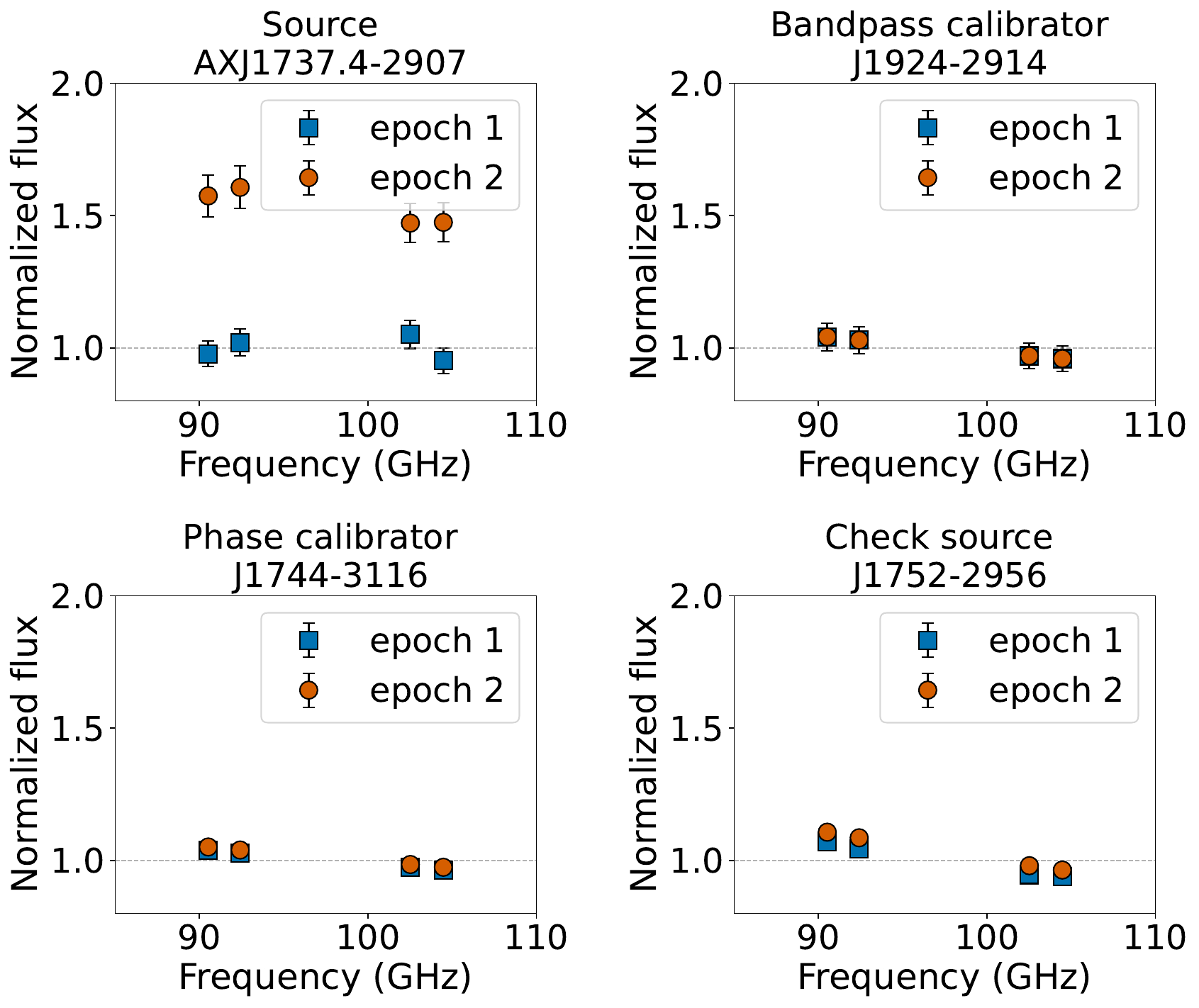}
        \caption{\textbf{AX\,J1737.4--2907}}
        \label{fig:calibrators_AX}
    \end{subfigure}
    \vspace{0.1cm}
    \begin{subfigure}[b]{0.49\textwidth}
        \includegraphics[width=\textwidth]{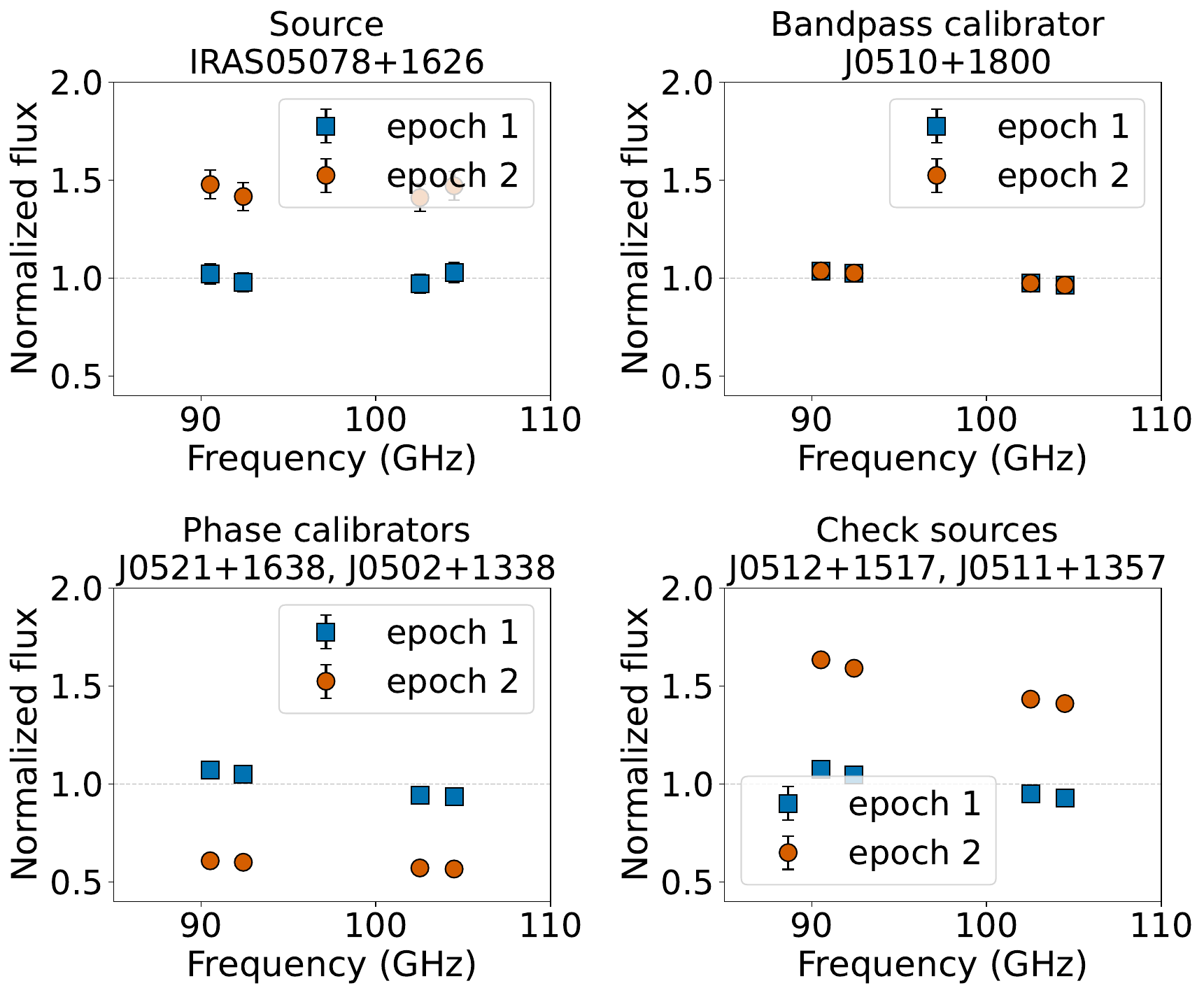}
        \caption{\textbf{IRAS\,05078+1626}}
        \label{fig:calibrators_IRAS}
    \end{subfigure}
    \caption{(Part 1 of 2) Normalized flux densities (relative to epoch\,1) of the target, bandpass calibrator, phase calibrator, and check source for AX\,J1737.4--2907, IRAS\,05078+1626, and NGC\,4593 in the two 100\,GHz epochs, for all four spectral windows. For IRAS\,05078+1626, the phase calibrators and check sources differ between epochs.}
    \label{fig:}
\end{figure*}
\begin{figure*}[ht!]
    \ContinuedFloat
    \centering
    \begin{subfigure}[b]{0.49\textwidth}
        \includegraphics[width=\textwidth]{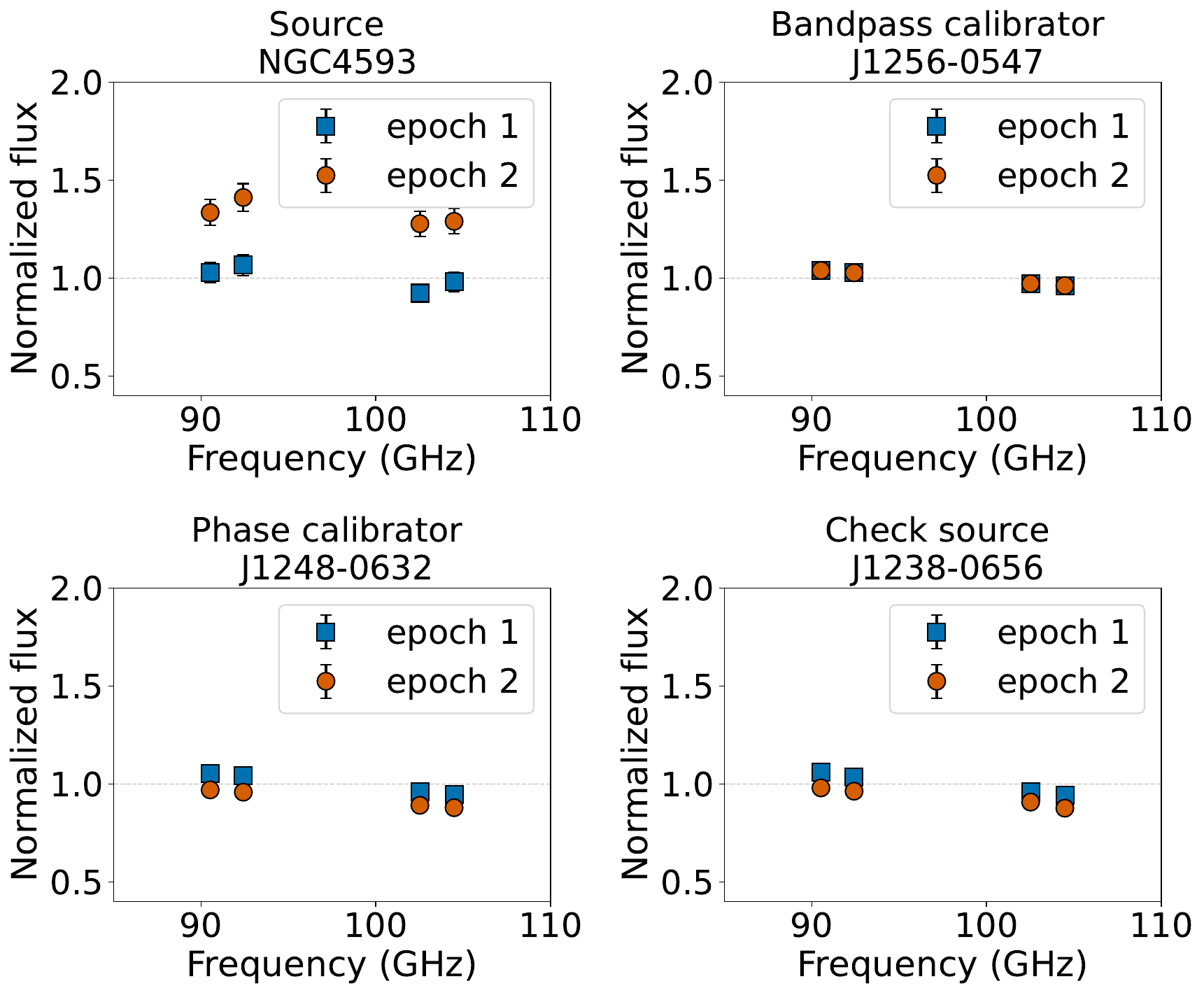}
        \caption{\textbf{NGC\,4593}}
        \label{fig:calibrators_NGC4593}
    \end{subfigure}
\caption{(Part 2 of 2) Normalized flux densities (relative to epoch\,1) of the target, bandpass calibrator, phase calibrator, and check source for AX\,J1737.4--2907, IRAS\,05078+1626, and NGC\,4593 in the two 100\,GHz epochs, for all four spectral windows. For IRAS\,05078+1626, the phase calibrators and check sources differ between epochs.}
\label{fig:calibrators_3sources}
\end{figure*}

For both AX\,J1737.4--2907 and NGC\,4593, the normalized flux densities increase between the two epochs, whereas the calibrators and check sources remain stable to within a few percent. This indicates that the observed variability is intrinsic to the sources.
For IRAS\,05078$+$1626, the normalized flux density increases by approximately 44\% from epoch\,1 to epoch\,2. Unlike the other two sources, the phase calibrators (J0521$+$1638, J0502$+$1338) and check sources (J0512$+$1517, J0511+1357) are different quasars in the two epochs, so the offsets in these panels reflect intrinsic flux differences between objects rather than variability. We therefore compared the measured flux density of each calibrator and check source with its monitored value in the ALMA Calibrator Source Catalogue\footnote{For the ALMA Calibrator Source Catalogue, see \url{https://almascience.eso.org/sc/}} near the corresponding observation date. All agree within the 5\% ALMA flux calibration uncertainty, indicating that the flux density increase in IRAS\,05078$+$1626 is intrinsic to the source.

\section{Molecular gas in Ark\,120}\label{AppMol}

\begin{figure}[h!]
    \centering
    \includegraphics[width=0.9\columnwidth]{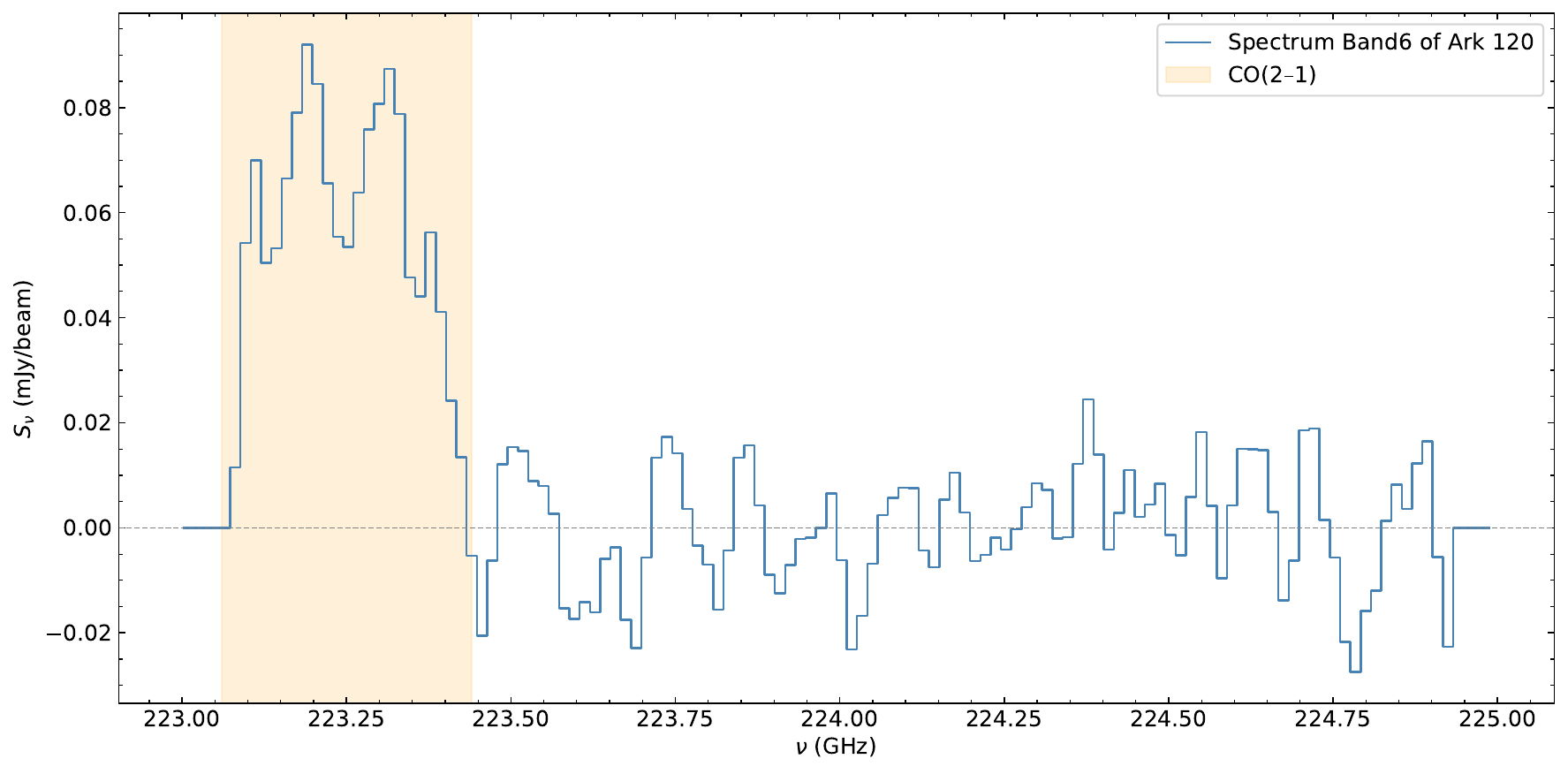}
    \caption{The continuum-subtracted spectrum of the ALMA 233\,GHz observation of Ark\,120 showing the presence of CO(2--1) emission in the form of two peaks. The region selected to subtract the spectrum has radii of $3.89 \arcsec \times3.59 \arcsec$, corresponding to $\rm 2626\,pc\times2423\,pc$}
    \label{fig:CO}
\end{figure}

The 233\,GHz ALMA observation of Ark\,120 showed an unexpectedly high flux density in its first SPW, which we determined was due to the presence of molecular gas.
Figure\,\ref{fig:CO} presents the spectrum of the first SPW in 233\,GHz, clearly showing the molecular line emission, which we determined to be CO(2--1). 
CO(2--1) has a rest frequency of $\nu^{\rm rest}_{\rm CO(2-1)} = 230.538$\,GHz, which means an observing frequency of $\nu^{\rm obs}_{\rm CO(2-1)}=\nu^{\rm rest}_{\rm CO(2-1)}/(1+z)=223.173$\,GHz in Ark\,120 ($z=0.033$).
We observe two peaks, suggesting a rotating disk or outflow of gas.

To estimate the molecular gas mass associated with this detection, we follow the standard framework for converting the velocity-integrated CO(2--1) line flux into the CO(2--1) luminosity, then into the CO(1--0) luminosity and, subsequently, into the $\rm H_2$ gas mass \citep[e.g.,][]{Solomon2005,Bolatto2013}.
We first determine the integrated flux density of the two peaks in Jy\,km\,s$^{-1}$ by fitting two Gaussians to the peaks, we obtain a value of $S_{\rm CO(2-1)}\textrm{d}v = 0.113 \pm 0.015$\,Jy\,beam$^{-1}$\,km\,s$^{-1}$.
This we multiply by the number of beams in the selected circular region in which the spectrum is taken, which has a radius of 0.294\,\arcsec ($\simeq 200$\,pc).
Then, we obtain the line luminosity through:
\begin{equation}
    \left( \frac{L'_{\rm CO(2-1)}}{{\rm K\,km\,s^{-1}\,pc^2}} \right) =  3.25\times10^{7}\,
    \left(\frac{S_{\rm CO(2-1)}\Delta v}{{\rm Jy\,km\,s^{-1}}}\right)
    \left(\frac{\nu_{\rm obs}}{{\rm GHz}}\right)^{-2}
    \left(\frac{D_{\rm L}}{{\rm Mpc}}\right)^{2}
    (1+z)^{-3},
    \label{eq:COlum}
\end{equation}
\citep[e.g.][]{Solomon2005}.
We obtain $L'_{\rm CO(2-1)}= 1.5\times 10^6\,\rm K\,km\,s^{-1}\,pc^2$, which can then be converted to $\rm H_2$ gas mass by obtaining the CO(1--0) line luminosity and, subsequently, converting this to $\rm H_2$ gas mass.
\begin{equation}
    M_{\rm gas} = \alpha_{\rm CO}\, L'_{\rm CO(1-0)}
     =  \alpha_{\rm CO}\, \frac{L'_{\rm CO(2-1)}}{R_{21}},
    \label{eq:Mgas}
\end{equation}
Here, we adopt values for the assumed CO(1--0)-to-H$_2$ conversion factor of $\alpha_{\rm CO}=2 \times 10^{20}\rm \,mol \,cm^{-2}(K \,km\,s^{-1})^{-1}$ \citep{Bolatto2013} and the CO(2--1)-to-CO(1--0) line ratio of $R_{21}=2.5$ \citep{Viti2014}, as described by e.g., \cite{GarciaBurillo2024}.
We find $M_{\rm gas}=14\times10^6\,M_{\odot}$ and a gas surface density of $\log(\Sigma_{\rm gas}/M_{\odot}\,\rm pc^2)=2.05$.
This value for $\log(\Sigma_{\rm gas})$ is in a similar range, $\log(\Sigma_{\rm gas}/M_{\odot}\,\rm pc^2)=0.80$--2.67 with an average of 2.09, as \cite{GarciaBurillo2024} found for a sample of 23 AGN with $\log(\lambda_{\rm Edd})<-1$.

\section{$\alpha_{\rm mm}$ versus AGN/host galaxy properties}\label{alpha_vs}

\begin{figure*}[h!]
    \centering
    \includegraphics[width=0.74\linewidth]{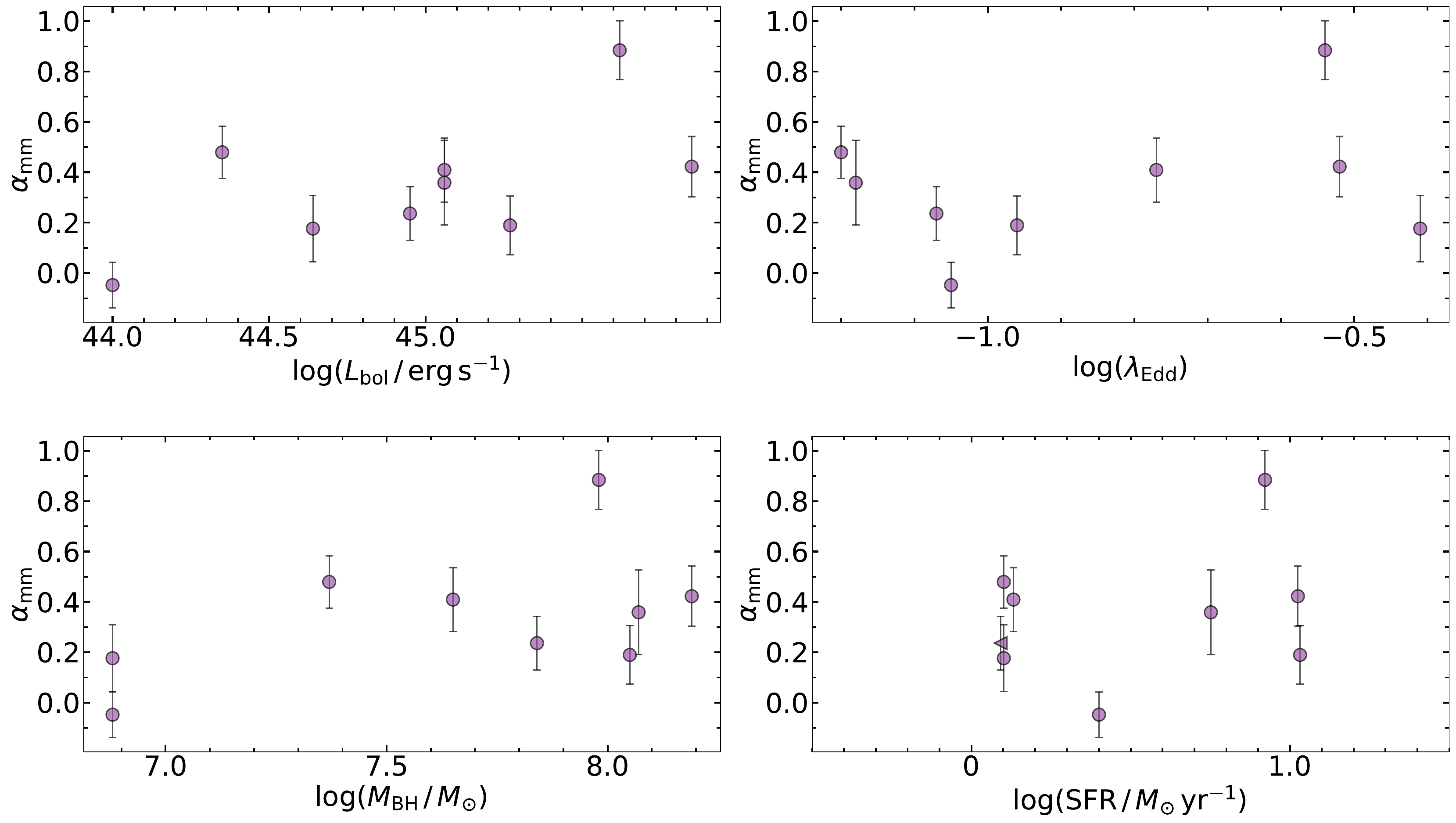}
    \caption{The spectral index ($\alpha_{\rm mm}$) versus $L_{\rm bol}$, $\lambda_{\rm Edd}$, $M_{\rm BH}$ and SFR for the sources in our sample. The values of $L_{\rm bol}$, $\lambda_{\rm Edd}$, $M_{\rm BH}$ and SFR are presented in Table\,\ref{tab:Sources2}. 
    The values for $\alpha_{\rm mm}$, which were obtained through fitting $S_{\nu}\propto\nu^{-\alpha}$ to the 100--233\,GHz ALMA data, is presented in Table\,\ref{tab:alpha}.
    We do not observe strong correlation between $\alpha_{\rm mm}$ and the specified parameters, obtaining $p$--values of 0.16, 0.89, 0.25 and 0.79, respectively.}
    \label{fig:alpha_vs_params}
\end{figure*}

In Figure\,\ref{fig:alpha_vs_params}, we present the millimeter spectral index ($\alpha_{\rm mm}$) versus $L_{\rm bol}$, $\lambda_{\rm Edd}$, $M_{\rm BH}$ and SFR.

\newpage
As described in Section\,\ref{synchr_origin} and as can be seen in all four figures, we do not observe any correlation between $\alpha_{\rm mm}$ and the parameters, obtaining $p$--values of 0.16, 0.89, 0.25 and 0.79, respectively. 
Future studies with larger samples will be needed to confirm the absence of these correlations.





\section{Intra- vs broadband spectral index comparison}\label{intrabroad}

\begin{figure*}[!b]
    \centering
    \begin{subfigure}[b]{0.48\textwidth}
        \includegraphics[width=\textwidth]{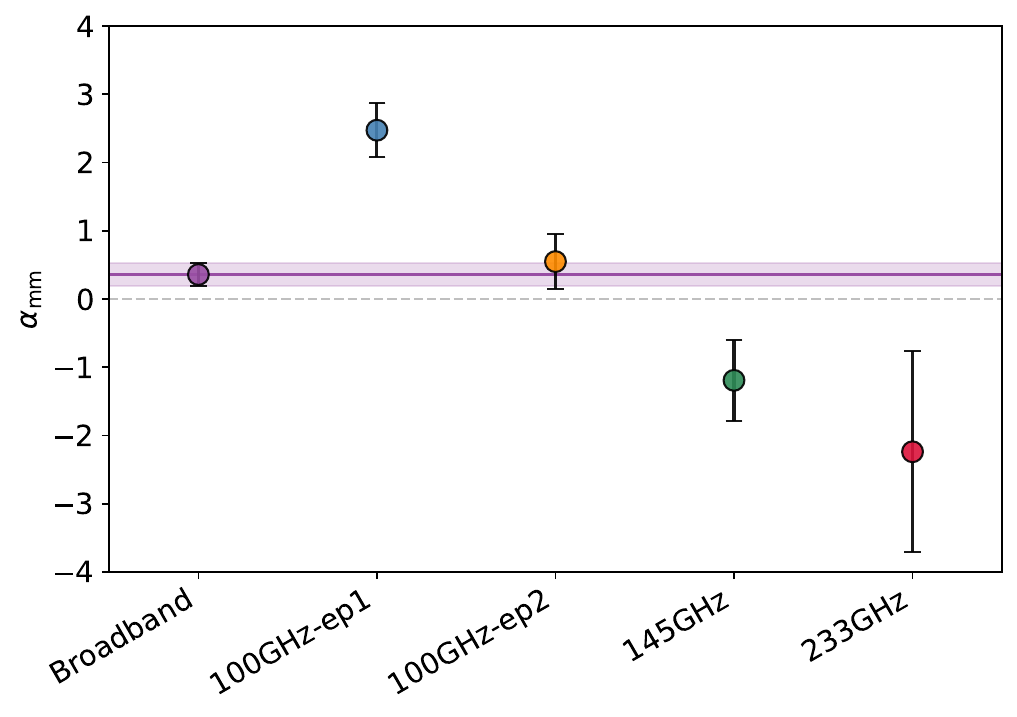}
        \caption{Ark\,120 }
        \label{fig:alpha_ark120}
    \end{subfigure}
    \hfill
    \begin{subfigure}[b]{0.48\textwidth}
        \includegraphics[width=\textwidth]{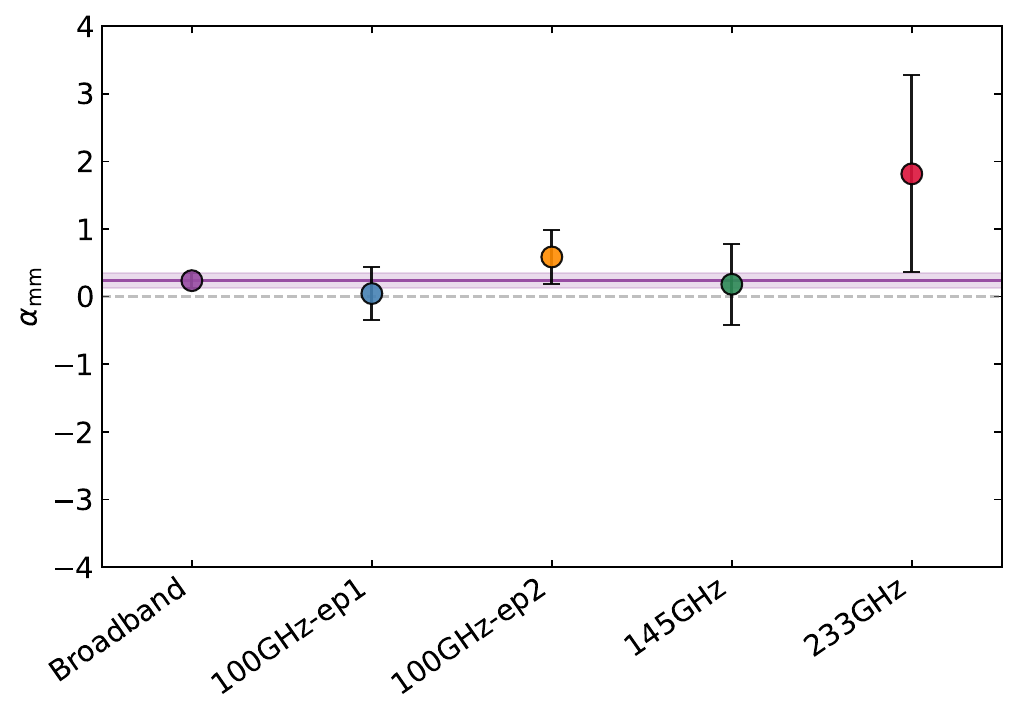}
        \caption{AX\,J1737.4--2907 }
        \label{fig:alpha_ax}
    \end{subfigure}
\vspace{0.5cm}
    \begin{subfigure}[b]{0.48\textwidth}
        \includegraphics[width=\textwidth]{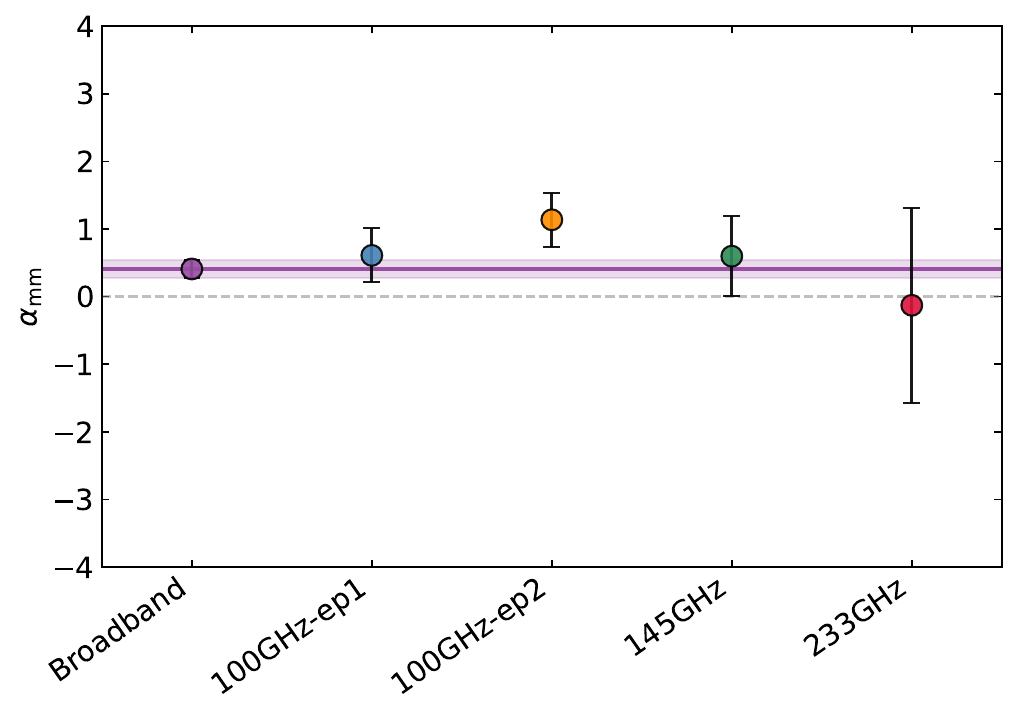}
        \caption{IC\,4329A}
        \label{fig:alpha_ic4329a}
    \end{subfigure}
    \hfill
    \begin{subfigure}[b]{0.48\textwidth}
        \includegraphics[width=\textwidth]{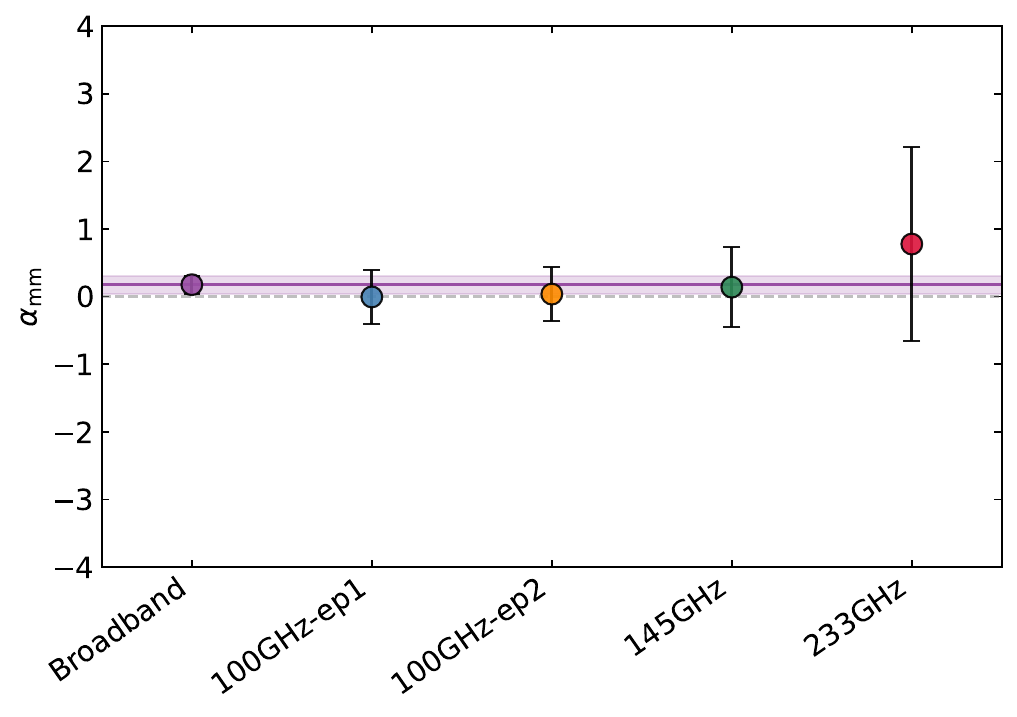}
        \caption{IRAS\,05078+1626}
        \label{fig:alpha_iras}
    \end{subfigure}
    \caption{(Part 1 of 2) Comparison of the spectral index ($\alpha_{\rm mm}$) obtained through broadband (100--233\,GHz) fitting and the intraband spectral indices. These spectral indices are based on the flux densities from the observations before beam matching and uv-clipping, as described in Section\,\ref{ALMAImaging}.}
\end{figure*}
\begin{figure*}[!p]
    \ContinuedFloat
    \centering
    \begin{subfigure}[b]{0.49\textwidth}
        \includegraphics[width=\textwidth]{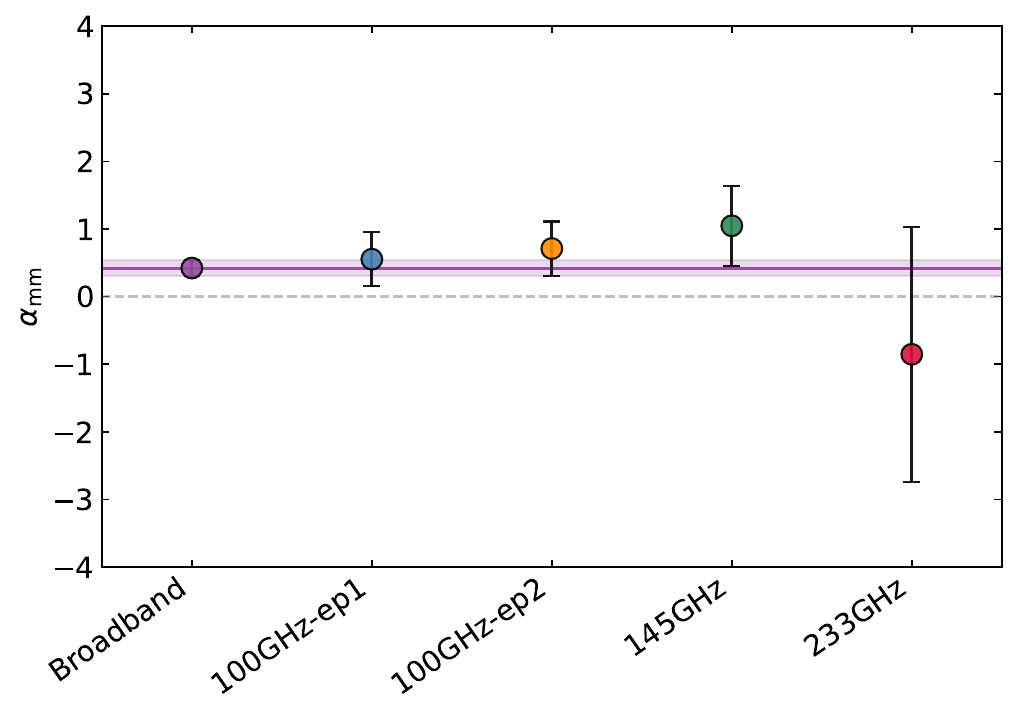}
        \caption{MR\,2251--178}
        \label{fig:alpha_mr2251}
    \end{subfigure}
    \hfill
    \begin{subfigure}[b]{0.49\textwidth}
        \includegraphics[width=\textwidth]{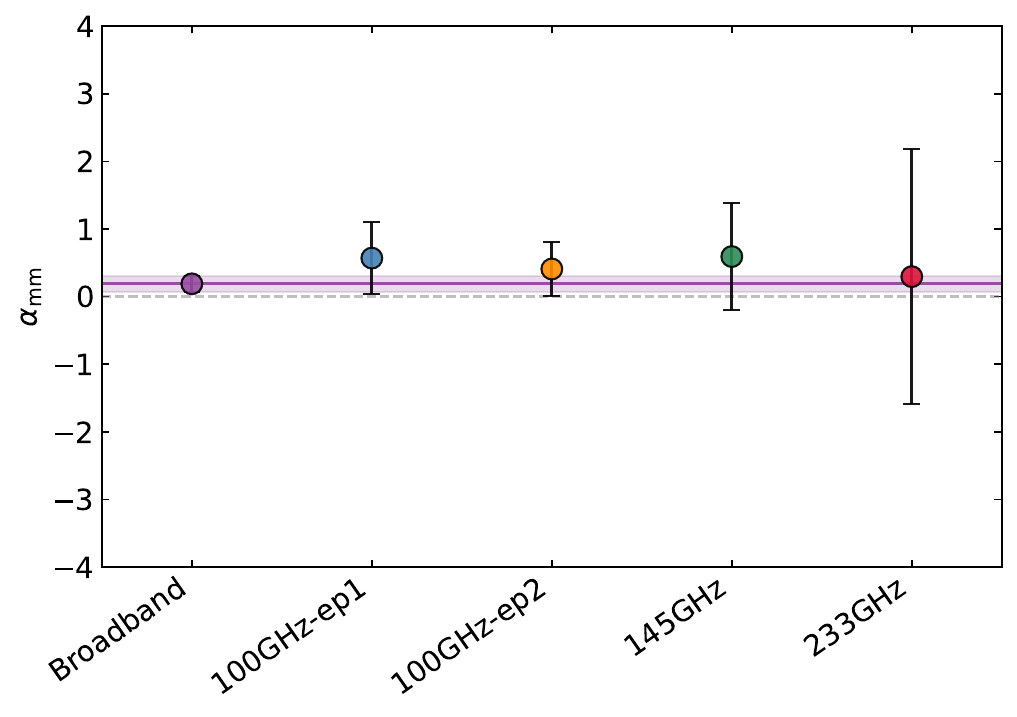}
        \caption{Mrk\,509}
        \label{fig:alpha_mrk509}
    \end{subfigure}
\vspace{0.5cm}
    \begin{subfigure}[b]{0.49\textwidth}
        \includegraphics[width=\textwidth]{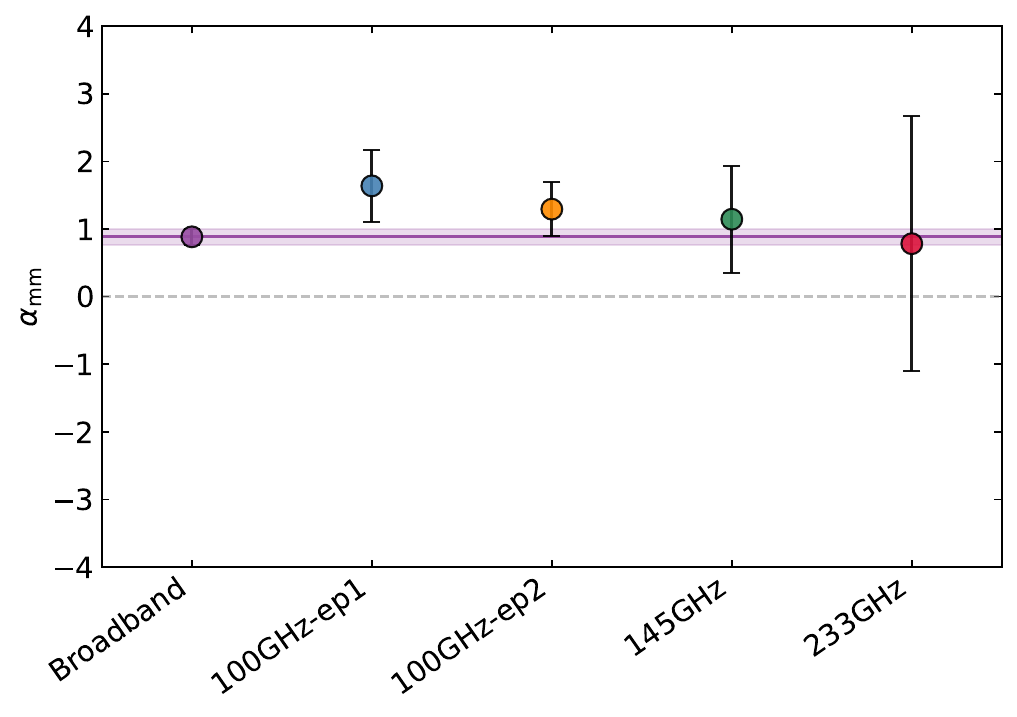}
        \caption{Mrk\,926}
        \label{fig:alpha_mrk926}
    \end{subfigure}
    \hfill
    \begin{subfigure}[b]{0.49\textwidth}
        \includegraphics[width=\textwidth]{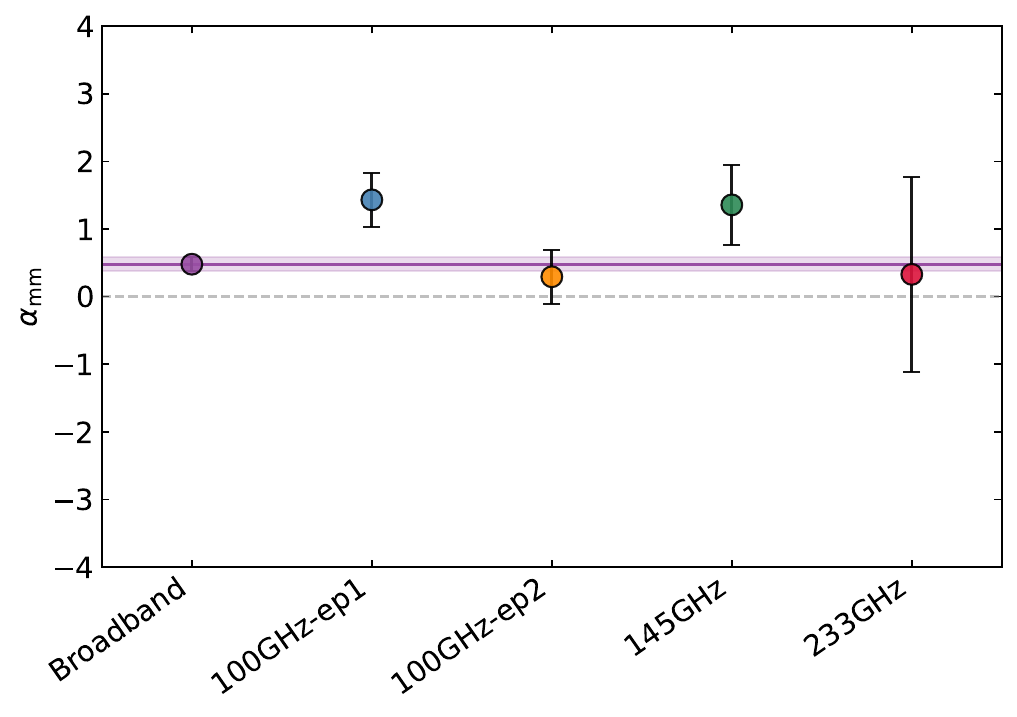}
        \caption{NGC\,3783}
        \label{fig:alpha_ngc3783}
    \end{subfigure}
\vspace{0.5cm}
    \begin{subfigure}[b]{0.49\textwidth}
        \includegraphics[width=\textwidth]{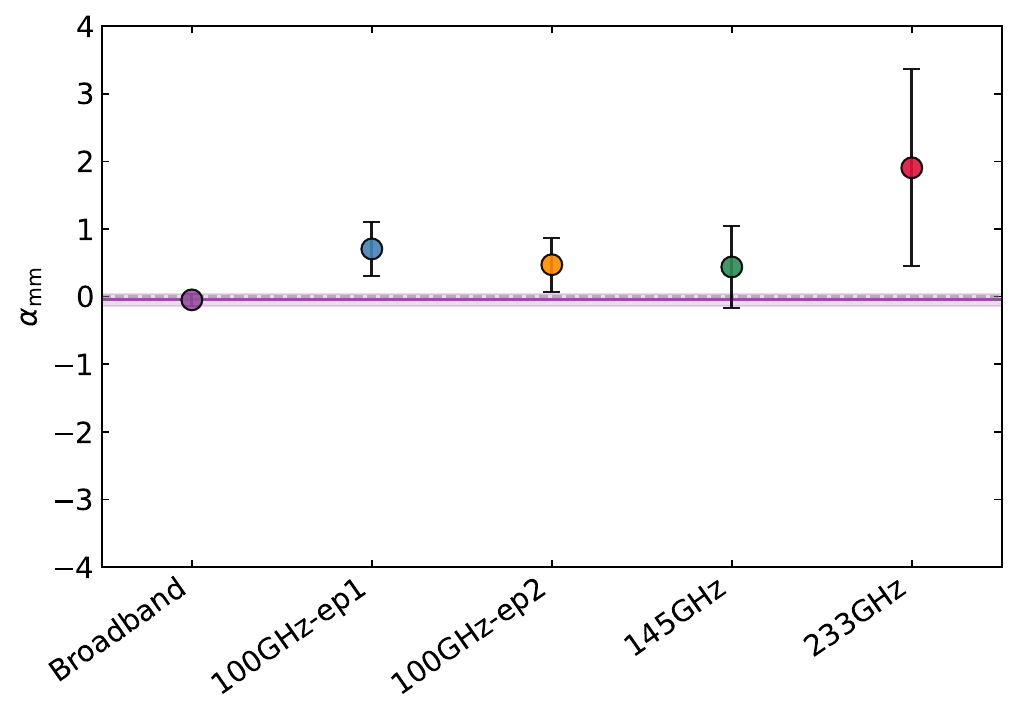}
        \caption{NGC\,4593}
        \label{fig:alpha_ngc4593}
    \end{subfigure}
    \caption{(Part 2 of 2) Comparison of the spectral index ($\alpha_{\rm mm}$) obtained through broadband (100--233\,GHz) fitting and the intraband spectral indices. These spectral indices are based on the flux densities from the observations before beam matching and uv-clipping, as described in Section\,\ref{ALMAImaging}.}
    \label{fig:spectralindex-comparison}
\end{figure*}

Intraband spectral indices ($\alpha_{\rm intraband}$) have often been used to determine the origin of the millimeter emission, though they generally carry large uncertainties \citep[e.g.,][]{kawamuro_bass_2022,shablovinskaya_joint_2024,Venselaar26}. 
When multi-band data are unavailable, these studies estimate the spectral index from the four SPWs within one single ALMA band. 
It has remained unclear, however, whether $\alpha_{\rm intraband}$ accurately represents the broadband index $\alpha_{\rm broadband}$ ($=\alpha_{\rm mm}$ in this work), since the limited frequency coverage of a single band and the scatter within individual SPWs can introduce significant uncertainties \citep{Francis2020}. 
To quantify the consistency between the two, we fitted a power-law across the four SPWs of each band, obtaining the $\alpha_{\rm intraband}$ at 100, 145 and 233\,GHz for each source. 
The $\alpha_{\rm intraband}$ values were derived without matching the angular resolutions between bands. 
Figure\,\ref{fig:spectralindex-comparison} compares the four intraband slopes with the broadband slope for each RQ AGN in our sample. 
From Figure\,\ref{fig:spectralindex-comparison}, it can be seen that the broadband uncertainties are significantly smaller, and most intraband values agree with the broadband value within their uncertainties. 
The observations exhibiting the largest discrepancies between $\alpha_{\rm broadband}$ and $\alpha_{\rm intraband}$ are Ark\,120, AX\,J1737.4--2907, NGC\,3783 and NGC\,4593 as illustrated in Figure\,\ref{fig:spectralindex-comparison}.
For example, in the 100\,GHz epoch\,1 data of Ark\,120, this is largely due to extensive flagging in the last SPW, where 35 of the 46 antennas were completely flagged, making the derived slope appear significantly steeper. 
In the 233\,GHz observation, the second SPW has significantly lower fluxes, which steepens the slope in the opposite direction.

\begin{figure}[ht!]
    \centering
    \includegraphics[width=0.5\columnwidth]{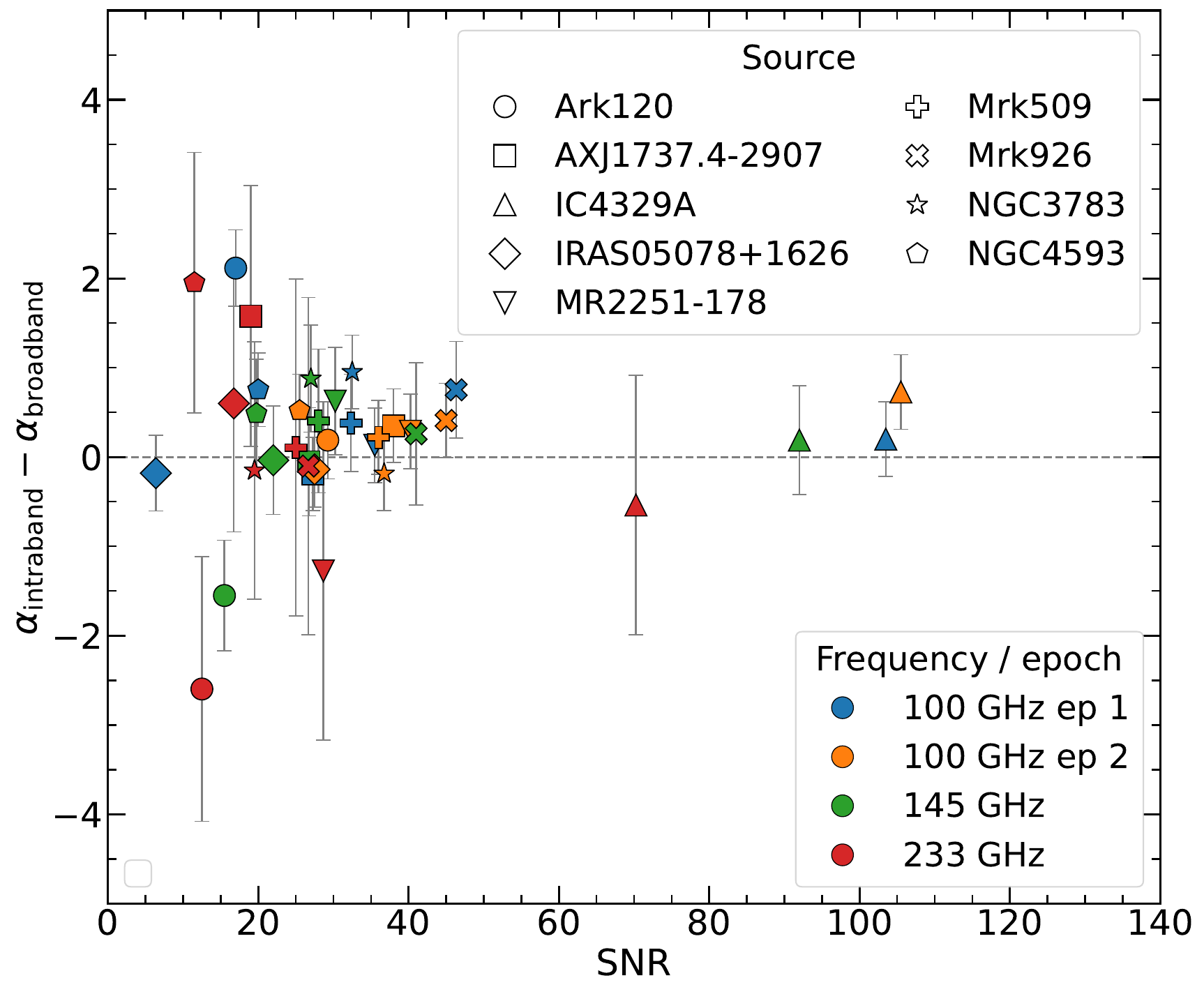}
    \caption{Scatter plot showing the difference ($\Delta\alpha$) between the intraband spectral index ($\alpha_{\rm intraband}$) and the broadband spectral index ($\alpha_{\rm broadband}$) for all sources and for all three ALMA frequencies and a second 100\,GHz epoch, plotted as a function of the mean SNR of all SPWs at each given frequency.}
    \label{fig:SNR_alpha}
\end{figure}

Furthermore, Figure\,\ref{fig:SNR_alpha} shows the difference between $\alpha_{\rm broadband}$ and $\alpha_{\rm intraband}$ as a function of the average SNR of the three or four SPWs measured at each frequency/epoch for each source. 
As illustrated in the figure, observations with a mean SNR below $\sim25$ exhibit a generally larger dispersion in $\Delta\alpha=\alpha_{\rm intraband}-\alpha_{\rm broadband}$. 
At $\rm SNR<25$, we obtain a median of $\Delta\alpha=0.49 \pm 0.47$, while at $\rm SNR>25$ this equals $\Delta\alpha=0.24 \pm 0.08$; the two values are statistically consistent, indicating that the effect of low SNR is to inflate mainly the scatter in $\Delta\alpha$.
This indicates that low SNR may affect the reliability of $\alpha_{\rm intraband}$. Consequently, caution is warranted when interpreting spectral indices derived from a single band.

\section{The Millimeter/X-ray relation}\label{AppmmXray}

\begin{figure*}[ht!]
    \centering
    \begin{subfigure}[b]{0.99\textwidth}
        \centering
        \includegraphics[width=\textwidth]{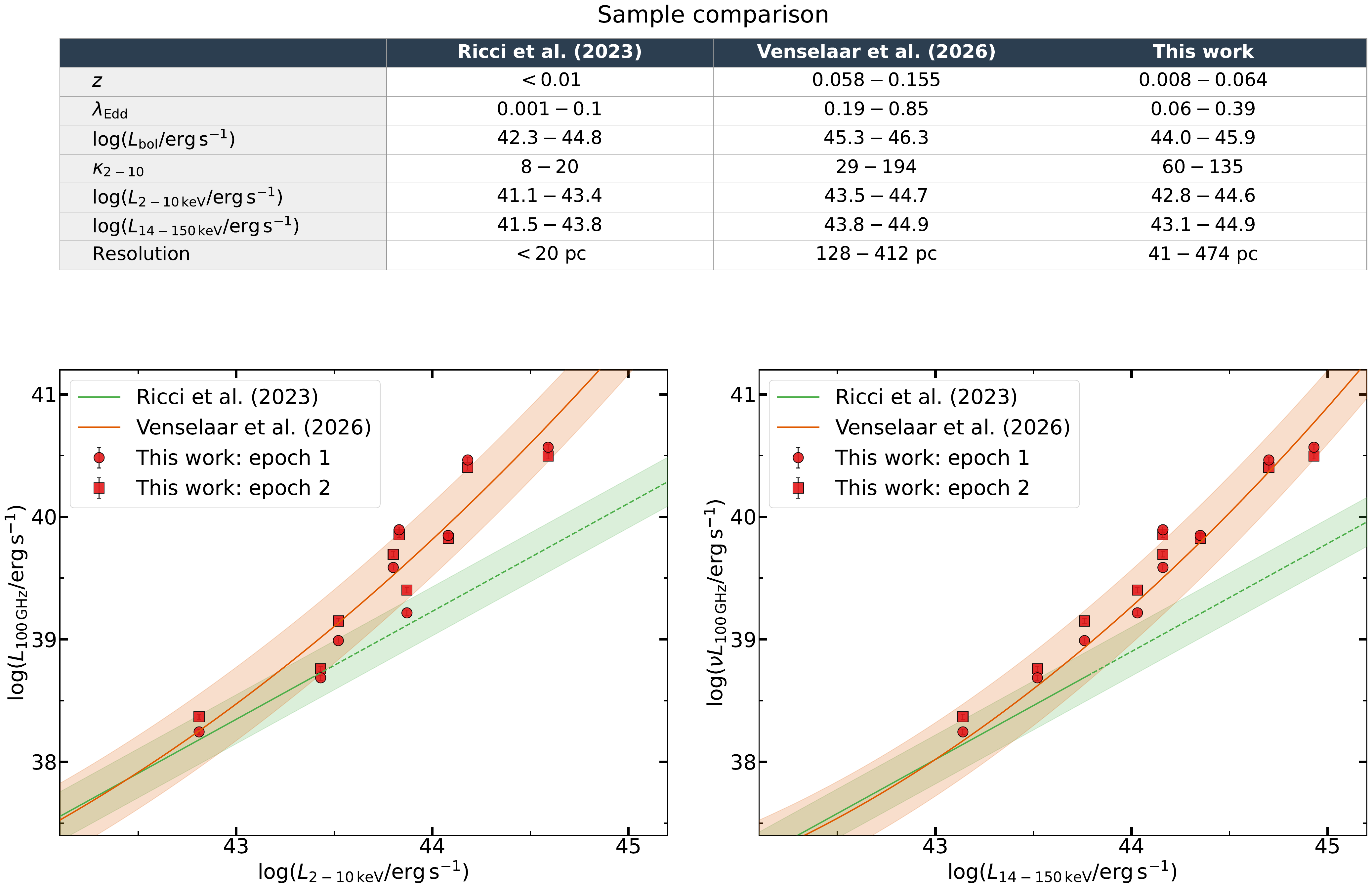}
        \caption{
        The 100\,GHz millimeter emission versus 2--10\,keV and 14--150\,keV X-ray emission. The relations derived by \cite{ricci_tight_2023} and \cite{Venselaar26} are shown for comparison. 
        Beyond the highest X-ray luminosity probed by \cite{ricci_tight_2023}, we extrapolate their relation (dashed line), which reveals a clear deviation from the \cite{Venselaar26} relation at higher luminosities. 
        Both epochs of our 100\,GHz observations are overplotted on these relations, and our sources are seen to follow the established trend closely. A table comparing the key parameters of the three samples is presented below.}
        \label{fig:band3}
    \end{subfigure}
    \begin{subfigure}[b]{0.95\textwidth}
        \centering
        \includegraphics[width=\textwidth]{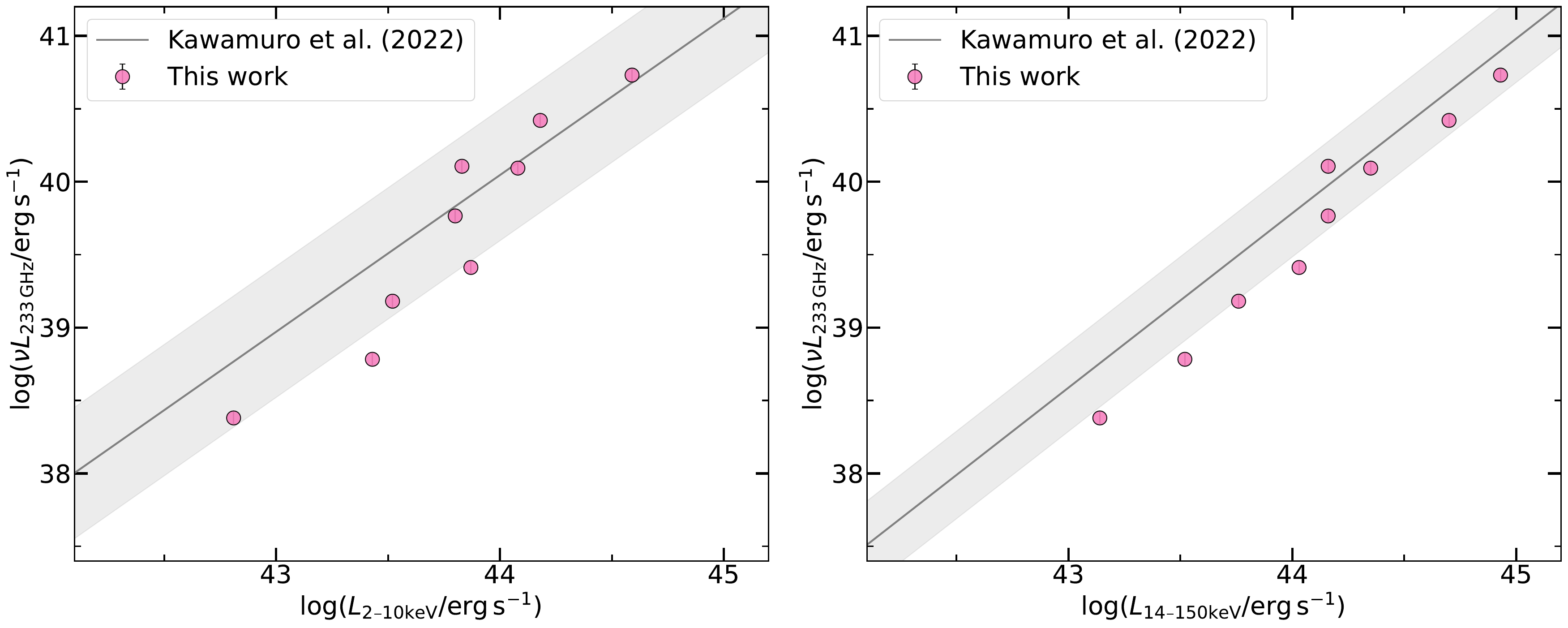}
        \caption{The 233\,GHz millimeter emission versus 2--10\,keV and 14--150\,keV X-ray emission. The relations derived by \cite{kawamuro_bass_2022} are shown for comparison. This ALMA data of our sources was part of the sample used by \cite{kawamuro_bass_2022} to determine these relations. }
        \label{fig:band6}
    \end{subfigure}

    \caption{The millimeter/X-ray relation for the three ALMA bands, spanning the frequencies 100\,GHz, 145\,GHz and 233\,GHz, versus 2--10\,keV and 14--150\,keV X-ray emission. Included are the relations obtained by \cite{ricci_tight_2023} and \cite{Venselaar26} at 100\,GHz, and by \cite{kawamuro_bass_2022} at 233\,GHz. We fitted new relations to the 145\,GHz versus X-ray emission, presented in Equations\,\ref{eq:L145_210} and \ref{eq:L145_14150}.}
    \label{fig:mmXray}
\end{figure*}

In Section\,\ref{mmXray}, we examined whether our sample is consistent with the millimeter/X-ray correlations previously established at 100\,GHz \citep{ricci_tight_2023,Venselaar26} and 233\,GHz \citep{kawamuro_bass_2022}.
Figure\,\ref{fig:mmXray} shows our measurements at 100\,GHz, 145\,GHz, and 233\,GHz overlaid on these existing relations.

At 100\,GHz (Figure\,\ref{fig:band3}), our sources trace the transition between the relations of \cite{ricci_tight_2023} and \cite{Venselaar26}, which differ as a function of $\log(L_{\rm bol})$, $\lambda_{\rm Edd}$, and $\kappa_{2-10}$; these differences are summarized in the table below.
In Figure\,\ref{fig:band4}, we provide the first relation between 145\,GHz millimeter emission and X-ray emission, obtaining a linear relation of the form as presented in Equations\,\ref{eq:L145_210} and \ref{eq:L145_14150}.
For completeness, we also show our 233\,GHz data in Figure\,\ref{fig:band6}, overlaid with the relation from \cite{kawamuro_bass_2022}. We note, however, that these particular ALMA observations were already used in establishing this 233\,GHz/X-ray relation.

\end{document}